\documentclass[reprint, superscriptaddress, amssymb, aps, prx]{revtex4-2}

\usepackage{graphicx}
\usepackage{mathtools}
\usepackage{dcolumn}
\usepackage{bm}
\usepackage[colorlinks=true,citecolor=blue,linkcolor=blue]{hyperref}
\usepackage[caption=false]{subfig}
\usepackage{bbm}
\usepackage{pifont}

\graphicspath{ {figures/}{figures/model_illus/}{figures/RG_numerics/} }

\renewcommand{\thesection}{\Roman{section}}
\renewcommand{\thesubsection}{\thesection.\Alph{subsection}}
\renewcommand{\thesubsubsection}{\arabic{subsubsection}}
\makeatletter
\def\p@subsection{}
\makeatother

\renewcommand{\vec}[1]{\boldsymbol{\mathbf{#1}}}

\global\long\def\diff{\mathop{}\!{\operatorfont d}}
\global\long\def\sgn{\operatorname{sgn}}
\global\long\def\diag{\operatorname{diag}}
\global\long\def\transpose{\top}

\newcommand*{\elemdontcare}{\ensuremath{\text{\scriptsize{\ding{54}}}}}

\newcommand*{\sump}{} 
\DeclareRobustCommand*{\sump}{
    \mathop{{\sum}^{\mathrlap{\prime}}}
}
\begin{document}

\title{Superconducting fractional quantum Hall edges via repulsive interactions}

\author{Barak A. Katzir}
\affiliation{\mbox{Physics Department, Technion, 320003 Haifa, Israel}}
\author{Ady Stern}
\affiliation{\mbox{Department of Condensed Matter Physics, Weizmann Institute of Science, Rehovot 76100, Israel}}
\author{Erez Berg}
\affiliation{\mbox{Department of Condensed Matter Physics, Weizmann Institute of Science, Rehovot 76100, Israel}}
\author{Netanel H. Lindner}
\affiliation{\mbox{Physics Department, Technion, 320003 Haifa, Israel}}

\date{\today}

\begin{abstract}
We study proximity coupling between a superconductor and counter-propagating gapless modes arising on the edges of Abelian fractional quantum Hall liquids with filling fraction $\nu=1/m$ (with $m$ an odd integer). This setup can be utilized to create non-Abelian parafermion zero-modes if the coupling to the superconductor opens an energy gap in the counter-propagating modes. However, when the coupling to the superconductor is weak an energy gap is opened only in the presence of sufficiently strong attractive interactions between the edge modes, which do not commonly occur in solid state experimental realizations. We therefore investigate the possibility of obtaining a gapped phase by increasing the strength of the proximity coupling to the superconductor. To this end, we use an effective wire construction model for the quantum Hall liquid and employ renormalization group methods to obtain the phase diagram of the system. Surprisingly, at strong proximity coupling we find a gapped phase which is stabilized for sufficiently strong repulsive interactions in the bulk of the quantum Hall fluids. We furthermore identify a duality transformation that maps between the weak coupling and strong coupling regimes, and use it to show that the gapped phases in both regimes are continuously connected through an intermediate proximity coupling regime.
\end{abstract}

\maketitle
    
\section{Introduction}

Topological quantum computation (TQC) utilizes non-local encoding of quantum information in a quantum many-body system. Such encoding protects the information from degradation due to interaction with an environment, and allows for logical operations with topologically robust precision \cite{Kitaev_2003, Nayak_2008}. In particular, TQC can be implemented using topologically ordered phases of matter of two-dimensional (2D) systems \cite{Leinaas_1977, Wilczek_1982, Wu_1984, Stern_2010}, which harbour quasiparticles with non-Abelian exchange statistics, called non-Abelian anyons. In such systems the quantum gates are executed by exchanges of well-separated anyons and their action on the quantum memory depends only on the topology of the space-time paths of the quasiparticles. In this manner the quantum memory and gates are robust to local errors and local decoherence processes, thus enjoying topological protection.

Topologically ordered phases exhibit a wide variety of non-Abelian anyons. The simplest non-Abelian topological order is the Ising topological order, which, along with its simplicity, does not admit a universal set of quantum gates which are topologically protected \cite{Freedman_2006}. The Ising topological order is predicted to occur in the fractional quantum Hall (FQH) effect at filling $\nu=5/2$ \cite{Moore_1991} and in 2D spin systems \cite{Kitaev_2006}. Topologically ordered phases featuring anyons that admit topologically protected universal gate sets have been predicted to occur in other FQH states \cite{Read_1996, Stern_2008}.

Closely related platforms for TQC can be realized using defects in 2D topological phases of matter \cite{Barkeshli_2013}. Certain types of defects are often referred to as ``non-Abelian''. For these types of defects, the ground state degeneracy grows exponentially with the number of defects, and topologically protected unitary operations in the ground state manifold can be performed by adiabatically changing the couplings between the defects. Notably, superconductivity play an important role in the realization of many types of non-Abelian defects, mainly due to the experimental accessibility of superconductors (SC) and the simplicity and success of the theory of superconductivity. The first and simplest examples of these are non-Abelian defects that bind local Majorana zero-modes. Notable predictions of Majorana zero-modes in SC systems are Abrikosov vortex cores of $p+ip$ superconductors \cite{Read_2000, Ivanov_2001}, topological insulators in proximity to superconductors \cite{Fu_2008, Fu_2009, Stanescu_2010} or ends of semiconductor nanowires \cite{Kitaev_2001, Oreg_2010, Lutchyn_2010, Cook_2011}. In recent years, experimental signatures of Majorana zero-modes have been accumulating \cite{Mourik_2012, Rokhinson_2012, Deng_2012, Churchill_2013, Das_2012, Nadj_Perge_2014, Albrecht_2016, Deacon_2017}. While these results are encouraging, the topologically protected transformations supported by Majorana zero-modes are closely related to those enabled by Ising anyons, and thus do not admit a universal set of gates. 
This motivates an ongoing search for TQC platforms which go beyond the Majorana paradigm.

An important route towards this goal involves superconducting defects in Abelian FQH states. Such defects were shown to bind parafermion zero-modes, which can be used to implement a richer set of topologically protected set of gates than their Majorana counterparts \cite{Lindner_2012_parafermions, Clarke_2013, Cheng_2012, Vaezi_2013, Alicea_2016}. For instance, a system of parafermions may implement an entangling gate (analogous to a controlled-NOT gate) using only exchange operations, which is not possible using Majorana zero-modes \cite{Clarke_2013}. While parafermions do not support a universal gate set, they can be used as building blocks for obtaining topologically ordered phases which do support universal TQC \cite{Mong_2014}.

The main ingredient in realizing superconducting defects in Abelian FQH are counter-propagating edge states proximity coupled to a superconductor. Realizing a defect requires that the coupling opens an energy gap in the edge states. The defect consists of a finite segment of induced superconductivity in the edge states, flanked by regions in which the edge states are gapped due to backscattering \cite{Lindner_2012_parafermions, Clarke_2013, Cheng_2012, Vaezi_2013} (which essentially removes the edge states in these regions). Parafermion zero-modes are bound to the ends of the superconducting segment. Recent experimental progress has pursued realizations of superconducting defects in a variety of topological phases: superconductors coupled to two edge states exposed in a trench of integer quantum Hall (IQH) states in graphene \cite{Lee_2017}, graphene bilayers \cite{Sanchez-Yamagishi_2016}, GaAs-heterostructure systems \cite{Ronen_2018} and most recently superconductors coupled to two edge states of FQH states in graphene \cite{Ronen_2020}. 
Other possible routes that utilize the counter-propagating edges have been proposed in 2D topological insulators \cite{Zhang_2014, Orth_2015} and quantum spin Hall insulators \cite{Fleckenstein_2019}.

Blueprints for parafermions based on superconducting defects in FQH edge states encapsulate an intrinsic and important difficulty, as the superconducting gap does not necessarily form under common conditions found in most condensed matter systems. The difficulty of inducing a superconducting gap can be understood by considering two limits. When the coupling between the SC and the FQH edge states is weak, repulsive interactions between the edge states impede a superconducting gap. Conversely, in the limit of strong coupling of the edge states to a SC, one might expect the region near the SC to join the superconducting condensate and push the FQH edge modes away from the system edge and further into the bulk. In both limits, the region proximity coupled to a SC may remain gapless, yielding a system that cannot host parafermion zero-modes. Therefore, an important question is whether it is possible to induce a superconducting gap in the FQH edge states, and what are the appropriate conditions for obtaining such a gap.

Here, we consider these questions for the case of FQH states at filling fraction $\nu=1/3$ (our results can be generalized to $\nu=1/m$ for odd $m$). We show that a superconducting gap can be induced in the edge states at this filling fraction under appropriate conditions, in the limit of strong proximity coupling. Surprisingly, the conditions we find involve a range of sufficiently strong repulsive interactions in the bulk of the FQH states. To obtain these results, we model the 2D system via an effective coupled-wires Hamiltonian \cite{Kane2002_LaughlinWire}. This allows us to probe the two limits of weak and strong coupling to the SC. In the limit of weak coupling to the SC we show that repulsive interactions thwart the superconducting gap, while a gap can be obtained for sufficiently strong attractive interactions. However, in the limit of strong coupling to the SC, even though the edge states are indeed pushed towards the bulk, a residual coupling between them opens a superconducting gap if the strength of the repulsive interactions in the bulk of the FQH states are within a specific range. 

Additionally, we identify a duality of our model. This duality allows us to show that the gapped phases that occurs in the limit of weak and strong proximity coupling to the SC (which require attractive and repulsive interactions, respectively) are adiabatically connected, and both are capable of hosting parafermion zero-modes.

The paper is organized as follows. In Sec.~\ref{sec:Summary} we give the physical picture and summary of our results. In Sec.~\ref{sec:Model} we derive a quasi-one dimensional model from a coupled-wires construction and discuss strategies of analyzing the two regimes of weak and strong proximity coupling. In Sec.~\ref{sec:Integer quantum Hall case} we discuss the IQH case and present most of the tools we will use in the FQH case. In Sec.~\ref{sec:Fractional quantum Hall case} we analyze and present our results for the FQH case of $\nu=1/3$, in the weak and strong proximity coupling regimes. In Sec.~\ref{sec:duality} we discuss the duality between the weak and strong proximity coupling regimes and use it determine the nature of the gapped phase we identify in the strong proximity coupling limit. In Sec.~\ref{sec:Conclusions} we discuss our results and relate them to experimental realizations and the limit studied numerically in Ref.~\cite{Repellin_2018}.

\section{Summary of main results}\label{sec:Summary}

\begin{figure}[!tp]
    \centering
    \subfloat[][]{
        \includegraphics[width=0.45\columnwidth]
            {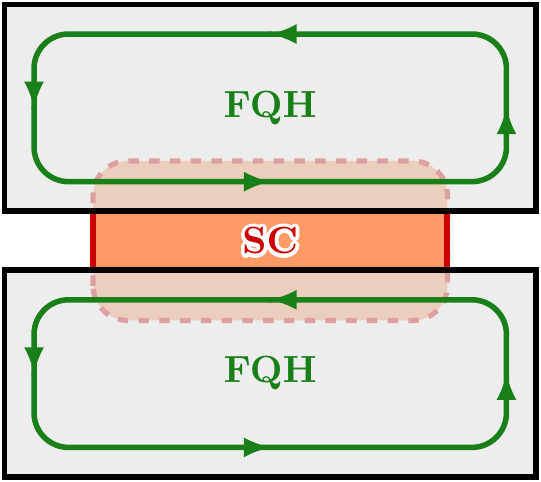}
        \label{subfig:weak_illus}
    }
    \hfill
    \centering
    \subfloat[][]{
        \includegraphics[width=0.45\columnwidth]
            {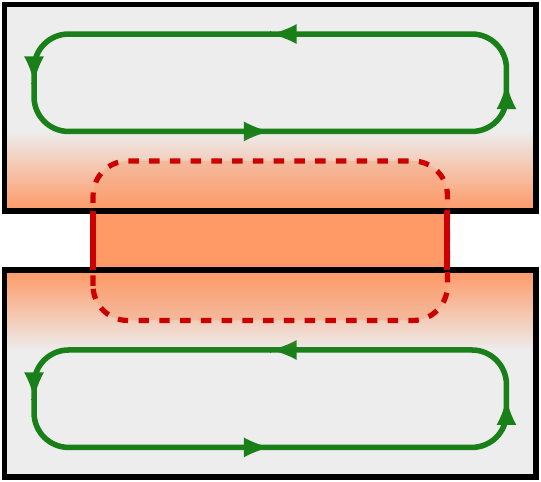}
        \label{subfig:strong_illus}
    }
    \caption{
        Illustration of two FQH slabs coupled by proximity to a SC. Panel \protect\subref{subfig:weak_illus} depicts the situation in the regime of weak proximity coupling to the SC, for which the edge modes remain gapless. Panel \protect\subref{subfig:strong_illus} illustrates a possible scenario in the regime of strong coupling to the SC, in which the region near the SC becomes  superconducting due to the proximity effect thereby ``pushing'' the gapless edge modes deeper into the bulk of the FQH slabs.
    }
    \label{fig:illustrations}
\end{figure}

The difficulty in inducing a superconducting gap in counter-propagating FQH edge modes can be understood by examining an effective one dimensional model for these modes. Consider two FQH droplets at filling $\nu=1/m$ for odd $m$ (also called the Laughlin states \cite{Laughlin_1983}) weakly coupled along their edges to a SC. Tunneling of electrons between the FQH edges and the SC leads to superconducting proximity coupling between the edges and the SC. By weak proximity coupling we mean the limit in which the energy scale characterizing the proximity coupling is much smaller than the gap of the FQH states, and we can treat the proximity coupling using a one dimensional description of the fractional counter-propagating edge state. The corresponding Hamiltonian density is
\begin{equation}\label{eq:1d model}
    \frac{um}{2\pi}\left[K_e(\partial_x\varphi_e)^2+K_e^{-1}(\partial_x\theta_e)^2\right]+\Delta \cos(2m\varphi_e)
\end{equation}
where $u$ is a velocity scale, $\Delta$ is the coefficient characterizing the strength of the superconducting proximity coupling and here and throughout the paper we take $\hbar=1$. The bosonic fields $\varphi_e,\theta_e$ obey the commutation relations $[\partial_x\varphi_e(x),\theta_e(x')]=2\pi i\delta(x-x')/m$ and relate to the fermionic edge modes via the bosonization identity $\psi_{R/L}(x)=e^{i m(\varphi_e\pm\theta_e)}/\sqrt{2\pi a}$, with $a$ a microscopic length-scale cutoff \cite{Delft1998_BosoBegginers}. In this model the Luttinger parameter $K_e$ indicates repulsive interactions between the two edge modes when $K_e<1$. The model can be viewed as a perturbed scale invariant theory in 1+1 dimensions. As such, we can ascertain a gap by examining the scaling dimension of the perturbation \cite[p.~38]{Di_Francesco_1997}, which we denote $d$. If $d<2$ the perturbation is relevant and opens an energy gap. The model \eqref{eq:1d model} conserves the difference between the number of right and left moving electrons, and it includes the local perturbation with the smallest scaling dimension that follows this conservation rule. This perturbation is the $\cos(2m\varphi_e)$ proximity coupling term, for which $d=m/K_e$. Therefore, for simple fractions of the form $\nu=1/m$ with $m>2$, a superconducting gap cannot form in the weak proximity coupling regime in the presence of repulsive interactions. This limit is illustrated schematically in Fig.~\ref{subfig:weak_illus}.

Now consider the limit in which the SC proximity coupling is larger than the bulk gap of the FQH state. In this limit, we expect the regions of the FQH liquids near the SC to join the superconducting condensate due to the proximity effect as illustrated in Fig.~\ref{subfig:strong_illus}. If the FQH liquids in the regions further away from the SC are weakly coupled to this larger SC, gapless edge states are formed in the interface between the enlarged SC condensate and the FQH droplets. As argued above, in the weak proximity coupling limit of filling fraction $m>2$, the edge modes remain gapless in the presence of any repulsive interactions between them. In this situation we can think of the gapless edge modes as simply being pushed deeper into the bulk, as illustrated in Fig.~\ref{subfig:strong_illus}. 

In the case that the counter-propagating modes are edge states of IQH liquids with $m=1$, this analysis yields the known results that the model is gapped under common conditions of condensed matter systems \cite{Fu_2008}. In the weak proximity coupling regime, the SC proximity coupling is always relevant for mild repulsion of $1/2<K_e$ between the edges and it induces a gap. In the regime of strong proximity coupling, we can again think of the electronic edge states as being pushed away from the edge and being weakly coupled being by proximity to the SC. Thus, in both weak and strong regime we expect a gapped phase for mild repulsive interactions in the IQH case.

An important conclusion from the above discussion is that an analysis of the proximity coupling away from the weak proximity coupling limit cannot be done using a strictly one dimensional model of the edge states. Rather, a proper analysis requires a 2D description of the FQH state. In order to allow an analytical study, we model the FQH state using an effective coupled-wires construction, an approach first utilized by Kane et.~al \cite{Kane2002_LaughlinWire}. In this description, which can be used for both IQH and FQH phases, each quantum Hall (QH) slab is modeled as a set of parallel electronic wires in the presence of perpendicular magnetic field, as shown in Fig.~\ref{fig:Wire-construction-model}. The wire construction also includes specific inter-wire interactions that induce the QH phase of filling fraction $\nu=1/m$. We also include a SC proximity coupling between the wires which are directly adjacent to the SC in the two QH slabs (see Fig.~\ref{fig:Wire-construction-model}). This approach has the benefit that we can keep track of the location of the gapless edge mode, i.e.,\ on which wire it is manifested. Using perturbative renormalization group (RG), we determine the resulting phase of the model in the two regimes of strong and weak proximity coupling. 
The wire construction approach in the case of $\nu=1$ reproduces the known results mentioned above.

In the weak proximity coupling regime of the $\nu=1/3$ case, we recover a gapless phase when the density-density interactions between the FQH liquids are repulsive, i.e.,\ $K_e<1$ and find a gapped phase when $K_e>3/2$. The critical value of $K^c_e$ for the transition between the gapped and the gapless phases depends on the coupling strength. In the limit of infinitely small coupling, $K^c_e$ approaches $3/2$, but as the coupling strength is increased $K^c_e$ decreases. As shown in Sec.~\ref{subsec:FQH weak pairing regime}, this occurs due to a renormalization of $K_e$ during the flow \footnote{
    This transition is reminiscent of the cascaded Kosterlitz-Thouless transition discussed by Podolsky et al.~\cite{Podolsky_2009}.
}. 

In the strong proximity coupling regime for filling fraction $\nu=1/3$, we find that the density-density repulsion \emph{in the bulk} of each of the FQH liquid plays a pivotal role. We incorporate this repulsion in a Luttinger parameter $K_b$ (see Subsec.~\ref{subsec:Weak and strong pairing regimes} below for definition). In the limit where the proximity coupling is much stronger than the interactions leading to the QH gap and much smaller than the band width, we find the phase diagram in terms of $K_b$. For mild repulsion in the bulk, which is nevertheless sufficient to open the FQH gap, $1/3<K_b<2/3$, we reproduce the situation discussed above, in which the edge modes remain gapless, but are ``pushed'' to wires deeper into the bulk. For stronger repulsion in the bulk, $2-\sqrt{3}<K_b<1/3$, we find a gapped phase that is characterized by superconducting long range order. Such a system can be used to stabilize parafermion zero-modes \cite{Lindner_2012_parafermions, Clarke_2013, Cheng_2012, Alicea_2016, Vaezi_2013}. For stronger repulsion, $K_b<2-\sqrt{3}$, we find that the strong proximity coupling regime, in which the edge modes are pushed away from the edge, is suppressed, and the system remains gapless.

Furthermore, we identify a $\mathbb{Z}_2$-duality which maps wire-construction models of the SC-FQH interface in the weak proximity coupling regime to wire-construction models of the interface in the strong proximity coupling regime and vice-versa. We use this duality to identify a class of continuously parameterized gapped models. We show that this class of models includes models both in the weak and strong proximity coupling regimes. This shows that the gapped phases that we find in the weak and strong proximity coupling regimes are the same and can be adiabatically connected without closing the gap. Thus, the SC-FQH heterostructure in the regime of strong proximity coupling can be used to bind parafermion zero-modes. Furthermore, the duality analysis also suggests that the phase diagram of the system features this gapped phase also at intermediate values of the coupling.

\section{Model\label{sec:Model}}

In this section, we describe the model which we analyze throughout the paper. As we discuss in detail below, the model is based on a description of a FQH state using an array of coupled one dimensional wires \cite{Kane2002_LaughlinWire}.  In Subsec.~\ref{subsec:Reduction to six chiral movers}, we present the coupled-wires construction including the proximity coupling to the SC, and show how the problem can be reduced to a model of perturbed six bosonic one-dimensional chiral fields. In Subsec.~\ref{subsec:reduced model of six chiral movers} we summarize the reduced model we will analyze throughout the paper and in Subsec.~\ref{subsec:Weak and strong pairing regimes} we discuss the two regimes of weak and strong proximity coupling and introduce appropriate degrees of freedom for analyzing the two regimes.

\subsection{Reduction to the model of six chiral movers\label{subsec:Reduction to six chiral movers}}

\begin{figure}[b!]
    \bigskip
    \includegraphics[width=1\columnwidth]{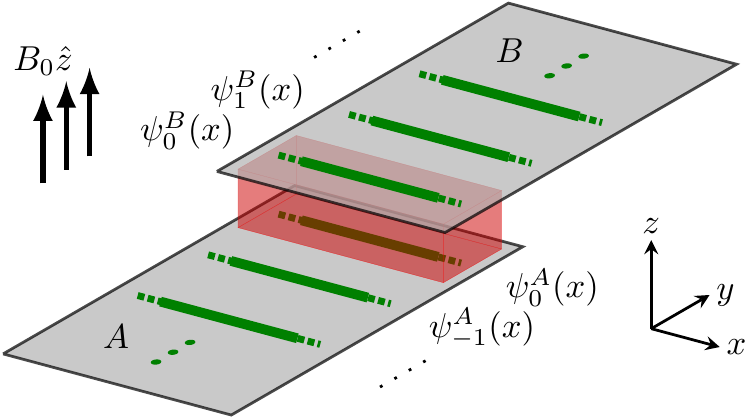}
    \caption{
        \label{fig:Wire-construction-model}
        Coupled-wire constructions of the two slabs $A$ and $B$. Each green line denotes an electronic wire. The electrons corresponding to $j=0$ wires in both slabs, $\psi_{j=0}^{A/B}$, are proximity-coupled to the superconductor (shown in red).
    }
\end{figure}

We consider two slabs of spin-polarized FQH states at filling $\nu=1/m$ where $m$ is odd, with counter-propagating edge states which are proximity coupled to a SC. We label the two slabs $A$ and $B$ and model the FQH states in each slab using a wire construction, following Kane et al.~\cite{Kane2002_LaughlinWire}. In this construction, shown in Fig.~\ref{fig:Wire-construction-model}, each slab consists of an array of wires laid parallel such that they are parallel to the $x$-axis at distance $\ell$ from each other in the $y$-axis. The wires are located at $y=\ell j$ with integer $j$, with the wires of the $A$ and $B$ slabs having $j\leq 0$ and $j\geq 0$ respectively. We denote by $\psi^{A/B}_{j}(x)$ the annihilation operators of electrons in the $j$-th wire in the $A$/$B$ slab. The charge density in each wire is $n_e = k_F/\pi$ and the array of wires is subjected to a perpendicular magnetic field $B_0\hat{z}$. We define an analogue to the two-dimensional filling fraction is $\nu = \frac{n_e/\ell}{B_0/\phi_0}$ with $\phi_0 = 2\pi c/e$ the flux quanta. Moreover, we define the useful parameter $b=B_0e\ell/c$, so that $\nu=2k_F/b$.

We work in Landau gauge, $\vec{A} = -B_0y\hat{x}$. In this gauge the electronic operators decouple at low-energies to left and right operators near the two Fermi points of each wire $\psi_j^s = e^{i(bj+k_F)x} \psi_{R,j}^s + e^{i(bj-k_F)x} \psi_{L,j}^s$ with $\psi_{R/L,j}^s$ the right/left fermionic fields and $s=A,B$. The Hamiltonian of the full wire construction is 
\begin{equation}\label{eq:Hfull}
    H_{\text{full}}=H_{\text{free}}+H_{n\text{-}n}+H_{\Delta}+H_{\text{QH,}2D}.
\end{equation}
The first term, $H_{\text{free}}$ describes free electrons in the different wires,
\begin{equation}\label{eq:Hfree fermion}
    H_{\text{free}} = \int\diff x\, \sum_{s,j}v_{F}\Bigl( -i{\psi_{R,j}^{s,\dagger}\partial_{x}\psi_{R,j}^{s}} +i{\psi_{L,j}^{s,\dagger}\partial_{x}\psi_{L,j}^{s}} \Bigr),
\end{equation}
where $v_{F}$ is the Fermi velocity, and $j$ is an integer compatible with $s=A,B$ (see Fig.~\ref{fig:Wire-construction-model}).

Density-density interactions between the electrons are described by
\begin{equation}\label{eq:Hnn fermion}
    H_{n\text{-}n} = \int\diff x\,\sum_{\alpha,\alpha'}V_{\alpha,\alpha'}\,{n_{\alpha}(x) n_{\alpha'}}(x)
\end{equation}
where $\alpha=(s,j,\rho)$ with $\rho=R,L$ and $s,j$ indices the same as in Eq.~\eqref{eq:Hfree fermion}, and $n_{\alpha}(x)={\psi_{\alpha}^\dagger(x) \psi_{\alpha}}(x)$ is the electronic density operators associated with $\alpha$. Below we show that $H_{n\text{-}n}$ is required to stabilize the QH phase in the coupled-wire construction.

The wires labeled by $j=0$ in both slabs are proximity-coupled to the SC. Assuming translation invariance is maintained along the interface with the SC, this proximity coupling leads to a pairing term involving one electron in each wire, given by
\begin{equation}\label{eq:electron proximity pairing}
    H_{\Delta} = \int\diff x\,\left( \tilde{\Delta}_{1}{\psi_{R,0}^{A}\psi_{L,0}^{B}}+\tilde{\Delta}_{2}{\psi_{L,0}^{A}\psi_{R,0}^{B}}\right)+\text{h.c.}
\end{equation}
Note that $\tilde{\Delta}_1$, $\tilde{\Delta}_2$ can take different values, since time-reversal is broken. Furthermore, we have assumed that the Cooper pairs in the SC are not strictly singlets, thus allowing the proximity coupling to pair electrons with parallel spin. We note that a pairing term of the form \eqref{eq:electron proximity pairing} can be achieved even for singlet SCs, if the electrons in the two slabs are not strictly polarized in the same direction (for possible experimental setups, see the discussion in Refs \cite{Lindner_2012_parafermions, Clarke_2013, Alicea_2016}).

Lastly, we include the QH gap-opening term tailored for the $\nu=1/m$ Laughlin state with odd $m$ by including \footnote{
    This term is discussed in \cite{Kane2002_LaughlinWire}. The operators $\psi_{\alpha}$, in $H_{\text{QH,}2D}$ must operate at separated yet close points to avoid cancellation due to their fermionic algebra.}
\begin{widetext}
\begin{equation}\label{eq:HQH fermion}
    H_{\text{QH,}2D} = \sum_{s,j} \int\diff{}x\,\tilde{J}_s{\left( \psi^{s\dagger}_{R,j+1} \psi^{s\dagger}_{R,j}
    \psi^s_{L,j+1} \psi^s_{L,j} \right)^{\frac{m-1}{2}} \psi^{s\dagger}_{R,j} \psi^s_{L,j+1}} + \rm{h.c.}
\end{equation}
\end{widetext}

Note that in the operators $H_\Delta$ and $H_{\text{QH,}2D}$ we have omitted irrelevant terms that oscillate as $e^{i n k_F x}$ with integer $n\neq 0$. For example, in the operator $H_\Delta$ we have neglected contributions from $e^{2ik_F x}\psi_{R,0}^A\psi_{R,0}^B$ that appear in the pairing $\psi_0^A\psi_0^B$. The irrelevance of these terms is due to their oscillatory nature. More specifically, they couple low-energy states to high-momentum (and energy) states, so in an effective low-energy description we may neglect them \cite{Schulz_1980, Haldane1983}. We neglect any terms that have this oscillatory nature in this work.

For each wire, we represent the fermionic fields $\psi_\alpha$ as exponential of a bosonic fields $\phi_\alpha$ by using the bosonization identity (see for example Ref.~\cite{Delft1998_BosoBegginers})
\begin{equation}\label{eq:bosonization id}
    \psi_{R/L,j}^{s}(x) = \left(2\pi a\right)^{-1/2} e^{\pm i\phi_{R/L,j}^{s}(x)},
\end{equation}
where $a$ is a short-distance cutoff. Using the bosonization identity, the electron density operators are rewritten as $n_\alpha = \partial_x\phi_\alpha/2\pi$. The bosonic fields are Hermitian, $\phi_\alpha(x)=\phi_\alpha^\dagger (x)$, and obey the commutation relations 
\begin{equation}
    [\phi^{s}_{R/L,j}(x),\phi^{s}_{R/L,j}(x')] = \pm i\pi \sgn(x-x'),
\end{equation}
and the commutation relation $[\phi^{s}_{\rho,j}(x),\phi^{s'}_{\rho',j'}(x')]$ is such that the following anticommutation relations hold
\begin{equation}
    \left\{e^{i\phi_{\rho,j}^{s}(x)},\,e^{i\phi_{\rho',j'}^{s'}(x')}\right\} = 0,
\end{equation}
if any of the indices $s,j,\rho$ differ from $s',j',\rho'$ (here $s,s'$ and $\rho,\rho'$ take the values from $A,B$ and $R,L$ respectively).

The bosonized form of the $H_{\text{free}}$ and $H_{n\text{-}n}$ parts of the Hamiltonian [Eqs.~\eqref{eq:Hfree fermion}  and \eqref{eq:Hnn fermion}], are written as
\begin{multline}
    H_{\text{free}}+H_{n\text{-}n} \\=
        \int\diff x\,\sum_{\alpha,\alpha'}
        \left(\frac{v_F\delta_{\alpha,\alpha'}}{4\pi}+\frac{V_{\alpha,\alpha'}}{4\pi^2}\right)
        \partial_{x}\phi_{\alpha}\,\partial_{x}\phi_{\alpha'},
    \label{eq:H_free+H_nn}
\end{multline}
and the pairing term is given by $H_\Delta=\int\diff{}x\,\mathcal{H}_\Delta$ with
\begin{equation}\label{eq:pairing density}
    \mathcal{H}_{\Delta} = 
            \Delta_{1} \cos\left(\phi_{R,0}^{A}-\phi_{L,0}^{B}\right) 
            + \Delta_{2} \cos\left(\phi_{R,0}^{B}-\phi_{L,0}^{A}\right),
\end{equation}
where $\Delta_1$ and $\Delta_2$ differ from $\tilde{\Delta}_1$ and $\tilde{\Delta}_2$ by a dimensionful normalization factor due to units and normal ordering,  $\tilde{\Delta}_j \propto a \Delta_j$ for $j=1,2$ (for a more detailed discussion see for example Ref.~\cite{Delft1998_BosoBegginers}).

The QH gap-opening term, Eq.~\eqref{eq:HQH fermion}, can be written down compactly by introducing the fractional chiral fields 
\begin{equation}\label{eq:eta def}
    \eta_{R/L,j}^{s}=\frac{m+1}{2m}\phi_{R/L,j}^{s}+\frac{m-1}{2m}\phi_{L/R,j}^{s},
\end{equation}
which obey commutation relations
\begin{equation}
    [\partial_x\eta_{R/L,j}^s(x),\eta_{R/L,j}^s(x')]=\pm 2\pi i\delta(x-x')/m,
\end{equation}
and the commutation relations of $\eta$ fields with different indices $[\eta_{\rho,j}^s(x),\eta_{\rho',j'}^{s'}(x')]$ is such that 
\begin{equation}
    \left\{e^{i m\eta_{\rho,j}^s(x)},e^{i m\eta_{\rho',j'}^{s'}(x')}\right\}=0,
\end{equation}
if $(s,j,\rho,x)$ differs from $(s',j',\rho',x')$.

Using these fields, Eq.~\eqref{eq:HQH fermion} can be written as 
\begin{equation} \label{eq:H_QH wire_const}
    H_{\text{QH,}2D} = \sum_{s,j}\int\diff x\,
        J_{s}\cos\left(m\eta_{R,j}^{s}+m\eta_{L,j+1}^{s}\right),
\end{equation}
where $\tilde{J}_s \propto a^3 J_s$.

Next, we discuss some of the properties we desire of the interactions, $H_{n\text{-}n}$. As shown by Kane et al.~\cite{Kane2002_LaughlinWire}, for the small perturbation $H_{\text{QH,}2D}$ to open a bulk gap (in the absence of the pairing term $H_\Delta$) an additional interaction term, quadratic in bosonic fields, is needed. For example, the term
\begin{equation} \label{eq:Kane n-n term}
    H_{n\text{-}n}=\int\diff x\,\sum_{s,j} 
    w\partial_x\eta_{R,j}^{s}\partial_x\eta_{L,j+1}^{s}
\end{equation} 
together with $H_{\text{QH},2D}$ opens the FQH bulk gap for $w>w_0>0$ for some critical $w_0$. In the model presented in Sec.~\ref{subsec:reduced model of six chiral movers} we require of that the quadratic boson term imply a gap in the FQH bulk in the absence of the pairing term. We will mainly be interested to relate the model to experimental situations in which we usually expect repulsive interactions, i.e.,\ $V_{\alpha,\alpha'}>0$ element-wise, but we will also consider attractive interactions.

The number of wires included in the model scales linearly with the width of the QH slabs. Focusing only on degrees of freedom that reside close to the edges of the two QH strips, we will analyze a reduced model that includes only six fields:
\begin{equation}\label{eq:DOF 3 wire}
    \vec{\Phi} = \left(\begin{array}{c c c c c c}
                    \eta_{R,-1}^{A}, & \phi_{L,0}^{A}, & \phi_{R,0}^{A}, & \phi_{L,0}^{B}, & \phi_{R,0}^{B}, & \eta_{L,1}^{B}
                  \end{array}\right)^{\transpose},
\end{equation}
which describe degrees of freedom on the wires $|j|\leq 1$.

The parameters of the density-density interactions [$V_{\alpha\alpha'}$ in Eq.~\eqref{eq:H_free+H_nn}] can be chosen such the Hamiltonian $H_\text{full}$, in Eq.~\eqref{eq:Hfull}, does not couple the six fields $\vec{\Phi}$ in Eq.~\eqref{eq:DOF 3 wire} to the rest of the $\eta$-fields. Note that this decoupling is evidently true for the $H_{\Delta}$ and $H_{\text{QH,}2D}$ terms, Eqs.~\eqref{eq:pairing density} and \eqref{eq:H_QH wire_const}. Moreover, parameter choices which allow this decoupling are consistent with a bulk gap in the QH slabs. An example of such parameter choice is given by Eq.~\eqref{eq:Kane n-n term}. 

With this decoupling, the Hamiltonian of the coupled wire-construction can be written as a sum of two commuting terms $H_\text{full}=H+H^c$ where $H$ involves only the $\vec{\Phi}$ degrees of freedom, i.e.,~those near the trench, and $H^c$ involves only $\eta$-fields outside \eqref{eq:DOF 3 wire}, i.e.,~involves degrees of freedom deeper into the bulk of the two FQH slabs. Assuming that the QH bulk is gapped means that in the absence of the proximity coupling $H_\Delta$, all of the cosine terms in Eq.~\eqref{eq:H_QH wire_const} open a gap and pin their respective fields, and the only gapless fields  are $\eta_{R/L,0}^{A/B}$. Consequently, the Hamiltonian $H^c$ is gapped even in the presence of proximity coupling. Thus, the Hamiltonian $H$ alone determines the existence of a gap in the full system,  \footnote{
    The only remnant of the deep-bulk degrees of freedom in the edge problem of $H$ is accounted for by specifying the fractional charge accumulated in the bulk, $\int\diff x\,\partial_x\eta_{R,-1}^{A}/2\pi$ and $\int\diff x\,\partial_x\eta_{L,1}^{B}/2\pi$ mod $1/m$.
} and we will focus on analyzing it in the reminder of this work.

\begin{figure}[t!]
    \bigskip
    \includegraphics{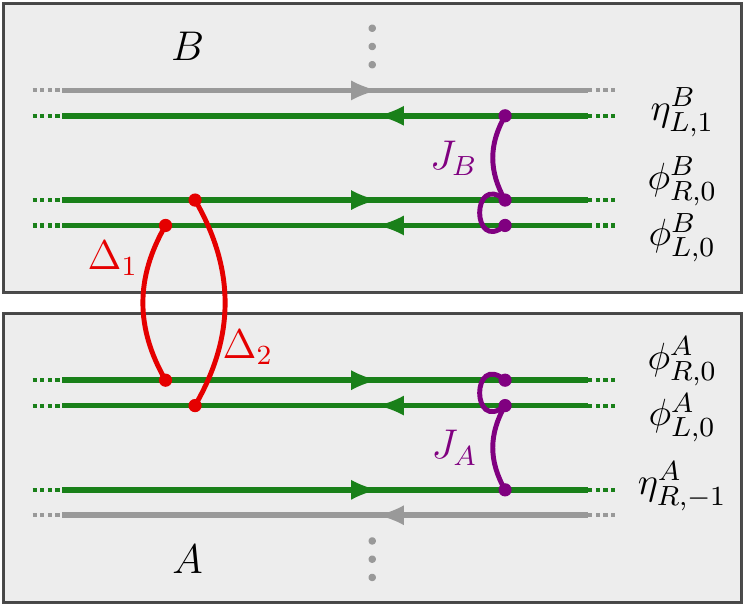}
    \caption{
        \label{fig:eff_model_ints}
        Illustration of the wire construction and the different couplings. In green are the degrees of freedom of the edge problem -- the six chiral fields in the model \eqref{eq:3_wire_model}. The red lines depict the pairing term \eqref{eq:pairing density} and the purple lines depict the FQH-inducing terms \eqref{eq:H_QH density}. with slight fine tuning, the edge degrees of freedom decouple from the bulk which are greyed out.
    }
\end{figure}

\begin{table*}[t!]
\centering
\begin{tabular}{|c | c | l|} 
    \hline
    Fields & Definition & Description \\ [0.5ex]
    \hline 
    
    $\psi_{\rho,j}^s$ & Eq.~\eqref{eq:Hfree fermion} &  chiral electronic fields of the coupled-wire construction \\ [0.2ex]
    
    $\phi_{\rho,j}^s$ & Eq.~\eqref{eq:bosonization id} & chiral bosonic fields of the coupled-wire construction\\ [0.2ex]
    
    $\eta_{\rho,j}^s$ & Eq.~\eqref{eq:eta def} & fractional chiral bosonic fields\\ [0.2ex]
    
    $\vec{\Phi}$ & Eq.~\eqref{eq:DOF 3 wire} & vector of all fields of reduced edge problem\\ [0.2ex]
    
    $\varphi_\alpha,\,\theta_\alpha$,\quad $\alpha=A,B,e$ & Eq.~\eqref{eq:weak-pair-DOF} & weak proximity coupling bosonic fields and duals \\ [0.2ex]
    
    $\varphi_\alpha,\,\theta_\alpha$,\quad $\alpha=1,2,3$ & Eq.~\eqref{eq:strong-pair-DOF} & strong proximity coupling bosonic fields and duals \\ [0.2ex]
    
    $\varphi_\alpha,\,\theta_\alpha$,\quad $\alpha=+,-,D$ & Eq.~\eqref{eq:def of dual fields} & bosonic fields and duals appropriate for duality\\ [0.5ex]
    
    \hline 
\end{tabular}
\caption{
    Table summarizing the different bases of fields used in this paper. In the three first entries the indices are $\rho=R,L$, $s=A,B$ and $j\in\mathbb{Z}$.}
\label{table:1}
\end{table*}

\subsection{Reduced model of six chiral movers}\label{subsec:reduced model of six chiral movers}

Our reduced Hamiltonian density $\mathcal{H}$ of the degrees of freedom $\vec{\Phi}$ is written as
\begin{equation}
    \mathcal{H} = \mathcal{H}_{0} + \mathcal{H}_{\Delta} + \mathcal{H}_{\text{QH}},
    \label{eq:3_wire_model}
\end{equation}
where $\mathcal{H}_\Delta$ is as given in Eq.~\eqref{eq:pairing density} and the QH gap-opening term, restricted to the degrees of freedom $\vec \Phi$, yields
\begin{equation}
\begin{aligned}
    \mathcal{H}_{\text{QH}} &= J_{A} \cos\left(
            m\eta_{R,-1}^{A} 
            +\tfrac{m+1}{2}\phi_{L,0}^{A}
            +\tfrac{m-1}{2}\phi_{R,0}^{A} \right) \\
            & + J_{B} \cos\left(
            m\eta_{L,1}^{B}
            +\tfrac{m+1}{2}\phi_{R,0}^{B}
            +\tfrac{m-1}{2}\phi_{L,0}^{B}\right).
\end{aligned}
\label{eq:H_QH density}
\end{equation}
i.e.,~$\mathcal{H}_{\text{QH}}$ is given by just the $j=0$ contributions to the sum in Eq.~\eqref{eq:H_QH wire_const}. The quadratic part of the Hamiltonian, is obtained by restricting Eq.~\eqref{eq:H_free+H_nn} to the degrees of freedom $\vec \Phi$, and is given by
\begin{equation}
    \mathcal{H}_{0} = \frac{1}{4\pi}
	\partial_x\vec{\Phi}^{\transpose}\vec U \partial_x\vec{\Phi},
    \label{eq:model_fixed_point}
\end{equation}
where $\vec{U}$ is a real, symmetric, and positive-definite $6\times 6$ matrix whose elements are determined by $v_F$ and $V_{\alpha,\alpha'}$.

The degrees of freedom $\vec\Phi$ are characterized by commutation relations  $\left[\partial_x\Phi_\alpha(x),\Phi_\beta(y)\right]=2\pi i(\vec{K}^{-1})_{\alpha\beta}\delta(x-y)$ with the $K$-matrix 
\begin{equation}
    \vec{K} = \diag{\left(\begin{array}{cccccc}
        +m, & -1, & +1, & -1, & +1, & -m
    \end{array}\right)}.
\end{equation}
In this paper, we will consider two cases: IQH case of $m = 1$ and then the Laughlin FQH state of $m = 3$.

Throughout this work we will assume that the system is symmetric under a $\pi$-rotation of the system plane
\begin{equation}
    \mathcal{R}:\ 
    \phi_{R/L,j}^{B} \leftrightarrow \phi_{L/R,-j}^{A}
    \text{ and }
    \eta_{R/L,j}^{B} \leftrightarrow \eta_{L/R,-j}^{A},
    \label{eq:pi-symmetry}
\end{equation}
which is equivalent to considering two identical FQH slabs. This symmetry imposes $J=J_A = J_B$ and restricts the form of the matrix $\vec{U}$, to $\vec{U} = \vec{R}^\transpose\vec{U}\vec{R}$, where the matrix $\vec{R}$ given by $\mathcal{R}\vec{\Phi}\mathcal{R}^{-1}=\vec{R}\vec{\Phi}$ or explicitly as
\begin{equation}
    \vec{R} = \begin{pmatrix}
    	  &   &   &   &   & 1 \\
		  &   &   &   & 1 &   \\
		  &   &   & 1 &   &   \\
		  &   & 1 &   &   &   \\
		  & 1 &   &   &   &   \\
		1 &   &   &   &   &  
	\end{pmatrix}.
\end{equation}

We will also simplify the analysis by considering systems that are Lorentz invariant with respect to a single velocity scale, $u$, or equivalently, ones where the following equation holds
\begin{equation}\label{eq:conf coupled wires}
    (\vec{K}^{-1}\vec{U})^2=u^2\mathbbm{1}.
\end{equation}
A proof that this condition is equivalent to Lorentz invariance with respect to velocity $u$ is given in Appendix \ref{apdx:coupled wires RG}.

The matrix $\vec{U}$ in Eq.~\eqref{eq:model_fixed_point} that corresponds to a model with $\mathcal{R}$ and Lorentz symmetry is parameterized by nine real numbers. For simplicity, we will focus in this work on models for which the $\vec U$ matrix belongs to two-parameter family, which is given explicitly in Subsec.~\ref{subsec:Weak and strong pairing regimes}. However, under the RG flow these models flow to effective low-energy descriptions corresponding to general $\vec{U}$ matrices. To take this into account, we keep the full $\vec{U}$ matrices in our formalism.

In this section and throughout this paper we introduce several different field bases to describe the dynamical fields. The different bases we introduce are summarized in Table \ref{table:1}.

\subsection{Weak and strong proximity coupling regimes}\label{subsec:Weak and strong pairing regimes}

In the RG analysis we will focus on two regimes: weak strong proximity coupling. These two regimes  correspond to the regimes of $|\Delta_1|,|\Delta_2|\ll|J|$ and $|J|\ll|\Delta_1|,|\Delta_2|$ respectively and are defined in the following subsection. 

First consider the weak proximity coupling regime. When $\Delta=0$, there are two gapless edge modes on the $j=0$ wires, which will then be weakly coupled to the SC for finite value of $\Delta$. In the strong proximity coupling regime, for $J=0$ there is a $\Delta$-gap on the inner $j=0$ wires and gapless edge modes on the $j=\pm 1$ wires. For finite value of $J$, these two edge modes are coupled to the SC. In both regimes we will find emergent interactions that appear in low-energy description that might gap the corresponding gapless modes.

In both regimes, we introduce new basis of fields, in which the analysis of each regime is simpler. We can then express the Hamiltonian using the field redefinitions. The expressions for the cosine terms using the different field bases are summarized in Appendix~\ref{apdx:basis-dictionary}.

\subsubsection{Weak proximity coupling regime}

We perform the analysis in the weak proximity coupling regime using the three bosonic fields and their duals
\begin{equation}\label{eq:weak-pair-DOF}
\begin{aligned}
    \varphi_A &= (\eta_{R,-1}^A-\eta_{L,0}^A)/2, & \theta_A &= (\eta_{R,-1}^A+\eta_{L,0}^A)/2, \\
    \varphi_B &= (\eta_{R,0}^B-\eta_{L,1}^B)/2, & \theta_B &= (\eta_{R,0}^B+\eta_{L,1}^B)/2, \\
    \varphi_e &= (\eta_{R,0}^A-\eta_{L,0}^B)/2, & \theta_e &= (\eta_{R,0}^A+\eta_{L,0}^B)/2,
\end{aligned}
\end{equation}
that obey the commutation relations
\begin{equation}
\begin{gathered}
    \left[\partial_x\varphi_{\alpha}(x),\theta_{\alpha'}(x')\right] = i \pi \delta_{\alpha,\alpha'} \delta(x-x')/m,\\
    \left[\partial_x\varphi_{\alpha}(x),\varphi_{\alpha'}(x')\right] = \left[\partial_x\theta_{\alpha}(x),\theta_{\alpha'}(x')\right] = 0.
\end{gathered}
\end{equation}

In terms of these fields, the QH gap-opening term \eqref{eq:H_QH density} is written
\begin{equation}\label{eq:HQH-weak-fields}
    \mathcal{H}_{\text{QH}} = J\cos(2m\theta_A) + J\cos(2m\theta_B).
\end{equation}
This leads us to our definition of the weak proximity coupling regime as the parameter regime for which the $\theta_A$, $\theta_B$ fields are pinned to one of the $m$ minima of cosines in $\mathcal{H}_\text{QH}$ of Eq.~\eqref{eq:HQH-weak-fields}. To determine if the Hamiltonian open a gap in this regime we will examine the remaining sector of the $\varphi_e$, $\theta_e$ fields.

In terms of the fields given in Eq.~\eqref{eq:weak-pair-DOF}, the form of $\mathcal{H}_0$, which we will focus on throughout most of this work (Sec.~\ref{sec:Integer quantum Hall case} and \ref{sec:Fractional quantum Hall case}), is
\begin{equation}\label{eq:3_LuttLiquids}
    \mathcal{H}_{0} = \sum_{\alpha=A,B,e} \frac{u m}{2\pi}\left[ K_{\alpha} (\partial_x \varphi_{\alpha})^2 + K_{\alpha}^{-1} (\partial_x \theta_{\alpha})^2 \right],
\end{equation}
where $K_A=K_B\equiv K_b$ holds due to $\mathcal{R}$-symmetry. The above form for $\mathcal{H}_0$ will be used both in the weak and strong proximity coupling regime. Note that Eq.~\eqref{eq:3_LuttLiquids} specifies the matrix $\vec{U}$ as per Eq.~\eqref{eq:model_fixed_point}.

Since we require a gapped QH bulk, we consider parameters $K_\alpha$ such that the QH gap-opening term, Eq.~\eqref{eq:H_QH density}, is initially relevant, and opens a gap in the absence of the proximity coupling, i.e.,~
\begin{equation}\label{eq:Kbulk_for_Hall_gap}
    m K_b < 2.
\end{equation}
Note that the Luttinger parameter $K_\alpha$ incorporates quadratic boson interactions both of the type appearing in Eq.~\eqref{eq:Kane n-n term} and other such as $(\partial_x\eta_{R/L,j}^s)^2$. Moreover, Eq.~\eqref{eq:3_LuttLiquids} describes repulsive density-density interactions only if \footnote{
    Condition \eqref{eq:K_repulsive} establishes that any density-density of the form $\partial_x\eta_{R,j}^s\partial_x\eta_{L,j+1}^s$ comes with a positive coefficient. For the rest of the coefficients in $V_{\alpha,\alpha'}$ to be positive, we also require $u(K_b+K_b^{-1})/2\geq (m+m^{-1})v_F$.}
\begin{equation}\label{eq:K_repulsive}
    0 < K_\alpha < 1,\qquad \alpha=A,B,e.
\end{equation}
Similarly, the condition on $K_e$ comes from inter-slab repulsive density-density interactions 

In the absence of the $\Delta_{1,2}$ terms, the parameter $K_b$ is related to the gap in the $\theta_{A,B}$ fields \cite{Zamolodchikov_1995} as $E_\text{gap}\propto \frac{u}{a}\left(\frac{Ja^2}{u}\right)^{1/(2-mK_b)}$.

\subsubsection{Strong proximity coupling regime}

In the strong proximity coupling regime we will use the set of three bosonic fields and their duals
\begin{equation}\label{eq:strong-pair-DOF}
\begin{aligned}
    \varphi_1 &= (\phi_{R,0}^A-\phi_{L,0}^B)/2, & \theta_1 &= (\phi_{R,0}^A+\phi_{L,0}^B)/2, \\
    \varphi_2 &= (\phi_{R,0}^B-\phi_{L,0}^A)/2, & \theta_2 &= (\phi_{R,0}^B+\phi_{L,0}^A)/2, \\
    \varphi_3 &= (\eta_{R,-1}^A-\eta_{L,1}^B)/2, & \theta_3 &= (\eta_{R,-1}^A+\eta_{L,1}^B)/2,
\end{aligned}
\end{equation}
that obey the commutation relations 
\begin{equation}
\begin{gathered}
    \left[\partial_x\varphi_{\alpha}(x),\theta_{\alpha'}(x')\right] = i\pi \delta_{\alpha,\alpha'} \delta(x-x')/m_\alpha,\\
    \left[\partial_x\varphi_{\alpha}(x),\varphi_{\alpha'}(x')\right] = \left[\partial_x\theta_{\alpha}(x),\theta_{\alpha'}(x')\right] = 0,
\end{gathered}
\end{equation}
with $m_{1,2}=1$ and $m_3=m$. We can use them to write the pairing interaction \eqref{eq:pairing density} as
\begin{equation}\label{eq:H_pairing}
    \mathcal{H}_{\Delta} = \Delta_{1}\cos\left(2\varphi_{1}\right)+\Delta_{2}\cos\left(2\varphi_{2}\right).
\end{equation}
In the strong proximity coupling regime $\mathcal{H}_\Delta$ opens a gap in the fields $\varphi_1$ and $\varphi_2$, leaving the field $\varphi_3$ as the low-energy degree of freedom. We define the strong proximity coupling regime as the parameter regime for which the fields $\varphi_1$ and $\varphi_2$ are pinned.

\section{Proximity coupling to Integer Quantum Hall Edges}\label{sec:Integer quantum Hall case}

We begin by analyzing SC proximity coupling to counter-propagating edge states in the IQH effect at filling fraction $\nu=1$. Our goal is to study the flow of the different couplings in the model within the renormalization group approach. We will treat $\mathcal{H}_0$ as the fixed point about which $\mathcal{H}_\Delta$ and $\mathcal{H}_\text{QH}$ act as small perturbations.

The field definitions \eqref{eq:eta def}, \eqref{eq:weak-pair-DOF} and \eqref{eq:strong-pair-DOF} reveal the identification of the fields
\begin{equation}
    \varphi_1=\varphi_e,\quad\theta_1=\theta_e.
\end{equation}
This identification, together with the form of the all the different Hamiltonian terms $\mathcal{H}_0$, $\mathcal{H}_\Delta$ and $\mathcal{H}_\text{QH}$ in Eqs.~\eqref{eq:3_LuttLiquids}, \eqref{eq:H_pairing} and  \eqref{eq:HQH-weak-fields}, establish that the $\varphi_1,\theta_1$ fields decouple from $\varphi_j,\theta_j$ with $j=2,3,A,B$. This decoupling will be reflected in our analysis in both the weak and strong proximity coupling regimes.

\subsection{IQH weak proximity coupling regime}

In the weak proximity coupling regime, we will use perturbative RG about the fixed point Eq.~\eqref{eq:3_LuttLiquids}, and give a condition for a gap and also a set of conditions for being in the weak proximity coupling regime. In this regime, we assume that the QH gap-opening term $\mathcal{H}_\text{QH}$ is initially a small perturbation but eventually flows to strong coupling. Thus, the system is gapped if the field $\varphi_e$ is gapped, i.e.,\ if $\Delta_1$ flows to strong coupling at low-energies.

We treat $\mathcal{H}_\text{QH}$ and $\mathcal{H}_\Delta$ as perturbations of the fixed point \eqref{eq:3_LuttLiquids} and apply an RG step by rescaling the cutoff $a\to e^\ell a$ and obtain $\ell$ dependent coefficients. The dominant behavior under RG flow of the different perturbations, which have coefficients $J$, $\Delta_1$ and $\Delta_2$, is characterized by their scaling dimensions, $d_J = K_b$, $d_{\Delta,1}=K_e^{-1}$ and $d_{\Delta,2}=(K_b+K_b^{-1})/2$ respectively. These scaling behaviour are most easily read from the form of the interactions in terms of the weak proximity coupling fields: the $J$ terms in Eq.~\eqref{eq:HQH-weak-fields}, $\Delta_1\cos(2\varphi_e)$ and $\Delta_2\cos(\varphi_A-\theta_A+\varphi_B+\theta_B)$.

We calculated higher order corrections to the RG equations in terms of the dimensionless coefficients $y_J=\pi a J/u$, $y_{\Delta,j}=\pi a\Delta_j/u$. To second order the RG flow equations are
\begin{equation}\label{eq:IQH-weak-flow-D1-Ke}
\begin{aligned}
    \frac{\diff y_{\Delta ,1}}{\diff\ell} & =(2-d_{\Delta,1})y_{\Delta ,1},\\
    \frac{\diff K_{e}}{\diff\ell} & =y_{\Delta ,1}^{2},
\end{aligned}
\end{equation}
and 
\begin{equation}\label{eq:IQH-weak-flow-J-Kb-D2}
\begin{aligned}
    \frac{\diff y_{J}}{\diff\ell} & =(2-d_{J})y_{J},\\
    \frac{\diff K_b}{\diff\ell} & = \frac{1-K_b^{2}}{4} y_{\Delta,2}^{2} -K_b^{2}y_{J}^{2},\\
    \frac{\diff y_{\Delta ,2}}{\diff \ell} & =\left(2-d_{\Delta,2}\right) y_{\Delta ,2}.
\end{aligned}
\end{equation}
The renormalization scheme and derivations of the RG flow equations are given in detail in Appendix \ref{apdx:coupled wires RG}.

In the RG analysis of the weak proximity coupling regime, we neglected any inter-species bosonic quadratic terms, e.g.\  $\partial_x\varphi_B\partial_x\varphi_A$, $\partial_x\theta_A\partial_x\varphi_e$. Neglecting these inter-species terms is justified, since they do not get renormalized considerably by the dominant $y_{\Delta,1},y_J$ couplings. We can verify this approximation after solving for the RG flow as explained below.

Equations \eqref{eq:IQH-weak-flow-D1-Ke} are the celebrated Kosterlitz-Thouless RG flow equations \cite{Kosterlitz_1974}. These RG equations predict a gap in the sector of the $\varphi_e,\theta_e$ fields if
\begin{equation}\label{eq:IQH-weak-gap}
    |y_{\Delta,1}| > 1 - 2 K_e.
\end{equation}

A quantitative condition for the system to be in the weak proximity coupling regime can be obtained by requiring that $y_J(\ell)$ flows to strong coupling, i.e.,~$|y_J(\ell^*)|\sim1$, and that the coupling $y_{\Delta,2}(\ell)$ remains small in comparison,
\begin{equation}\label{eq:IQH-weak_ness-init-cond}
    |y_{\Delta,2}(\ell)|<y_\text{thresh}\ll1,\quad\text{for }\ell<\ell^*.
\end{equation}
To translate condition \eqref{eq:IQH-weak_ness-init-cond} to an initial condition, we use an approximate solution, $y^{(1)}(\ell)$, of Eq.~\eqref{eq:IQH-weak-flow-J-Kb-D2} with initial values $d_{r}^{(0)}$ and $y_{r}^{(0)}$ of the scaling dimensions and dimensionless couplings, where index $r$ goes over the labels of the different perturbative couplings. A suitable approximate solution is given by the scaling behavior
\begin{equation}\label{eq:scaling behavior}
    y_{r}^{(1)}(\ell) = y_{r}^{(0)} e^{(2-d_{r}^{(0)})\ell}.
\end{equation}
This expression yields the sufficient condition for the system to be in the weak proximity coupling regime
\begin{equation}\label{eq:IQH-weak-pair-cond}
    \left(\frac{|y_{\Delta,2}^{(0)}|^{(2-d_{\Delta,2}^{(0)})^{-1}}}{|y_{J}^{(0)}|^{(2-d_{J}^{(0)})^{-1}}}\right)^{ 2-d_{\Delta,2}^{(0)}} 
    = |y_{\Delta,2}^{(1)}(\ell^{*})|
    < y_\text{thresh}\ll 1.
\end{equation}

Demanding \eqref{eq:IQH-weak-pair-cond} and \eqref{eq:Kbulk_for_Hall_gap} guarantees that $\mathcal{H}_\text{QH}$ is minimized in low-energy states and that the fields $\theta_{A,B}$ are pinned to the minimum of each cosine in Eq.~\eqref{eq:HQH-weak-fields}. Together with \eqref{eq:IQH-weak-gap} $\varphi_e$ is pinned as well and the system is gapped.

In the non-interacting case, $K_{b,e}=1$, the condition \eqref{eq:IQH-weak-gap} for gap in $\varphi_e$ is satisfied. Furthermore, in the non-interacting case, the conditions of being in the weak proximity coupling regime, \eqref{eq:Kbulk_for_Hall_gap}, yield $|y_{\Delta,2}|\ll|y_{J}|$. 

We can also verify that the assumption that the inter-species bosonic quadratic terms are negligible is sound. Along the flow such terms have an associated dimensionless coupling $y_\text{is}$ with an RG flow equation $\frac{\diff y_\text{is}}{\diff\ell}\propto C y_{\Delta,2}^2$, with $C$ of order 1. Inserting the scaling behavior \eqref{eq:scaling behavior} of $r=\Delta_2$ we find
\begin{equation}
    y_\text{is}(\ell^*)\approx C\frac{(y_{\Delta,2}^{(1)}(\ell^*))^2-(y_{\Delta,2}^{(0)})^2}{2(2-d_{\Delta,2}^{(0)})},
\end{equation}
and thus show that $y_\text{is}$ remains small during the RG flow.

\subsection{IQH strong proximity coupling regime}
\label{subsec:IQH strong proximity coupling regime}

The strong proximity coupling regime is characterized by phases in which the pairing term, in Eq.~\eqref{eq:H_pairing}, is dominant. We can use the RG equations \eqref{eq:IQH-weak-flow-D1-Ke} to analyze also the strong proximity coupling regime, because as in the weak proximity coupling regime the fields $\varphi_1,\theta_1$ decouple from $\varphi_{2,3},\theta_{2,3}$. Consequently, the condition Eq.~\eqref{eq:IQH-weak-gap} is a sufficient condition for the field $\varphi_1$ to be pinned.

We can  use a similar scaling analysis to that employed in the IQH weak proximity coupling regime to obtain a sufficient condition on the initial parameters, under which $\varphi_2$ is pinned. This condition is 
\begin{equation}\label{eq:IQH-strong-pair-cond}
    |y_{\Delta,2}|^{2-d_{\Delta,2}}\gg|y_J|^{2-d_{J}}.
\end{equation}
Together with \eqref{eq:IQH-weak-gap}, these two conditions guarantee that the system is in the strong proximity coupling regime, and that the fields $\varphi_{1,2}$ are pinned.

Tracing out the pinned fields leaves the $\varphi_3,\theta_3$ fields, which describe a non-chiral fermion (with Luttinger parameter $K_b$) perturbed by a $\cos(2\varphi_3)$ term induced by the $J$ and $\Delta_2$ terms. As discussed in Sec.~\ref{sec:Summary}, this model is gapped. In the following, we reestablish a gap in an approach that will reoccur in analysis of the FQH case. 

We examine different emergent terms that arise during the RG flow, and can lead to an energy gap. A general term that achieves this is
\begin{equation}\label{eq:IQH-strong-pairing-gap-term}
    \cos\left(2\varphi_3 + M\,2\varphi_2\right),
\end{equation}
with $M$ integer. Note that in Eq.~\eqref{eq:IQH-strong-pairing-gap-term} we did not include $\theta_2$ and $\theta_3$ in the argument of the cosine. The field $\theta_2$ is absent, since we require that the term \eqref{eq:IQH-strong-pairing-gap-term} commutes with the $\Delta_2$ term. Furthermore, the model \eqref{eq:3_wire_model} has the conserved quantity $\sum_{j\leq 0} n_j^A - \sum_{j\geq 0} n_j^B$, which prohibits appearance of $\theta_3$ (this symmetry is not used later in the fractional case). 

The scaling dimension of the term \eqref{eq:IQH-strong-pairing-gap-term} in the initial fixed point [Eq.~\eqref{eq:3_LuttLiquids}] is
\begin{equation}
    d_{M} = (M-1)^2 K_b/2 + (M+1)^2 / (2 K_b).
\end{equation}
If the term \eqref{eq:IQH-strong-pairing-gap-term} is initially relevant, i.e.,~$d_{M}<2$, then the model will be gapped. Since this model has non-commuting terms (namely $\mathcal{H}_\Delta$ and $\mathcal{H}_\text{QH}$), one might worry of competing terms that will spoil the gap established via scaling analysis. However, the emergent perturbation \eqref{eq:IQH-strong-pairing-gap-term} commutes with other perturbations of the fixed point, $\mathcal{H}_\Delta$ and $\mathcal{H}_\text{QH}$, and as such no competing terms arise and change the relevant scaling behavior of term \eqref{eq:IQH-strong-pairing-gap-term} along the flow. More rigorously, $d_M(\ell)$ is a monotonic decreasing function of the RG-time, $\ell$, up to corrections of order $y^3$. The proof of this statement is given in Appendix \ref{apdx:coupled wires RG}.

The simplest emergent term is the term with $M=0$, which is a pairing term between the $j=\pm1$ wires. This term is initially relevant if
\begin{equation}
    |K_b-2|<\sqrt{3}.
\end{equation}
This is satisfied in the non-interacting case of $K_b=1$, or conversely is impeded by repulsive inter-wire interaction obeying $K_b<2-\sqrt{3}$. The $M=0$ emergent term can be thought of as a second order perturbative correction with coupling of order $\sim J^2\Delta_2/E_\text{gap}^2$.

Another emergent term we may consider is the $M=-1$ term
\begin{equation}
    \cos\left(2\theta_A-2\theta_B\right) = \cos\left(2\varphi_3-2\varphi_2\right).
\end{equation}
This emergent term is first order perturbative term with coupling of magnitude $\sim J^2/E_\text{gap}$ and is electrically uncharged. Its scaling dimension is $d_{M=-1} = 2 K_b$, so it is relevant if $K_b < 1$.

Lastly, consider the emergent term corresponding to $M=+1$, which is relevant if $K_b>1$, i.e.,~if the system has underlying attractive interactions. From the above analysis of the scaling dimension of the term \eqref{eq:IQH-strong-pairing-gap-term} with $M=-1,0,+1$, we see that for \emph{any} initial $K_b$ we can find some integer $M$ such that $d_{M}<2$. Thus, in the strong proximity coupling regime IQH case, the $\varphi_{1,2,3}$ fields are pinned and model \eqref{eq:3_wire_model} is fully gapped.

\section{Proximity coupling to Fractional Quantum Hall Edges}\label{sec:Fractional quantum Hall case}

We now consider superconducting proximity coupling to fractional QH edges. We will focus on filling fraction $\nu=1/3$, but the analysis can be easily extended to other filling fractions of the form $\nu=1/m$ with odd $m$. A main difference between the FQH and the IQH cases is that in the IQH the initial fixed point Eq.~\eqref{eq:3_LuttLiquids} allows decoupling of the  $\varphi_{e},\theta_{e}$ degrees of freedom from the other degrees of freedom. In contrast, in the FQH case this decoupling is spoiled by the non-commuting structure of $\mathcal{H}_\text{QH}$ and $\mathcal{H}_\Delta$ as shown schematically in Fig.~\ref{fig:eff_model_ints}.

\subsection{FQH weak proximity coupling regime} \label{subsec:FQH weak pairing regime}

In the weak proximity coupling limit the fields $\theta_A, \theta_B$ are pinned by the $\mathcal{H}_\text{QH}$ term and we want to determine whether the $\varphi_e,\theta_e$ fields are gapped. The simplest emergent term that can open a gap in the $\varphi_e$ field is a co-tunneling of a Cooper pair, in which one electron tunnels into each of the counter-propagating edges. This term commutes with $\mathcal{H}_{\text{QH}}$. In terms of the bosonic fields, this term is given be the Hamiltonian density term
\begin{equation}
    \mathcal{H}_{\Delta,g} = \Delta_g \cos(6\varphi_e).
    \label{eq:weak pairing gapping term}
\end{equation}
This term appears in second order perturbation theory in $\Delta_1$, $\Delta_2$, as can be read from the relation $6\varphi_e = 4\varphi_1 + 2\varphi_2$ and the form of $\mathcal{H}_\Delta$ in Eq.~\eqref{eq:H_pairing}. If the term \eqref{eq:weak pairing gapping term} is relevant, i.e.,~
\begin{equation}\label{eq:Weak FQH gap condition}
    d_{\Delta,g} < 2,
\end{equation}
it will open a gap in the $\varphi_e$ field, since it commutes with both perturbation $\mathcal{H}_\text{QH}$ and $\mathcal{H}_\Delta$ of the fixed point $\mathcal{H}_0$ of the microscopic model in Eq.~\eqref{eq:3_wire_model}. For the fixed point \eqref{eq:3_LuttLiquids} this scaling dimension is 
\begin{equation}\label{eq:scaling Delta_g in std fixed point}
    d_{\Delta,g}=3/K_e,
\end{equation}
and the condition \eqref{eq:Weak FQH gap condition} is satisfied for a sufficiently strong electron-electron attraction, i.e.,\ 
\begin{equation}\label{eq:Weak FQH gap init cond}
    K_e > 3/2.
\end{equation}
Importantly, even if the $H_{\Delta,g}$ is not initially relevant, its scaling dimension can get renormalized during the flow. We will now describe a mechanism which can turn the term $\mathcal{H}_g$ to be relevant at some point along the flow. 

A term which plays a crucial role in this mechanism is the electronic interaction term 
\begin{equation}
    \mathcal{H}_{\Delta,B} = \tilde{\Delta}_B \psi_{L,0}^{B\dagger} \psi_{R,0}^{A\dagger} \psi_{L,0}^A \psi_{R,0}^B + \text{h.c.}
\end{equation}
In terms of the weak proximity coupling fields it can be written in the form
\begin{equation} \label{eq:DB_term}
    \mathcal{H}_{\Delta,B} = \Delta_B  \cos\left(-2\varphi_e+\varphi_A-\theta_A+\varphi_B+\theta_B\right).
\end{equation}

We can compare the initial scaling dimensions of the cosine terms in $\mathcal{H}_\Delta$ [Eq.~\eqref{eq:H_pairing}] and $\mathcal{H}_{\Delta,B}$
\begin{equation}\label{eq:FQH - pairing scaling dim}
\begin{aligned}
    d_{\Delta,1} &= 4K_e^{-1}/3 + \bigl(K_b+K_b^{-1}\bigr)/6, \\
    d_{\Delta,2} &= K_e^{-1}/3  + 4\bigl(K_b+K_b^{-1}\bigr)/6, \\
    d_{\Delta,B} &= K_e^{-1}/3 + \bigl(K_b+K_b^{-1}\bigr)/6.
\end{aligned}
\end{equation}

With Eq.~\eqref{eq:3_LuttLiquids} as the fixed point, the RG-flow equations for the couplings are
\begin{equation}
\begin{aligned}
    \frac{\diff y_{J}}{\diff\ell} &= (2-d_{J})y_{J},\\
    \frac{\diff y_{\Delta,1}}{\diff\ell} &=
        (2-d_{\Delta,1})y_{\Delta,1}-y_{\Delta,2}y_{\Delta,B},\\
    \frac{\diff y_{\Delta,2}}{\diff\ell} &=
        (2-d_{\Delta,2})y_{\Delta,2}-y_{\Delta,1}y_{\Delta,B},\\
    \frac{\diff y_{\Delta,B}}{\diff\ell} &=
        (2-d_{\Delta,B})y_{\Delta,B}-y_{\Delta,1}y_{\Delta,2},
\end{aligned}
\label{eq:Weak FQH RG y-s}
\end{equation}
and for the Luttinger parameters
\begin{align}
    \frac{\diff K_b}{\diff\ell} &= 
        \frac{1-K_b^{2}}{12} \left(
            y_{\Delta,1}^{2} + 4 y_{\Delta,2}^{2} + y_{\Delta,B}^{2}\right)
        -3K_b^{2} y_{J}^{2},\nonumber\\
    \frac{\diff K_{e}}{\diff\ell} &=
        \frac{1}{3} \left(
            4 y_{\Delta,1}^{2} + y_{\Delta,2}^{2} + y_{\Delta,B}^{2}\right).
    \label{eq:Weak FQH RG K-s}
\end{align}
As in the IQH case, in the RG flow equations \eqref{eq:Weak FQH RG K-s} we omit the inter-species bosonic quadratic terms [off diagonal terms of the $\vec{U}$ matrix in the basis of weak proximity coupling fields, Eq.~\eqref{eq:weak-pair-DOF}].

For simplicity we do not include the coupling of the emergent term $\Delta_g$ in the flow equations. This is justified in the regime of interest $K_e<3/2$ for which the term $\Delta_g$ is irrelevant at the beginning and throughout most of the flow and thus does not change the flow of the parameters in Eqs.~\eqref{eq:Weak FQH RG y-s} and \eqref{eq:Weak FQH RG K-s} considerably during the flow to strong coupling (this property of the flow was verified numerically). Sufficiently far into the flow, when the flow approaches strong coupling, the $\Delta_g$ term might become relevant which would indicate that the system is gapped. Therefore, to determine whether the system is in a gapped phase we track the scaling dimension of $\Delta_g$ throughout the flow.

In particular, for the case of repulsive interactions $K_e<1$ we do not expect that $K_e$ will renormalize enough for the highly irrelevant $\Delta_g$ to become relevant. Interestingly, below we show that in the weak proximity coupling regime the system does exhibit an energy gap for values of $K_e$ smaller than $3/2$ (but larger than $1$). In particular, the Luttinger parameter $K_e$ is renormalized by the $\Delta_1$, $\Delta_2$ and $\Delta_B$ terms, as seen in Eq.~\eqref{eq:Weak FQH RG K-s}. Thus, the gapped phase in the weak proximity coupling regime corresponds to a flow such that the terms $\Delta_1$, $\Delta_2$ and $\Delta_B$, that do not commute with $\mathcal{H}_\text{QH}$, become irrelevant and flow to weak coupling, while sufficiently renormalizing $K_e$ such that $K_e>3/2$.

\begin{figure}[!htp]
    \centering
    \subfloat[][]{
        \includegraphics[scale=0.64]
            {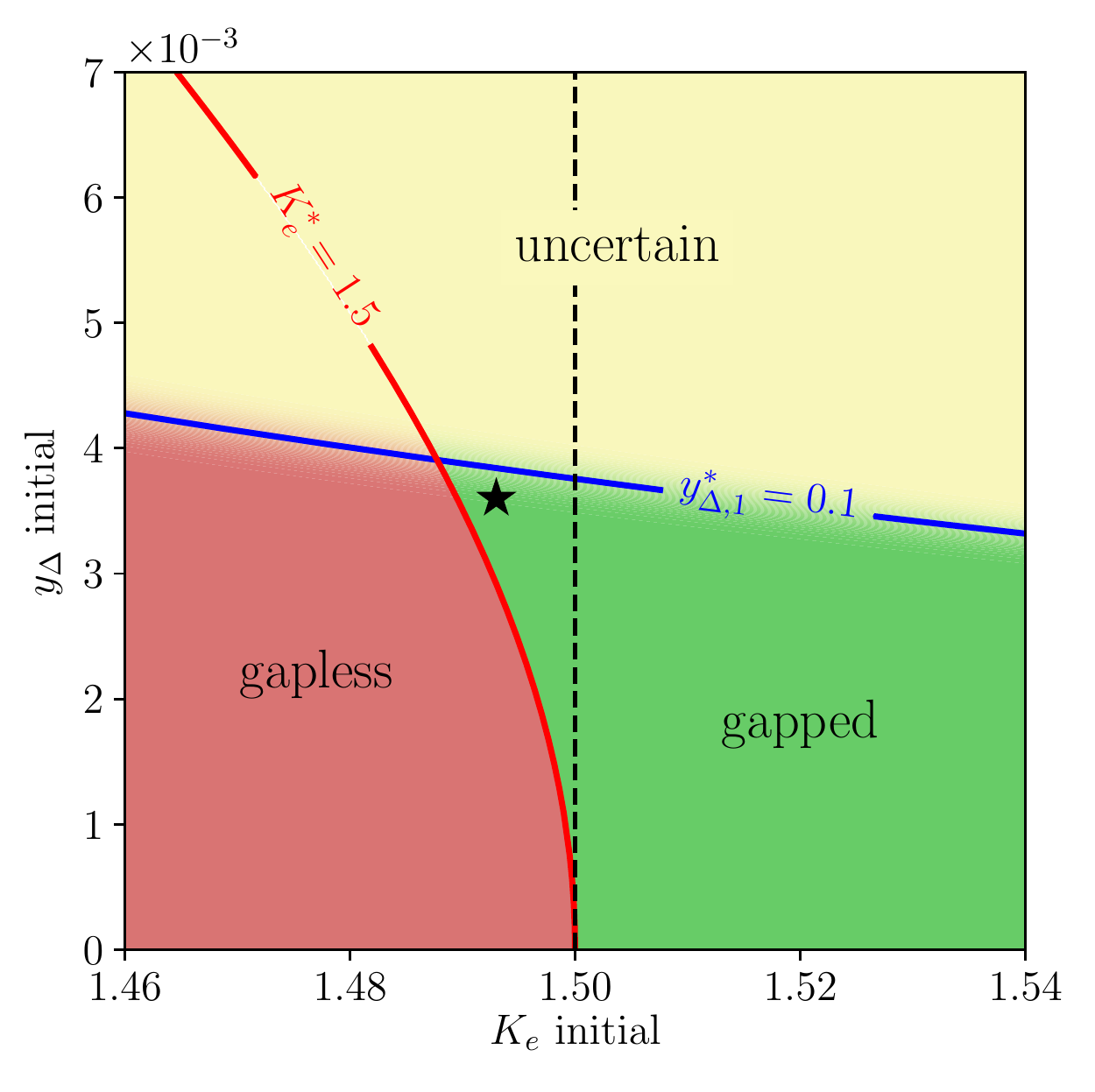}
        \label{subfig:weak pairing phase map}
    
    }
    \qquad
    \subfloat[][]{
        \includegraphics[scale=0.64]
            {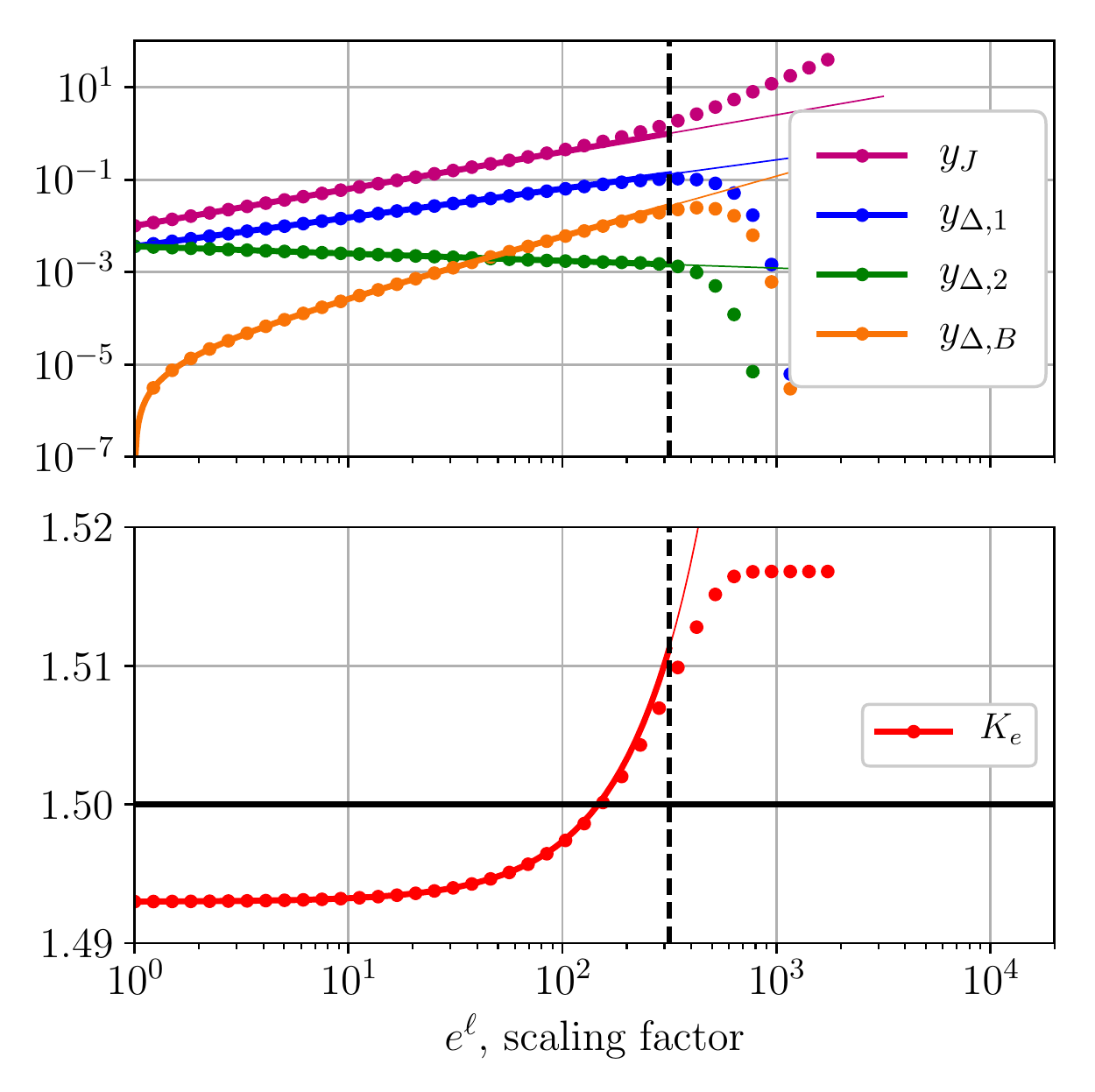}
        \label{subfig:weak pairing flow plot}
    }
    \caption{
        A phase diagram and a plot of an RG flow, both obtained numerically from RG flow Eqs.~\eqref{eq:Weak FQH RG y-s}, \eqref{eq:Weak FQH RG K-s}. Panel ~\protect\subref{subfig:weak pairing phase map} is the phase diagram as function of initial $y_\Delta\equiv y_{\Delta,1}=y_{\Delta,2}$ and $K_e$ for initial values $K_b=0.4$, $y_J=10^{-6}$ and $y_{\Delta,B}=0$. The flow is solved numerically until strong coupling is reached, with $|y_J^*|=1$. At the uncertain region, the RG analysis is unreliable as we find that the couplings of competing terms become large,  $|y_{\Delta,\alpha}|>y_\text{thresh}\equiv 0.1$ for $\alpha=1,2,B$. The gapped phase corresponds to $K_e^*>3/2$. Panel~\protect\subref{subfig:weak pairing flow plot} depicts the RG flow for the initial values at the starred point in panel~\protect\subref{subfig:weak pairing phase map}. The dotted lines are numerical solution of the flow equations and the solid lines are the approximate solutions, \eqref{eq:approx(1)}, \eqref{eq:approx DB(2)} and \eqref{eq:approx Ke}. The vertical dashed lines depict the point $\ell=\ell^*$. The values of the couplings and Luttinger parameters at this point are used to draw the phase diagram.
    }
    \label{fig:weak pairing phase map & flow plot}
\end{figure}

Numerically, we search for a flow such that the $J$ term flows to strong coupling at some RG-time $\ell^*$ at which $|y_J^*|=|y_J(\ell^*)|=1$, and examine whether the Luttinger parameter satisfies $K_e^*=K_e(\ell^*)>3/2$ at the end of the flow. Notably, $K_e(\ell)$ is monotonically increasing in $\ell$, as seen in Eq.~\eqref{eq:Weak FQH RG K-s}, so the behavior does not change to $K_e(\ell)<2/3$ at later RG-times. Caution is needed in interpreting the resulting flow when non-commuting terms are involved. Our reasoning is valid if throughout the entire flow, $|y_{\Delta,\alpha}|<y_\text{thresh}\ll1$ with $\alpha=1,2,B$ with an appropriately chosen $y_\text{thresh}$. Under this condition, we can confidently map the phase diagram. The resulting phase diagram is shown in Fig.~\ref{subfig:weak pairing phase map}. It consists of three regions: a gapped phase, a gapless phase and an uncertain region. In the uncertain region, $|y_{\Delta,\alpha}|>y_\text{thresh}$ at some $\ell$ less than $\ell^*$ and the result of the perturbative RG treatment is unclear. The criterion for a gapped phase is $d_{\Delta,g}(\ell^*)<2$, and for the gapless phase is $d_{\Delta,g}(\ell^*)>2$. A typical RG flow in the gapped phase is shown in Fig.~\ref{subfig:weak pairing flow plot}. The gapped phase in the phase diagram lies in a region of the initial model parameters which correspond to strong attractive interactions $K_e\approx3/2$. Note that the gapped phase may actually occupy a larger region of the phase diagram, but the perturbative RG method does not allow to determine the exact location of the phase transition. 

We can qualitatively find the form of the gapped phase boundary by employing an approximate iterative solution of the flow equations. For initial $y_{r}^{(0)}$ and initial scaling dimensions $d_r^{(0)}=2-\alpha_r$, we solve the RG Eqs.~\eqref{eq:Weak FQH RG y-s} to linear order in the initial couplings, yielding
\begin{equation}
    y_{r}^{(1)}(\ell) = y_{r}^{(0)} e^{\alpha_r\ell},
    \label{eq:approx(1)}
\end{equation}
where $r=\Delta_1,\Delta_2,J$, i.e.,~the couplings for which the initial value is non-zero. We must go at least one step beyond linear order to accommodate for initial $y_{\Delta,B}^{(0)}=0$. Reinserting $y_{r}^{(1)}(\ell)$ from Eq.~\eqref{eq:approx(1)} into Eqs.~\eqref{eq:Weak FQH RG y-s} yields
\begin{equation}
    \frac{\diff y_{\Delta,B}^{(2)}}{\diff\ell} = \alpha_{\Delta,B} y_{\Delta,B}^{(2)} - y_{\Delta,1}^{(1)} y_{\Delta,2}^{(1)}
\end{equation}
The solution to which is
\begin{equation}
    y_{\Delta,B}^{(2)}(\ell) = y^\prime(0) 
            e^{\alpha_{\Delta,B}\ell}-y^\prime(\ell),
    \label{eq:approx DB(2)}
\end{equation}
with $y^\prime = y_{\Delta,1}^{(1)} y_{\Delta,2}^{(1)}/(\alpha_{\Delta,1}+\alpha_{\Delta,2}-\alpha_{\Delta,B})$. We can continue to reinsert the solution back into the flow equations to obtain more accurate solutions, but for our purpose the expressions \eqref{eq:approx(1)} and \eqref{eq:approx DB(2)} will suffice.

In this perturbative approximation, the renormalized value $K_e^*$ is given by
\begin{multline}
    \Delta K_e(\ell) = K_e(\ell)- K_{e}^{(0)} \\
    \approx \intop_0^{\ell} \diff\ell\,
    \frac{1}{3}\left[
        4 (y_{\Delta,1}^{(1)})^{2} 
        + (y_{\Delta,2}^{(1)})^{2} 
        + (y_{\Delta,B}^{(2)})^{2}
    \right].\label{eq:approx Ke}
\end{multline}
Using expression \eqref{eq:approx Ke} we can ascertain  whether the condition for the gapped phase [Eq.~\eqref{eq:Weak FQH gap init cond}] is satisfied, given the initial parameters of the model. We can compare the approximate solution of the flow equations and find that they mostly agree with numerical solutions for $\ell<\ell^*$, as shown in Fig.~\ref{subfig:weak pairing flow plot}. 

In the weak proximity coupling regime we also require $\left|y_{\Delta,j}(\ell)\right|<y_\text{thresh}\ll 1$ for $j=1,2,B$ throughout $0<\ell<\ell^*$, where $y_\text{thresh}$ is a chosen threshold value. Equivalently, we can write
\begin{equation}\label{eq:approx FQH weak condition}
    |y_{\Delta,j}^{(0)}|,\,|x_{\Delta,j}|^{\alpha_{\Delta,j}} < y_\text{thresh}\ll 1,\qquad j=1,2,B
\end{equation}
where
\begin{equation}
    x_B = \frac{|y^\prime(0)|^{\alpha_{\Delta,B}^{-1}}}{|y_{J}^{(0)}|^{\alpha_{J}^{-1}}},\quad 
    x_j = \frac{|y_{\Delta,j}^{(0)}|^{\alpha_{\Delta,j}^{-1}}}{|y_{J}^{(0)}|^{\alpha_{J}^{-1}}},\quad j=1,2.
\end{equation}
In these limits and using the initial condition $K_e^{(0)}<3/2$, we can bound 
\begin{equation}
\begin{aligned}
    \Delta K_{e} &< \frac{y_{\text{thresh}}^{2}}{3}\left(\frac{1}{2|\alpha_{\Delta,B}|}+\frac{4}{|\alpha_{\Delta,1}|}+\frac{1}{2|\alpha_{\Delta,2}|}\right) \\ &\hphantom{<}+O(y_{\text{thresh}}^{3}),
\end{aligned}
\end{equation}
indicating that the gapped phase in the weak proximity coupling regime only occupies a small portion of the phase diagram. This can also be seen in the numerically obtained phase diagram, see Fig.~\ref{subfig:weak pairing phase map}.

\subsection{FQH strong proximity coupling regime}

In the strong proximity coupling regime, the $\varphi_1$ and $\varphi_2$ fields are pinned. In this section, we first use a harmonic approximation on the pinned fields which predicts that a gapped phase occurs at sufficiently strong repulsive interactions. We then use perturbative RG to map out the phase diagram. We determine the conditions for the system to be in the strong proximity coupling regime, and for an energy gap to occur. Our analysis shows that if the repulsive interactions are too strong, they suppress the proximity coupling to the SC and yield a gapless phase.

\subsubsection{Harmonic approximation analysis}

In the strong proximity coupling regime, we can use harmonic approximation to replace the terms $\cos(2\varphi_{j})$ in the Hamiltonian, with $\varphi_{j}^2$ where $j=1,2$. Below, we give a sketch of the derivation of the effective field theory, and leave the technical details to Appendix \ref{apdx:coupled wires projection}. Within the harmonic approximation, the term $\mathcal{H}_\text{QH}$ couples low-energy to high-energy states, and is thus negligible in the low-energy description. The Hamiltonian becomes quadratic in the bosonic fields with the fields $\varphi_1$, $\varphi_2$ having mass terms. Integrating out the two massive fields yields an effective low-energy model of a single Luttinger liquid with some effective Luttinger parameter, $K_\text{eff}$.

To determine $K_\text{eff}$, some care is needed since the dual fields $\theta_{1}$, $\theta_{2}$ fluctuate wildly and can affect the low-energy description due to their coupling to the low-energy modes via terms such as $\partial_x\theta_{1}\partial_x\theta_3$ and $\partial_x\theta_{2}\partial_x\varphi_3$. To properly integrate out the massive fields, a canonical transformation is needed to a new set of fields in which such coupling are absent. A general formulation of such transformations is detailed in Appendix \ref{apdx:coupled wires projection}. Here we report the resulting low-energy effective theory. 

When the above procedure is applied to the fixed point \eqref{eq:3_LuttLiquids}, it yields the effective low-energy Hamiltonian density 
\begin{equation}\label{eq:eff strong cp lim Hamiltonian}
    \mathcal{H}_\text{eff} = \frac{3u}{2\pi}\left( K_\text{eff} (\partial_x\tilde{\varphi}_3)^2 + K_\text{eff}^{-1} (\partial_x\tilde{\theta}_3)^2\right),
\end{equation}
where $\tilde{\varphi}_3$ and $\tilde{\theta}_3$ are fractional fields with commutation relations $[\partial_x\tilde\theta_3(x),\tilde\varphi_3(y)]=i\pi\delta(x-y)/3$, and
\begin{equation}
    K_\text{eff}=(K_b+K^{-1}_b)/2.
\end{equation}
The fields $\tilde{\varphi}_3,\tilde{\theta}_3$ are equal to $\varphi_3,\theta_3$ up to addition of terms linear in the expectation values of the fields $\varphi_1,\varphi_2$, which we integrate out.

In the effective Hamiltonian, we can consider the superconducting perturbation ${\Delta}_\text{eff}\int\diff x\,\cos(6\tilde{\varphi}_3)$. This perturbation has scaling dimension $3/K_\text{eff}$, and thus is relevant if
\begin{equation}\label{eq:Kbulk strong repulsion Harmonic approx}
	K_b < (3-\sqrt{5})/2\approx 0.382.
\end{equation}
This condition implies a superconducting gap is induced by sufficiently strong repulsive density-density interactions \emph{in the bulk} of the FQH liquids.

\subsubsection{Perturbative RG}

The harmonic approximation leaves a couple of unanswered questions. First, what are the conditions for the system to be in the strong proximity coupling regime? Second, what is the microscopic mechanism leading to the gapped phase in the strong proximity coupling regime? Here we will use a perturbative RG approach to address these two questions.

We begin by using a scaling analysis based on the initial scaling dimensions. We first determine the conditions for the existence of relevant operators that can pin the $\varphi_1$ and $\varphi_2$ fields. The initial scaling dimensions $d_{\Delta,1}$, $d_{\Delta,2}$ and $d_{\Delta,B}$ [see Eqs.~\eqref{eq:FQH - pairing scaling dim}] satisfy $d_{\Delta,B}<d_{\Delta,j}$ for $j=1,2$. Thus, if either the $\Delta_1$ or $\Delta_2$ term is relevant, then so is the $\Delta_B$ term and the fields $\varphi_1$ and $\varphi_2$ are pinned as required in the strong proximity coupling regime. Conversely, if both the $\Delta_1$ and $\Delta_2$ terms are irrelevant and flow to weak coupling, then the system is not in the strong proximity coupling regime. The latter flow can result from sufficiently strong repulsive interactions $K_b<2-\sqrt{3}$.

To fully understand the different possible phases in the strong proximity coupling regime, we consider an emergent perturbation that can open a gap in the system. We require that this perturbation commutes with the $\Delta_B$ term. Furthermore, we require that it conserves momentum and is relevant in a repulsive setting $K_e,K_b\leq1$. These conditions admit only a few possible perturbations, which are worked out in Appendix \ref{apdx: rel ops in strong pairing}. Other than the terms $\Delta_1$, $\Delta_2$ and $\Delta_B$, the only admitted perturbation is of the form $e^{i n(6\varphi_3+2\varphi_1-4\varphi_2)}$ with integer $n$ \footnote{Note that, in contrast to the IQH case, we have not used the conserved quantity $\sum_{j\leq 0} n_j^A - \sum_{j\geq 0} n_j^B$ present in our model. The addition of backscattering terms, that explicitly break this conservation rule, to the model do not change our arguments in the strong proximity coupling regime of the FQH case.}. The simplest perturbation of this type corresponds to $n=1$, and we denote it by 
\begin{equation}\label{eq:strong pair gapping term}
    \mathcal{H}_{J,g} = J_g \cos\left(6\varphi_3+2\varphi_1-4\varphi_2\right).
\end{equation}
The term $\mathcal{H}_{J,g}$ appears as a first order perturbative term with coupling of magnitude $J_g\sim J^2/\Delta$. 

In strong proximity coupling regime, when $\varphi_1$ and $\varphi_2$ are pinned, the emergent term \eqref{eq:strong pair gapping term} can pin the $\varphi_3$ and open an energy gap in the system. In terms of the original microscopic fermions of the wire construction, the $\mathcal{H}_{J,g}$ is a 6-body interaction (compared to the $\mathcal{H}_\text{QH}$ which is a 3-body interaction). Peculiarly, the term $\mathcal{H}_{J,g}$ is uncharged, but still induces a superconducting gap.

Note that $\mathcal{H}_{J,g}$ commutes with both $\mathcal{H}_\Delta$ and $\mathcal{H}_\text{QH}$. This implies that its scaling dimension $d_{J,g}(\ell)$ is a monotonic decreasing functions of the RG-time, $\ell$, to second order in all dimensionless couplings $y_r$ (proof of this is detailed in Appendix \ref{apdx:coupled wires RG}). Thus, in the strong proximity coupling regime, if along the RG flow we find that $\mathcal{H}_{J,g}$ is relevant then we can conclude that the system is fully gapped.

The initial scaling dimension of $\mathcal{H}_{J,g}$ is
\begin{equation}\label{eq:scaling of strong gap term}
    d_{J,g} = 6 K_b.
\end{equation}
Thus, the system is gapped under the initial condition
\begin{equation}
    K_b < 1/3,
    \label{eq:bulk strong repulsion}
\end{equation}
which is more restrictive than the condition \eqref{eq:Kbulk strong repulsion Harmonic approx}, obtained via harmonic approximation, but still shows that sufficiently strong repulsive density-density interactions in the bulk induce an energy gap. However, if the repulsive interactions are too strong, $K_b<2-\sqrt{3}$, then the $\Delta_1$, $\Delta_2$ terms are both irrelevant and the system is not in the strong proximity coupling regime. Thus we establish a gapped phase that occurs at a finite range of repulsive interactions. Curiously, in the case that the $\Delta_B$ term is relevant and flows to strong coupling, but the $\Delta_1$, $\Delta_2$ terms are irrelevant, we find a gapless phase as even if the $\Delta_B$ and $J_g$ terms pin their respective fields, there are not any relevant cosine terms left to pin remaining fields.

\begin{figure*}[!htp]
    \centering
    \subfloat[][]{
        \includegraphics[scale=0.66]
            {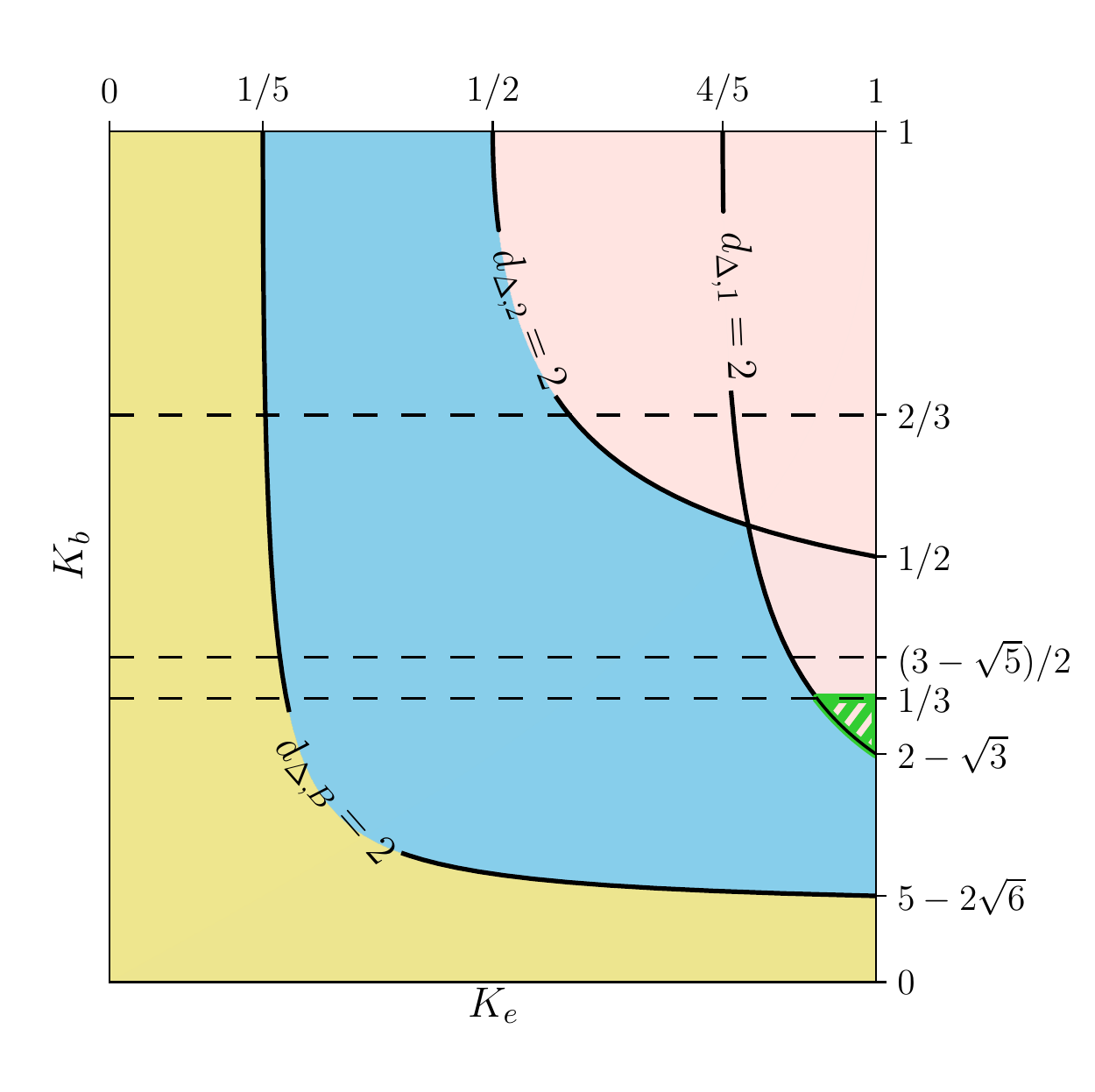}
        \label{subfig:banana}
    }
    \hfill
    \subfloat[][]{
        \includegraphics[scale=0.66]
            {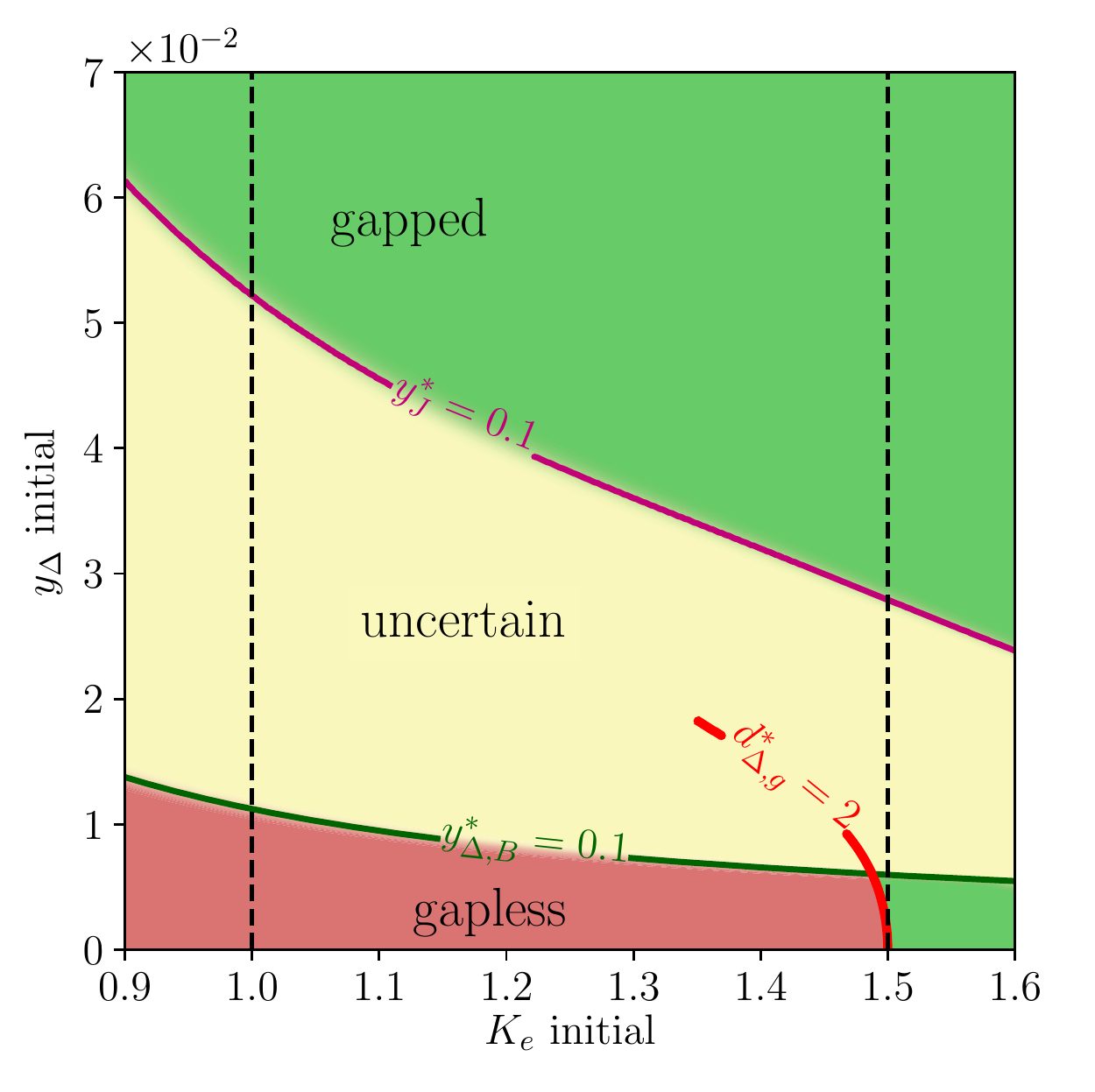}
        \label{subfig:yD-Ke strong repulsion}
    }
    \newline
    \subfloat[][]{
        \includegraphics[scale=0.66]
            {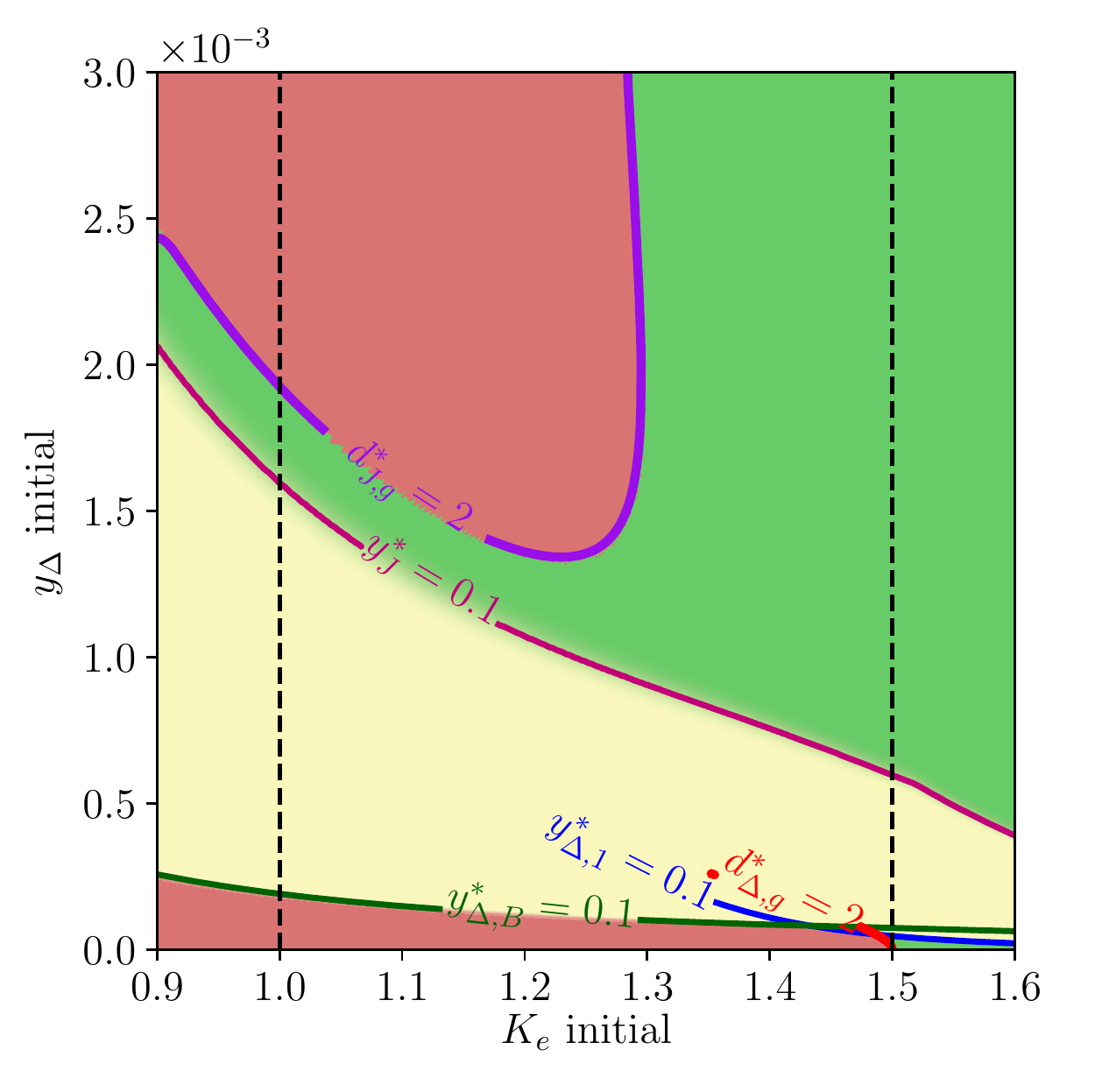}
        \label{subfig:yD-Ke weak repulsion}
    }
    \hfill
    \subfloat[][]{
        \includegraphics[scale=0.66]
            {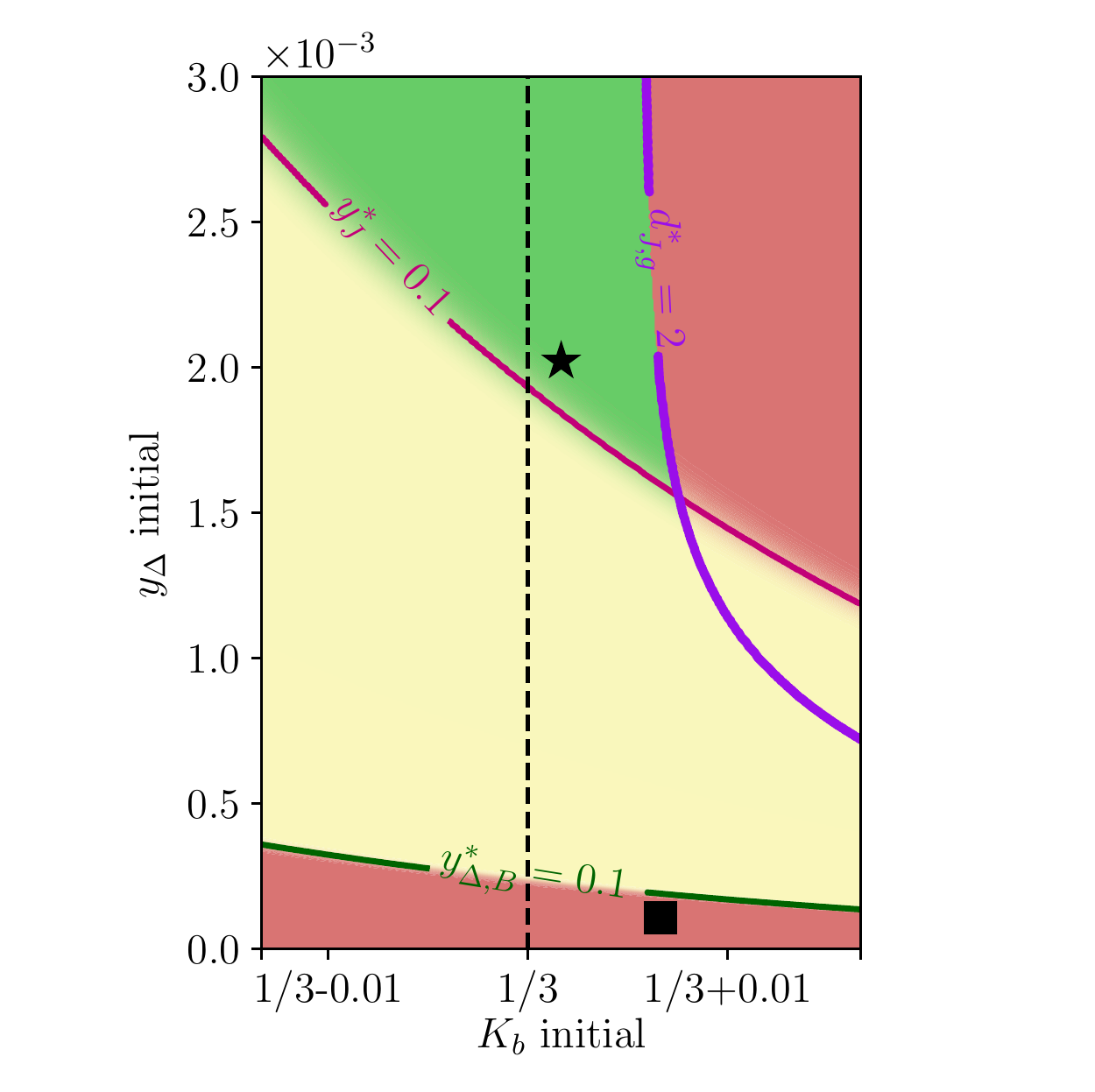}
        \label{subfig:yD-Kb}
    }
    \caption{
        \protect\subref{subfig:banana} Phase diagram based on scaling dimensions for the strong proximity coupling regime in the limit $|y_{\Delta,j}|\gg|y_J|$ for $j=1,2$. In the yellow region the system is in the weak proximity coupling regime. In the blue region the system is gapless, with $\varphi_1-\varphi_2$ pinned, but $\varphi_1+\varphi_2$ constitute a gapless mode. In the pink or green-hatched regions the model is in the strong proximity coupling regime. The green-hatched region is gapped. Panels \protect\subref{subfig:yD-Ke strong repulsion}-\protect\subref{subfig:yD-Kb} show phase diagrams obtained by solving the RG flow Eqs.~\eqref{eq:FQH RG unify} numerically with initial dimensionless couplings $y_r$ of $y_{J}=y_{J,A}=y_{J,B}=10^{-6}$, $y_\Delta=y_{\Delta,1}=y_{\Delta,2}$ and all the other y's are set to zero. These phase diagrams are suitable for both the weak and strong proximity coupling regimes. The flow is solved until $|y_r(\ell^*)|=1$ for some $r$. The uncertain regions are those where $|y_p(\ell^*)|>y_\text{thresh}=0.1$ for a $y_p$ term that does not commute with the dominant $y_r$ term. Panels~\protect\subref{subfig:yD-Ke strong repulsion} and \protect\subref{subfig:yD-Ke weak repulsion} depict a phase diagram as a function of the initial parameters $y_\Delta$ and $K_e$, as in Fig.~\ref{subfig:weak pairing phase map} (but for a broader range of initial values). In \protect\subref{subfig:yD-Ke strong repulsion} and \protect\subref{subfig:yD-Ke weak repulsion} we set an initial value $K_b=0.2$ and $K_b=0.34$ respectively. Panel \protect\subref{subfig:yD-Kb} shows the phase diagram as a function of the initial values of $y_\Delta$ and $K_b$, with initial condition $K_e=1$. A gapped phase is stabilized in the strong proximity coupling regime for sufficiently strong bulk repulsion. 
    }
    \label{fig:yD-Ke_Phase-map_fullRG_w/_strong_repulsion}
\end{figure*}

The scaling behaviour considerations lead to Fig.~\ref{subfig:banana}, which is the phase diagram in the limits of $\left|y_{J}\right| \ll \left|y_{\Delta,j}\right| \ll 1$ for $j=1,2$.
For finite values of $y_{\Delta,j}, y_J$, a phase diagram can be obtained by considering higher order perturbative RG analysis. In particular, we find a gapped phase even when the initial value of $K_b$ is larger than $1/3$. For $K_b>1/3$, although the term $\mathcal{H}_g$ is initially irrelevant, it becomes relevant along the flow, i.e.,~$d_{J,g}^* < 2$, yielding a fully gapped system. This mechanism of gap-opening is similar to the gapless-gapped phase transition in the $m=3$ weak proximity coupling regime discussed in subsection \ref{subsec:FQH weak pairing regime}. We now describe the second order perturbative RG analysis.

As in the weak proximity coupling regime, we need to identify terms that may appear and become relevant along the RG flow. Such a term is
\begin{equation}\label{eq;J2 term}
    \mathcal{H}_{J,2}= J_2\cos(6\theta_A+6\theta_B).
\end{equation}
This term does not commute with $\mathcal{H}_\Delta$, and is thus expected to flow to weak coupling at the strong proximity coupling regime. Despite flowing to weak coupling, the $J_2$ term still plays an important role as it is initially relevant and it renormalizes the fixed point $\mathcal{H}_0$ favorably for the emergent term $\mathcal{H}_{J,g}$, i.e.,\ it aids in opening a gap. This renormalization can be considerable if the system has underlying repulsive interactions, since $\mathcal{H}_{J,2}$ has an initial scaling dimension $d_{J,2}=6K_b$ (same as $\mathcal{H}_{J,g}$).

We can write the RG equations in a compact manner denoting the effective Hamiltonian density as $\mathcal{H}=\mathcal{H}_0+\mathcal{H}_{\text{pert}}$, with $\mathcal{H}_0$ as the fixed point in Eq.~\eqref{eq:model_fixed_point} and 
\begin{equation}\label{eq:Hpert}
    \mathcal{H}_{\text{pert}} = \sum_r \frac{u y_r}{\pi a^2} 
        \cos(\vec{\lambda}_r\cdot \vec{\Phi}).
\end{equation}
The index $r$ goes over the different terms with dimensionless couplings $y_{J,\alpha}$ and $y_{\Delta,\beta}$ where $\alpha=A,B,2,g$ and $\beta=1,2,B,g$ [Eqs.~\eqref{eq:H_QH density}, \eqref{eq;J2 term}, \eqref{eq:strong pair gapping term}, \eqref{eq:H_pairing},  \eqref{eq:DB_term} and \eqref{eq:weak pairing gapping term}].

At first order in the couplings $y_r$, the RG equations are determined by the scaling dimensions of the operators $\cos(\vec{\lambda}_r\cdot\vec{\Phi})$, given by
\begin{equation}
    d_r = \vec{\lambda}_r^\transpose (u\vec{U}^{-1}) \vec{\lambda}_r /2,
\end{equation}
where $\vec{U}$ is the matrix defined in the fixed point Eq.~\eqref{eq:model_fixed_point}, $u$ is the velocity scale of the model [see Eq.~\eqref{eq:conf coupled wires}], and $\vec{\lambda}^\transpose$ denotes matrix transpose of the column vector $\vec{\lambda}$.

To compactly write the $y^2$-order corrections to the RG flow equations, it is useful to introduce the fusions coefficient $C^r_{p q}$. These appear in the so called operator product expansion and commonly denoted by (see Appendix \ref{apdx:RG scheme})
\begin{equation}\label{eq:fusionrules}
    \cos(\vec{\lambda}_p\cdot \vec{\Phi}) \star
    \cos(\vec{\lambda}_q\cdot \vec{\Phi}) =
    \sum_r  C^r_{p q}
    \cos(\vec{\lambda}_r\cdot \vec{\Phi})+\cdots,
\end{equation} 
where the $\cdots$ includes less relevant terms in the expansion. The fusion coefficients are given by $C^r_{p q}=1/2$ if $\epsilon\vec{\lambda}_p+\epsilon^\prime\vec{\lambda}_q=\vec{\lambda}_r$ for some $\epsilon,\epsilon^\prime\in\{\pm1\}$ and $C^r_{p q}=0$ otherwise.

The RG flow equations are
\begin{subequations}
\begin{align}
    \frac{\diff y_{r}}{\diff\ell} &= \left(2-d_{r}\right) y_{r}
    -\sum_{p,q} C^r_{p q} y_{p} y_{q} ,
    \label{eq:FQH RG unify y}
    \\
    \vec{U}^{-1}\frac{\diff \vec{U}}{\diff \ell} &= \frac{1}{2}
        \Bigl[
            u^{-1}\vec{K}^{-1}\vec{U}\,,\ 
            \sum_r y_r^2 
            \vec{K}^{-1}\vec{\lambda}_r\vec{\lambda}_r^\transpose
        \Bigr].
        \label{eq:FQH RG unify U}
\end{align}\label{eq:FQH RG unify}
\end{subequations}
In Eq.~\eqref{eq:FQH RG unify U} $\vec{\lambda}_r$ are column vectors (and correspondingly $\vec{\lambda}_r\vec{\lambda}_r^\transpose$ are square matrices). The derivation of the RG equations \eqref{eq:FQH RG unify}, as well as an overview of the RG scheme, are given in Appendix \ref{apdx:coupled wires RG}.

The RG equations \eqref{eq:FQH RG unify} are general and can be used to analyze both the weak and the strong proximity coupling regimes. In the weak proximity coupling regime, they generalize the RG equations given in Eqs.~\eqref{eq:Weak FQH RG y-s} and \eqref{eq:Weak FQH RG K-s}, by removing the assumption that the matrix $\vec{U}$ is diagonal in the basis corresponding to \eqref{eq:weak-pair-DOF}. By solving Eqs.~\eqref{eq:FQH RG unify}, we plot a phase diagram that captures both the weak and strong proximity coupling regimes. Several examples of RG flows are shown in Appendix \ref{apdx:RG flow examples}.

We numerically solve the flow until any of the perturbative dimensionless couplings reaches $|y_r^*|=1$ at RG time $\ell^*$. If the first terms that reaches strong coupling are $y_{J,A}$ and $y_{J,B}$ (which are equal in our model), we consider the flow to belong to the weak proximity coupling regime. Conversely, if the first term to reach strong coupling is one of the set $y_{\Delta,1}$, $y_{\Delta,2}$ or $y_{\Delta,B}$, and an additional term is relevant at the same value of $\ell$, then the flow belongs to the strong proximity coupling regime. 
A gapped phase is established in the weak and strong proximity coupling regimes if at the end of the flow the scaling dimension of the corresponding emergent gap-opening term is small enough, i.e.,~$d_{\Delta,g}^*<2$ and $d_{J,g}^*<2$ respectively. 
We consider the RG flow inconclusive if $|y_p(\ell)|>y_\text{thresh}$ along the flow $\ell<\ell^*$ for the coupling of a perturbation $y_p$ that does not commute with the $y_r$ term. Several phase diagrams obtained in this manner are shown in Fig.~\ref{fig:yD-Ke_Phase-map_fullRG_w/_strong_repulsion}.

\section{Crossover from weak to strong proximity coupling}
\label{sec:duality}

We have identified a gapped phase both in the weak and in the strong proximity coupling regimes. The two phases are the same topological phases, which is uniquely characterized by the fact that the anyon condensate along the edge preserves charge only mod $2e$ \cite{Kapustin_2014, Barkeshli_2013_abel}. In this section we show that the two gapped phases are also continuously connected. To that end, we identify a $\mathbb{Z}_2$-duality, and using it we establish an energy gap in family of models that continuously connect the weak and strong proximity coupling regimes.

We start by constructing the $\mathbb{Z}_2$-duality as a linear map $\mathcal{M}:\vec{\Phi}\to\vec{M}\vec{\Phi}$, where $\vec M$ is a $6\times6$ matrix that satisfies
\begin{equation}\label{eq: Z2 duality}
    \vec{M}^2=\mathbbm{1},    
\end{equation}
and $\vec{\Phi}$ is given in Eq.~\eqref{eq:DOF 3 wire}. The duality maps the relevant terms of the weak proximity coupling regime, i.e.,\ those with couplings $J_A$, $J_B$ and $J_2$, to the relevant terms in the strong proximity coupling regime, whose couplings are $\Delta_1$, $\Delta_B$, $\Delta_2$ respectively [the different terms are given in Eqs.~\eqref{eq:H_QH density}, \eqref{eq;J2 term}, \eqref{eq:H_pairing}, and \eqref{eq:DB_term}]. Explicitly, we require that $\mathcal{M}$ maps the arguments of the appropriate cosine operators 
\begin{equation}\label{eq:M map ops}
    \mathcal{M}:\quad 
    6 \theta_A\leftrightarrow 2 \varphi_1
    \ \text{and}\ 
    6 \theta_B \leftrightarrow -2\varphi_1+2\varphi_2.
\end{equation} 
Consequently, the emergent terms that played a crucial role in the weak and strong proximity coupling regimes, with couplings $\Delta_g$ and $J_g$, are mapped onto one another. Lastly, we require the commutation relations
\begin{equation}\label{eq:comm rels M}
    \left[\partial_x(\vec{M\Phi})_\alpha(x),\, (\vec{M\Phi})_{\alpha'}(x')\right] =\left[\partial_x\Phi_\alpha(x),\, \Phi_{\alpha'}(x')\right].
\end{equation}
Alternatively, the condition \eqref{eq:comm rels M} is given in matrix form by $\vec{K}=\vec{M}^\transpose\vec{K}\vec{M}$.

The set of matrices that satisfy Eq.~\eqref{eq: Z2 duality}, \eqref{eq:M map ops} and \eqref{eq:comm rels M} are given by:
\begin{equation}
    \vec{M}=\begin{pmatrix}
        1-3p &  1-2p &   1-p &    p &    2p &    3p \\
        6p-3 &  3p-2 &    -1 &   -1 &   -3p &   -6p \\
        3-3p &     1 &    3p & 1-3p &     0 &    3p \\
         -3p &    -1 &  3p-1 & 1-3p &     0 &    3p \\
          6p &    3p &     0 &    0 &  2-3p &  3-6p \\
         -3p &   -2p &    -p &    p &  2p-1 &  3p-2 \\
    \end{pmatrix}
\end{equation}
with $p$ any real number. We fix the parameter $p$ by considering a fixed point $\mathcal{H}_0$ [Eq.~\eqref{eq:model_fixed_point}] that is both self-dual under $\mathcal{M}$ and symmetric under a $\pi$-rotation, $\mathcal{R}$ [defined in Eq.~\eqref{eq:pi-symmetry}]. Requiring that the corresponding $\vec{U}$ matrix has no zero eigenvalues (which corresponds to a vanishing velocity scale) limits the value of $p$ to $p=1/4.$

The self-dual fixed point is more easily studied in the basis of fields (recall the different field definitions in Table~\ref{table:1})
\begin{equation}\label{eq:def of dual fields}
\begin{aligned}
    \varphi_\pm &= \varphi_e\pm(\theta_A-\theta_B), \ \\
    \varphi_D &= \varphi_e-\eta_{L,0}^A+\eta_{R,0}^B, \\
    \theta_\pm &= \theta_e-\bigl(\theta_A+\theta_B\bigr)/2\pm\bigl(\eta_{R,-1}^A+\eta_{L,1}^B\bigr)/2, \\
    \theta_D &= 3(\theta_A+\theta_B),
\end{aligned}
\end{equation}
that satisfy fractional commutation relations 
\begin{equation}\label{eq:dual comm rels}
\begin{gathered}
    \left[\partial_x \varphi_\alpha(x),\,\theta_{\alpha'}(x')\right] = \pi i \delta_{\alpha,\alpha'} \delta(x-x')/m_\alpha, \\
    \left[\partial_x \varphi_\alpha(x),\,\varphi_{\alpha'}(x')\right] = \left[\partial_x \theta_\alpha(x),\,\theta_{\alpha'}(x')\right] = 0.
\end{gathered}
\end{equation}
In Eq.~\eqref{eq:dual comm rels} $\alpha,\alpha'=\pm,D$ and $m_\pm=3/2$, $m_D=1/2$. Under $\mathcal{M}$ (with $p=1/4$) and $\mathcal{R}$ the fields labeled by $\pm,\,D$ transform as
\begin{equation}
    \begin{aligned}
        \mathcal{M}:\quad &
        \varphi_\pm\to\pm\varphi_\pm,\ \ 
        \theta_\pm\to\pm\theta_\pm,\ \ 
        \varphi_D\leftrightarrow\theta_D,\\
        \mathcal{R}:\quad &
        \varphi_{\pm,D}\to-\varphi_{\pm,D},\ \ 
        \theta_{\pm,D}\to\theta_{\pm,D}.
    \end{aligned}
\end{equation}

Demanding self-duality under $\mathcal{M}$ and $\mathcal{R}$-symmetry restricts the fixed point to the form
\begin{equation}\label{eq:selfdual fixed pt}
    \mathcal{H}_0 = \frac{u}{2\pi}\sum_{\alpha=\pm,D}m_\alpha
    \left[K_\alpha(\partial_x\varphi_\alpha)^2+\frac{1}{K_\alpha}(\partial_x\theta_\alpha)^2\right],
\end{equation} 
with $K_D=1$ and $K_\pm>0$. This fixed point is compatible with the form of $\mathcal{H}_0$ in Eq.~\eqref{eq:model_fixed_point} [but generally differs from the fixed point in Eq.~\eqref{eq:3_LuttLiquids}]. The different perturbations and emergent terms of the effective model can be rewritten using the fields labeled $\pm$, $D$ as
\begin{widetext}
\begin{equation}\label{eq:Hpert duality}
\begin{aligned}
    \mathcal{H}_{\text{pert}} &= 
            J_A\cos\Bigl(\theta_D+\frac{3}{2}(\varphi_+-\varphi_-)\Bigr) +
            J_B\cos\Bigl(\theta_D-\frac{3}{2}(\varphi_+-\varphi_-)\Bigr) +
            J_2\cos\bigl(2\theta_D\bigr) +
            \Delta_g\cos\bigl(3\varphi_++3\varphi_-\bigr) 
            \\& +
            \Delta_1\cos\Bigl(\varphi_D+\frac{3}{2}(\varphi_++\varphi_-)\Bigr) +
            \Delta_B\cos\Bigl(\varphi_D-\frac{3}{2}(\varphi_++\varphi_-)\Bigr) +
            \Delta_2\cos\bigl(2\varphi_D\bigr) +
            J_g\cos\bigl(3\varphi_+-3\varphi_-\bigr).
\end{aligned}
\end{equation}
\end{widetext}
We further restrict the different couplings by
\begin{equation}
\begin{gathered}
    g_{1,\varphi}\equiv\Delta_1=\Delta_B,
    \qquad
    g_{1,\theta}\equiv J_A=J_B,\\
    g_3\equiv\Delta_g=J_g,
    \qquad
    g_2\equiv J_2=\Delta_2.
\end{gathered}
\end{equation}
This assures the model is symmetric under $\mathcal{R}$, and if $g_{1,\theta}=g_{1,\varphi}$ then the model is also self-dual under $\mathcal{M}$. Note that the $g_3$ term has scaling dimension $d_3=3(K_+^{-1}+K_-^{-1})/2$, the $g_{1}$ terms have scaling dimension $d_1=d_3/4+1/2$ and the $g_2$ term is marginal. 

We now show that in certain limits the model $\mathcal{H}_0+\mathcal{H}_\text{pert}$, defined by Eqs.~\eqref{eq:selfdual fixed pt} and \eqref{eq:Hpert duality}, is gapped. Neglecting the marginal $g_2$ term, $\mathcal{H}_{\text{pert}}$ can be written as
\begin{equation}\label{eq:selfdual-pert}
\begin{aligned}
    \mathcal{H}_{\text{pert}} &\approx
     2 g_3 \cos(3\varphi_+)\cos(3\varphi_-) \\
     &\hphantom{\approx}+ 2 g_{1,\theta}\cos(\theta_D)\cos\bigl(3(\varphi_+-\varphi_-)/2\bigl) \\ 
     &\hphantom{\approx}+2 g_{1,\varphi}\cos(\varphi_D)\cos\bigl(3(\varphi_++\varphi_-)/2\bigr).
\end{aligned}
\end{equation}
The interesting limit for which the model is gapped occurs for $d_3<2$, and
\begin{equation}\label{eq:dual-limit}
    g_{3}^{1/(2-d_1)} \gg g_{1,\chi}^{1/(2-d_2)},\qquad\chi=\varphi,\theta.
\end{equation}

Under these conditions, the fields $\varphi_+$ and $\varphi_-$ are both pinned by the $g_3$ term in Eq.~\eqref{eq:selfdual-pert}. The limit \eqref{eq:dual-limit} allows us to ignore the $g_{1}$ terms in the calculation of the expectation values involving only the $\varphi_\pm, \theta_\pm$ fields. In particular, the two expectation values
\begin{equation}
\begin{aligned}
    C_+ &\equiv \langle\cos(3(\varphi_++\varphi_-)/2)\rangle, \\
    C_- &\equiv \langle\cos(3(\varphi_+-\varphi_-)/2)\rangle,
\end{aligned}
\end{equation}
determined by the pinned values of the fields $\varphi_\pm$ are non-vanishing and satisfy $C_+=\pm C_-$. 
This allows us to trace out the $\varphi_\pm, \theta_\pm$ fields and obtain an effective model of $\varphi_D$ and $\theta_D$ written as
\begin{multline}\label{eq:Z1 selfdual sG}
    \mathcal{H}_\text{eff} = \frac{u}{4\pi}\bigl[(\partial_x\varphi_D)^2+(\partial_x\theta_D)^2\bigr] \\+ 2g_{1,\theta}C_-\cos\theta_D + 2g_{1,\varphi}C_+\cos\varphi_D.
\end{multline}

Assuming $\mathcal{M}$-self-duality, i.e.,~$g_{1,\theta}=g_{1,\varphi}$, the effective Hamiltonian density \eqref{eq:Z1 selfdual sG} is identical to that of the $\mathbb{Z}_1$ self-dual sine-Gordon model, which is known to be gapped \cite{Lecheminant_2002}. Furthermore, the model \eqref{eq:Z1 selfdual sG} remains gapped even if the couplings of the $\cos(\theta_D)$ and $\cos(\varphi_D)$ are not equal, allowing us to relate the gap also to models with $g_{1,\theta}\neq g_{1,\varphi}$.

There is a subtlety in the fact that not all pining values for $\varphi_\pm$ are physically distinct. This subtlety becomes apparent when noting that while $3(\varphi_{+}\pm\varphi_{-})$ has compactification radius of $2\pi$, the term $3(\varphi_{+}\pm\varphi_{-})/2$ that appears in $C_\pm$ has compactification radius of $\pi$. This subtlety is resolved by noting that there are two indistinguishable fields configurations which are related by the two $\mathbb{Z}_2$ gauge symmetries 
\begin{equation}
\begin{gathered}
   \mathbb{Z}_2^\varphi: \frac{3}{2}(\varphi_++\varphi_-)\mapsto \frac{3}{2}(\varphi_++\varphi_-)+\pi 
   \text{, }
   \varphi_D\mapsto\varphi_D+\pi,\\
   \mathbb{Z}_2^\theta: \frac{3}{2}(\varphi_+-\varphi_-)\mapsto \frac{3}{2}(\varphi_+-\varphi_-)+\pi 
   \text{, }
   \theta_D\mapsto\theta_D+\pi.
\end{gathered}
\end{equation}
The presence of these indistinguishable field configurations does not change our conclusion that the model $\mathcal{H}_0+\mathcal{H}_\text{pert}$ is gapped under our assumptions.

We can set $|g_{1,\varphi}|\ll|g_{1,\theta}|$ or $|g_{1,\varphi}|\gg|g_{1,\theta}|$, to obtain models that are ``close" to the model discussed previously in the weak or strong proximity coupling regimes respectively, which are discussed in Sec.~\ref{sec:Fractional quantum Hall case}. Tuning the ratio $|g_{1,\varphi}|/|g_{1,\theta}|$ allows us to tune the system from the weak to the strong proximity coupling regime. Therefore these gapped models can be continuously deformed to each other while retaining the energy gap.

\section{Conclusions and Discussion}\label{sec:Conclusions}

In this work, we considered a system composed of two $\nu=1/3$ Laughlin FQH liquids coupled along their edges by proximity to a SC. We employed a coupled-wires construction and bosonization techniques to write an effective model for the system. Using these techniques we determined conditions for a stable gapped phase in a system with underlying repulsive interactions. We focus mainly on the two regimes of weak and strong proximity coupling compared to the bulk gap of the FQH liquid, and analyzed both limits using perturbative RG. In the weak proximity coupling regime, a sufficiently strong attractive electronic interaction is required to obtain a gap resulting from the coupling of the superconductor to the edges [see Eq.~\eqref{eq:Weak FQH gap init cond}]. In the strong proximity coupling regime, we identify an emergent term \eqref{eq:strong pair gapping term} that opens a gap in the system for sufficiently strong repulsive interactions in the bulk of the FQH fluids, Eq.~\eqref{eq:bulk strong repulsion}.

We believe that the repulsion-induced gapped phase in the strong proximity coupling regime might be of experimental relevance, since electronic systems naturally tend to be repulsive. The Luttinger parameter $K_b$ depends on the ratio between the scale of the repulsive interactions in the bulk and the Fermi velocity, which can be controlled, for example, by changing the slope of confining potential. 

The gapped phase in the strong proximity coupling regime is obtained by a competition between the proximity coupling to the SC, which pushes the fractional edge modes towards the bulk, and the repulsive interactions within the bulk, which give a high energy penalty to charge fluctuations. This effect of the repulsive interactions suppresses the penetration of the edge modes into the bulk and thus enhances their coupling to the superconductor. This scenario is compatible with the numerical findings of Ref.~\cite{Repellin_2018}.
However, when the repulsive interactions are too strong, the coupling to the superconductor is irrelevant and the strong proximity coupling regime is not attained, resulting in a gapless edge. Consequently, a gapped edge is obtained for an intermediate range of repulsive interactions.

In both regimes, we characterized the phase transition between the gapped and gapless phases as a cascaded Kosterlitz-Thouless transition of the type discussed in \cite{Podolsky_2009}. Near the transition the gapped phases have perturbations that flow to weak coupling at low-energies, but renormalize the scaling behavior of the gapping perturbations, causing them to flow to strong coupling.

Lastly, we use a $\mathbb{Z}_2$-duality to identify a family of gapped models that connect the weak and strong proximity coupling regimes. Thus, we show that the gapped phases in the weak and strong proximity coupling regimes are the same phase and are continuously connected.

\begin{acknowledgments} 
BK is grateful to Daniel Podolsky and Joseph Avron for useful discussions.
This research was supported by the Israel Science Foundation Quantum Science and Technology grant no. 2074/19. 
BK and NL acknowledge support from the European Research Council (ERC) under the European Union Horizon 2020 Research and Innovation Programme (Grant Agreement No. 639172) and from the Israeli Center of Research Excellence (I-CORE) ``Circle of Light''.
EB and AS acknowledge support from CRC 183 of the Deutsche Forschungsgemeinschaft. 
AS was supported by the European Research Council (Project LEGOTOP) and the Israeli Science Foundation. 

\end{acknowledgments}

\bibliographystyle{apsrev4-2}
\bibliography{main}

\providecommand{\noopsort}[1]{}\providecommand{\singleletter}[1]{#1}%
\begin{thebibliography}{62}%
\makeatletter
\providecommand \@ifxundefined [1]{%
 \@ifx{#1\undefined}
}%
\providecommand \@ifnum [1]{%
 \ifnum #1\expandafter \@firstoftwo
 \else \expandafter \@secondoftwo
 \fi
}%
\providecommand \@ifx [1]{%
 \ifx #1\expandafter \@firstoftwo
 \else \expandafter \@secondoftwo
 \fi
}%
\providecommand \natexlab [1]{#1}%
\providecommand \enquote  [1]{``#1''}%
\providecommand \bibnamefont  [1]{#1}%
\providecommand \bibfnamefont [1]{#1}%
\providecommand \citenamefont [1]{#1}%
\providecommand \href@noop [0]{\@secondoftwo}%
\providecommand \href [0]{\begingroup \@sanitize@url \@href}%
\providecommand \@href[1]{\@@startlink{#1}\@@href}%
\providecommand \@@href[1]{\endgroup#1\@@endlink}%
\providecommand \@sanitize@url [0]{\catcode `\\12\catcode `\$12\catcode
  `\&12\catcode `\#12\catcode `\^12\catcode `\_12\catcode `\%12\relax}%
\providecommand \@@startlink[1]{}%
\providecommand \@@endlink[0]{}%
\providecommand \url  [0]{\begingroup\@sanitize@url \@url }%
\providecommand \@url [1]{\endgroup\@href {#1}{\urlprefix }}%
\providecommand \urlprefix  [0]{URL }%
\providecommand \Eprint [0]{\href }%
\providecommand \doibase [0]{https://doi.org/}%
\providecommand \selectlanguage [0]{\@gobble}%
\providecommand \bibinfo  [0]{\@secondoftwo}%
\providecommand \bibfield  [0]{\@secondoftwo}%
\providecommand \translation [1]{[#1]}%
\providecommand \BibitemOpen [0]{}%
\providecommand \bibitemStop [0]{}%
\providecommand \bibitemNoStop [0]{.\EOS\space}%
\providecommand \EOS [0]{\spacefactor3000\relax}%
\providecommand \BibitemShut  [1]{\csname bibitem#1\endcsname}%
\let\auto@bib@innerbib\@empty
\bibitem [{\citenamefont {Kitaev}(2003)}]{Kitaev_2003}%
  \BibitemOpen
  \bibfield  {author} {\bibinfo {author} {\bibfnamefont {A.}~\bibnamefont
  {Kitaev}},\ }\href {https://doi.org/10.1016/s0003-4916(02)00018-0} {\bibfield
   {journal} {\bibinfo  {journal} {Ann. Phys.}\ }\textbf {\bibinfo {volume}
  {303}},\ \bibinfo {pages} {2} (\bibinfo {year} {2003})}\BibitemShut {NoStop}%
\bibitem [{\citenamefont {Nayak}\ \emph {et~al.}(2008)\citenamefont {Nayak},
  \citenamefont {Simon}, \citenamefont {Stern}, \citenamefont {Freedman},\ and\
  \citenamefont {Sarma}}]{Nayak_2008}%
  \BibitemOpen
  \bibfield  {author} {\bibinfo {author} {\bibfnamefont {C.}~\bibnamefont
  {Nayak}}, \bibinfo {author} {\bibfnamefont {S.~H.}\ \bibnamefont {Simon}},
  \bibinfo {author} {\bibfnamefont {A.}~\bibnamefont {Stern}}, \bibinfo
  {author} {\bibfnamefont {M.}~\bibnamefont {Freedman}},\ and\ \bibinfo
  {author} {\bibfnamefont {S.~D.}\ \bibnamefont {Sarma}},\ }\href
  {https://doi.org/10.1103/revmodphys.80.1083} {\bibfield  {journal} {\bibinfo
  {journal} {Rev. Mod. Phys.}\ }\textbf {\bibinfo {volume} {80}},\ \bibinfo
  {pages} {1083} (\bibinfo {year} {2008})}\BibitemShut {NoStop}%
\bibitem [{\citenamefont {Leinaas}\ and\ \citenamefont
  {Myrheim}(1977)}]{Leinaas_1977}%
  \BibitemOpen
  \bibfield  {author} {\bibinfo {author} {\bibfnamefont {J.~M.}\ \bibnamefont
  {Leinaas}}\ and\ \bibinfo {author} {\bibfnamefont {J.}~\bibnamefont
  {Myrheim}},\ }\href {https://doi.org/10.1007/bf02727953} {\bibfield
  {journal} {\bibinfo  {journal} {Il Nuovo Cimento B}\ }\textbf {\bibinfo
  {volume} {37}},\ \bibinfo {pages} {1} (\bibinfo {year} {1977})}\BibitemShut
  {NoStop}%
\bibitem [{\citenamefont {Wilczek}(1982)}]{Wilczek_1982}%
  \BibitemOpen
  \bibfield  {author} {\bibinfo {author} {\bibfnamefont {F.}~\bibnamefont
  {Wilczek}},\ }\href {https://doi.org/10.1103/physrevlett.49.957} {\bibfield
  {journal} {\bibinfo  {journal} {Phys. Rev. Lett.}\ }\textbf {\bibinfo
  {volume} {49}},\ \bibinfo {pages} {957} (\bibinfo {year} {1982})}\BibitemShut
  {NoStop}%
\bibitem [{\citenamefont {Wu}(1984)}]{Wu_1984}%
  \BibitemOpen
  \bibfield  {author} {\bibinfo {author} {\bibfnamefont {Y.-S.}\ \bibnamefont
  {Wu}},\ }\href {https://doi.org/10.1103/physrevlett.52.2103} {\bibfield
  {journal} {\bibinfo  {journal} {Phys. Rev. Lett.}\ }\textbf {\bibinfo
  {volume} {52}},\ \bibinfo {pages} {2103} (\bibinfo {year}
  {1984})}\BibitemShut {NoStop}%
\bibitem [{\citenamefont {Stern}(2010)}]{Stern_2010}%
  \BibitemOpen
  \bibfield  {author} {\bibinfo {author} {\bibfnamefont {A.}~\bibnamefont
  {Stern}},\ }\href {https://doi.org/10.1038/nature08915} {\bibfield  {journal}
  {\bibinfo  {journal} {Nature}\ }\textbf {\bibinfo {volume} {464}},\ \bibinfo
  {pages} {187} (\bibinfo {year} {2010})}\BibitemShut {NoStop}%
\bibitem [{\citenamefont {Freedman}\ \emph {et~al.}(2006)\citenamefont
  {Freedman}, \citenamefont {Nayak},\ and\ \citenamefont
  {Walker}}]{Freedman_2006}%
  \BibitemOpen
  \bibfield  {author} {\bibinfo {author} {\bibfnamefont {M.}~\bibnamefont
  {Freedman}}, \bibinfo {author} {\bibfnamefont {C.}~\bibnamefont {Nayak}},\
  and\ \bibinfo {author} {\bibfnamefont {K.}~\bibnamefont {Walker}},\ }\href
  {https://doi.org/10.1103/physrevb.73.245307} {\bibfield  {journal} {\bibinfo
  {journal} {Phys. Rev. B}\ }\textbf {\bibinfo {volume} {73}},\ \bibinfo
  {pages} {245307} (\bibinfo {year} {2006})}\BibitemShut {NoStop}%
\bibitem [{\citenamefont {Moore}\ and\ \citenamefont
  {Read}(1991)}]{Moore_1991}%
  \BibitemOpen
  \bibfield  {author} {\bibinfo {author} {\bibfnamefont {G.}~\bibnamefont
  {Moore}}\ and\ \bibinfo {author} {\bibfnamefont {N.}~\bibnamefont {Read}},\
  }\href {https://doi.org/10.1016/0550-3213(91)90407-o} {\bibfield  {journal}
  {\bibinfo  {journal} {Nucl. Phys. B}\ }\textbf {\bibinfo {volume} {360}},\
  \bibinfo {pages} {362} (\bibinfo {year} {1991})}\BibitemShut {NoStop}%
\bibitem [{\citenamefont {Kitaev}(2006)}]{Kitaev_2006}%
  \BibitemOpen
  \bibfield  {author} {\bibinfo {author} {\bibfnamefont {A.}~\bibnamefont
  {Kitaev}},\ }\href {https://doi.org/10.1016/j.aop.2005.10.005} {\bibfield
  {journal} {\bibinfo  {journal} {Ann. Phys.}\ }\textbf {\bibinfo {volume}
  {321}},\ \bibinfo {pages} {2} (\bibinfo {year} {2006})}\BibitemShut {NoStop}%
\bibitem [{\citenamefont {Read}\ and\ \citenamefont
  {Rezayi}(1996)}]{Read_1996}%
  \BibitemOpen
  \bibfield  {author} {\bibinfo {author} {\bibfnamefont {N.}~\bibnamefont
  {Read}}\ and\ \bibinfo {author} {\bibfnamefont {E.}~\bibnamefont {Rezayi}},\
  }\href {https://doi.org/10.1103/physrevb.54.16864} {\bibfield  {journal}
  {\bibinfo  {journal} {Phys. Rev. B}\ }\textbf {\bibinfo {volume} {54}},\
  \bibinfo {pages} {16864} (\bibinfo {year} {1996})}\BibitemShut {NoStop}%
\bibitem [{\citenamefont {Stern}(2008)}]{Stern_2008}%
  \BibitemOpen
  \bibfield  {author} {\bibinfo {author} {\bibfnamefont {A.}~\bibnamefont
  {Stern}},\ }\href {https://doi.org/10.1016/j.aop.2007.10.008} {\bibfield
  {journal} {\bibinfo  {journal} {Ann. Phys.}\ }\textbf {\bibinfo {volume}
  {323}},\ \bibinfo {pages} {204} (\bibinfo {year} {2008})}\BibitemShut
  {NoStop}%
\bibitem [{\citenamefont {Barkeshli}\ \emph
  {et~al.}(2013{\natexlab{a}})\citenamefont {Barkeshli}, \citenamefont {Jian},\
  and\ \citenamefont {Qi}}]{Barkeshli_2013}%
  \BibitemOpen
  \bibfield  {author} {\bibinfo {author} {\bibfnamefont {M.}~\bibnamefont
  {Barkeshli}}, \bibinfo {author} {\bibfnamefont {C.-M.}\ \bibnamefont
  {Jian}},\ and\ \bibinfo {author} {\bibfnamefont {X.-L.}\ \bibnamefont {Qi}},\
  }\href {https://doi.org/10.1103/physrevb.87.045130} {\bibfield  {journal}
  {\bibinfo  {journal} {Phys. Rev. B}\ }\textbf {\bibinfo {volume} {87}},\
  \bibinfo {pages} {045130} (\bibinfo {year} {2013}{\natexlab{a}})}\BibitemShut
  {NoStop}%
\bibitem [{\citenamefont {Read}\ and\ \citenamefont {Green}(2000)}]{Read_2000}%
  \BibitemOpen
  \bibfield  {author} {\bibinfo {author} {\bibfnamefont {N.}~\bibnamefont
  {Read}}\ and\ \bibinfo {author} {\bibfnamefont {D.}~\bibnamefont {Green}},\
  }\href {https://doi.org/10.1103/physrevb.61.10267} {\bibfield  {journal}
  {\bibinfo  {journal} {Phys. Rev. B}\ }\textbf {\bibinfo {volume} {61}},\
  \bibinfo {pages} {10267} (\bibinfo {year} {2000})}\BibitemShut {NoStop}%
\bibitem [{\citenamefont {Ivanov}(2001)}]{Ivanov_2001}%
  \BibitemOpen
  \bibfield  {author} {\bibinfo {author} {\bibfnamefont {D.~A.}\ \bibnamefont
  {Ivanov}},\ }\href {https://doi.org/10.1103/physrevlett.86.268} {\bibfield
  {journal} {\bibinfo  {journal} {Phys. Rev. Lett.}\ }\textbf {\bibinfo
  {volume} {86}},\ \bibinfo {pages} {268} (\bibinfo {year} {2001})}\BibitemShut
  {NoStop}%
\bibitem [{\citenamefont {Fu}\ and\ \citenamefont {Kane}(2008)}]{Fu_2008}%
  \BibitemOpen
  \bibfield  {author} {\bibinfo {author} {\bibfnamefont {L.}~\bibnamefont
  {Fu}}\ and\ \bibinfo {author} {\bibfnamefont {C.~L.}\ \bibnamefont {Kane}},\
  }\href {https://doi.org/10.1103/physrevlett.100.096407} {\bibfield  {journal}
  {\bibinfo  {journal} {Phys. Rev. Lett.}\ }\textbf {\bibinfo {volume} {100}},\
  \bibinfo {pages} {096407} (\bibinfo {year} {2008})}\BibitemShut {NoStop}%
\bibitem [{\citenamefont {Fu}\ and\ \citenamefont {Kane}(2009)}]{Fu_2009}%
  \BibitemOpen
  \bibfield  {author} {\bibinfo {author} {\bibfnamefont {L.}~\bibnamefont
  {Fu}}\ and\ \bibinfo {author} {\bibfnamefont {C.~L.}\ \bibnamefont {Kane}},\
  }\href {https://doi.org/10.1103/physrevb.79.161408} {\bibfield  {journal}
  {\bibinfo  {journal} {Phys. Rev. B}\ }\textbf {\bibinfo {volume} {79}},\
  \bibinfo {pages} {161408(R)} (\bibinfo {year} {2009})}\BibitemShut {NoStop}%
\bibitem [{\citenamefont {Stanescu}\ \emph {et~al.}(2010)\citenamefont
  {Stanescu}, \citenamefont {Sau}, \citenamefont {Lutchyn},\ and\ \citenamefont
  {Sarma}}]{Stanescu_2010}%
  \BibitemOpen
  \bibfield  {author} {\bibinfo {author} {\bibfnamefont {T.~D.}\ \bibnamefont
  {Stanescu}}, \bibinfo {author} {\bibfnamefont {J.~D.}\ \bibnamefont {Sau}},
  \bibinfo {author} {\bibfnamefont {R.~M.}\ \bibnamefont {Lutchyn}},\ and\
  \bibinfo {author} {\bibfnamefont {S.~D.}\ \bibnamefont {Sarma}},\ }\href
  {https://doi.org/10.1103/physrevb.81.241310} {\bibfield  {journal} {\bibinfo
  {journal} {Phys. Rev. B}\ }\textbf {\bibinfo {volume} {81}},\ \bibinfo
  {pages} {241310(R)} (\bibinfo {year} {2010})}\BibitemShut {NoStop}%
\bibitem [{\citenamefont {Kitaev}(2001)}]{Kitaev_2001}%
  \BibitemOpen
  \bibfield  {author} {\bibinfo {author} {\bibfnamefont {A.}~\bibnamefont
  {Kitaev}},\ }\href {https://doi.org/10.1070/1063-7869/44/10s/s29} {\bibfield
  {journal} {\bibinfo  {journal} {Phys.-Uspekhi}\ }\textbf {\bibinfo {volume}
  {44}},\ \bibinfo {pages} {131} (\bibinfo {year} {2001})}\BibitemShut
  {NoStop}%
\bibitem [{\citenamefont {Oreg}\ \emph {et~al.}(2010)\citenamefont {Oreg},
  \citenamefont {Refael},\ and\ \citenamefont {von Oppen}}]{Oreg_2010}%
  \BibitemOpen
  \bibfield  {author} {\bibinfo {author} {\bibfnamefont {Y.}~\bibnamefont
  {Oreg}}, \bibinfo {author} {\bibfnamefont {G.}~\bibnamefont {Refael}},\ and\
  \bibinfo {author} {\bibfnamefont {F.}~\bibnamefont {von Oppen}},\ }\href
  {https://doi.org/10.1103/physrevlett.105.177002} {\bibfield  {journal}
  {\bibinfo  {journal} {Phys. Rev. Lett.}\ }\textbf {\bibinfo {volume} {105}},\
  \bibinfo {pages} {177002} (\bibinfo {year} {2010})}\BibitemShut {NoStop}%
\bibitem [{\citenamefont {Lutchyn}\ \emph {et~al.}(2010)\citenamefont
  {Lutchyn}, \citenamefont {Sau},\ and\ \citenamefont {Sarma}}]{Lutchyn_2010}%
  \BibitemOpen
  \bibfield  {author} {\bibinfo {author} {\bibfnamefont {R.~M.}\ \bibnamefont
  {Lutchyn}}, \bibinfo {author} {\bibfnamefont {J.~D.}\ \bibnamefont {Sau}},\
  and\ \bibinfo {author} {\bibfnamefont {S.~D.}\ \bibnamefont {Sarma}},\ }\href
  {https://doi.org/10.1103/physrevlett.105.077001} {\bibfield  {journal}
  {\bibinfo  {journal} {Phys. Rev. Lett.}\ }\textbf {\bibinfo {volume} {105}},\
  \bibinfo {pages} {077001} (\bibinfo {year} {2010})}\BibitemShut {NoStop}%
\bibitem [{\citenamefont {Cook}\ and\ \citenamefont {Franz}(2011)}]{Cook_2011}%
  \BibitemOpen
  \bibfield  {author} {\bibinfo {author} {\bibfnamefont {A.}~\bibnamefont
  {Cook}}\ and\ \bibinfo {author} {\bibfnamefont {M.}~\bibnamefont {Franz}},\
  }\href {https://doi.org/10.1103/physrevb.84.201105} {\bibfield  {journal}
  {\bibinfo  {journal} {Phys. Rev. B}\ }\textbf {\bibinfo {volume} {84}},\
  \bibinfo {pages} {201105(R)} (\bibinfo {year} {2011})}\BibitemShut {NoStop}%
\bibitem [{\citenamefont {Mourik}\ \emph {et~al.}(2012)\citenamefont {Mourik},
  \citenamefont {Zuo}, \citenamefont {Frolov}, \citenamefont {Plissard},
  \citenamefont {Bakkers},\ and\ \citenamefont {Kouwenhoven}}]{Mourik_2012}%
  \BibitemOpen
  \bibfield  {author} {\bibinfo {author} {\bibfnamefont {V.}~\bibnamefont
  {Mourik}}, \bibinfo {author} {\bibfnamefont {K.}~\bibnamefont {Zuo}},
  \bibinfo {author} {\bibfnamefont {S.~M.}\ \bibnamefont {Frolov}}, \bibinfo
  {author} {\bibfnamefont {S.~R.}\ \bibnamefont {Plissard}}, \bibinfo {author}
  {\bibfnamefont {E.~P. A.~M.}\ \bibnamefont {Bakkers}},\ and\ \bibinfo
  {author} {\bibfnamefont {L.~P.}\ \bibnamefont {Kouwenhoven}},\ }\href
  {https://doi.org/10.1126/science.1222360} {\bibfield  {journal} {\bibinfo
  {journal} {Science}\ }\textbf {\bibinfo {volume} {336}},\ \bibinfo {pages}
  {1003} (\bibinfo {year} {2012})}\BibitemShut {NoStop}%
\bibitem [{\citenamefont {Rokhinson}\ \emph {et~al.}(2012)\citenamefont
  {Rokhinson}, \citenamefont {Liu},\ and\ \citenamefont
  {Furdyna}}]{Rokhinson_2012}%
  \BibitemOpen
  \bibfield  {author} {\bibinfo {author} {\bibfnamefont {L.~P.}\ \bibnamefont
  {Rokhinson}}, \bibinfo {author} {\bibfnamefont {X.}~\bibnamefont {Liu}},\
  and\ \bibinfo {author} {\bibfnamefont {J.~K.}\ \bibnamefont {Furdyna}},\
  }\href {https://doi.org/10.1038/nphys2429} {\bibfield  {journal} {\bibinfo
  {journal} {Nat. Phys.}\ }\textbf {\bibinfo {volume} {8}},\ \bibinfo {pages}
  {795} (\bibinfo {year} {2012})}\BibitemShut {NoStop}%
\bibitem [{\citenamefont {Deng}\ \emph {et~al.}(2012)\citenamefont {Deng},
  \citenamefont {Yu}, \citenamefont {Huang}, \citenamefont {Larsson},
  \citenamefont {Caroff},\ and\ \citenamefont {Xu}}]{Deng_2012}%
  \BibitemOpen
  \bibfield  {author} {\bibinfo {author} {\bibfnamefont {M.~T.}\ \bibnamefont
  {Deng}}, \bibinfo {author} {\bibfnamefont {C.~L.}\ \bibnamefont {Yu}},
  \bibinfo {author} {\bibfnamefont {G.~Y.}\ \bibnamefont {Huang}}, \bibinfo
  {author} {\bibfnamefont {M.}~\bibnamefont {Larsson}}, \bibinfo {author}
  {\bibfnamefont {P.}~\bibnamefont {Caroff}},\ and\ \bibinfo {author}
  {\bibfnamefont {H.~Q.}\ \bibnamefont {Xu}},\ }\href
  {https://doi.org/10.1021/nl303758w} {\bibfield  {journal} {\bibinfo
  {journal} {Nano Lett.}\ }\textbf {\bibinfo {volume} {12}},\ \bibinfo {pages}
  {6414} (\bibinfo {year} {2012})}\BibitemShut {NoStop}%
\bibitem [{\citenamefont {Churchill}\ \emph {et~al.}(2013)\citenamefont
  {Churchill}, \citenamefont {Fatemi}, \citenamefont {Grove-Rasmussen},
  \citenamefont {Deng}, \citenamefont {Caroff}, \citenamefont {Xu},\ and\
  \citenamefont {Marcus}}]{Churchill_2013}%
  \BibitemOpen
  \bibfield  {author} {\bibinfo {author} {\bibfnamefont {H.~O.~H.}\
  \bibnamefont {Churchill}}, \bibinfo {author} {\bibfnamefont {V.}~\bibnamefont
  {Fatemi}}, \bibinfo {author} {\bibfnamefont {K.}~\bibnamefont
  {Grove-Rasmussen}}, \bibinfo {author} {\bibfnamefont {M.~T.}\ \bibnamefont
  {Deng}}, \bibinfo {author} {\bibfnamefont {P.}~\bibnamefont {Caroff}},
  \bibinfo {author} {\bibfnamefont {H.~Q.}\ \bibnamefont {Xu}},\ and\ \bibinfo
  {author} {\bibfnamefont {C.~M.}\ \bibnamefont {Marcus}},\ }\href
  {https://doi.org/10.1103/physrevb.87.241401} {\bibfield  {journal} {\bibinfo
  {journal} {Phys. Rev. B}\ }\textbf {\bibinfo {volume} {87}},\ \bibinfo
  {pages} {241401(R)} (\bibinfo {year} {2013})}\BibitemShut {NoStop}%
\bibitem [{\citenamefont {Das}\ \emph {et~al.}(2012)\citenamefont {Das},
  \citenamefont {Ronen}, \citenamefont {Most}, \citenamefont {Oreg},
  \citenamefont {Heiblum},\ and\ \citenamefont {Shtrikman}}]{Das_2012}%
  \BibitemOpen
  \bibfield  {author} {\bibinfo {author} {\bibfnamefont {A.}~\bibnamefont
  {Das}}, \bibinfo {author} {\bibfnamefont {Y.}~\bibnamefont {Ronen}}, \bibinfo
  {author} {\bibfnamefont {Y.}~\bibnamefont {Most}}, \bibinfo {author}
  {\bibfnamefont {Y.}~\bibnamefont {Oreg}}, \bibinfo {author} {\bibfnamefont
  {M.}~\bibnamefont {Heiblum}},\ and\ \bibinfo {author} {\bibfnamefont
  {H.}~\bibnamefont {Shtrikman}},\ }\href {https://doi.org/10.1038/nphys2479}
  {\bibfield  {journal} {\bibinfo  {journal} {Nat. Phys.}\ }\textbf {\bibinfo
  {volume} {8}},\ \bibinfo {pages} {887} (\bibinfo {year} {2012})}\BibitemShut
  {NoStop}%
\bibitem [{\citenamefont {Nadj-Perge}\ \emph {et~al.}(2014)\citenamefont
  {Nadj-Perge}, \citenamefont {Drozdov}, \citenamefont {Li}, \citenamefont
  {Chen}, \citenamefont {Jeon}, \citenamefont {Seo}, \citenamefont {MacDonald},
  \citenamefont {Bernevig},\ and\ \citenamefont {Yazdani}}]{Nadj_Perge_2014}%
  \BibitemOpen
  \bibfield  {author} {\bibinfo {author} {\bibfnamefont {S.}~\bibnamefont
  {Nadj-Perge}}, \bibinfo {author} {\bibfnamefont {I.~K.}\ \bibnamefont
  {Drozdov}}, \bibinfo {author} {\bibfnamefont {J.}~\bibnamefont {Li}},
  \bibinfo {author} {\bibfnamefont {H.}~\bibnamefont {Chen}}, \bibinfo {author}
  {\bibfnamefont {S.}~\bibnamefont {Jeon}}, \bibinfo {author} {\bibfnamefont
  {J.}~\bibnamefont {Seo}}, \bibinfo {author} {\bibfnamefont {A.~H.}\
  \bibnamefont {MacDonald}}, \bibinfo {author} {\bibfnamefont {B.~A.}\
  \bibnamefont {Bernevig}},\ and\ \bibinfo {author} {\bibfnamefont
  {A.}~\bibnamefont {Yazdani}},\ }\href
  {https://doi.org/10.1126/science.1259327} {\bibfield  {journal} {\bibinfo
  {journal} {Science}\ }\textbf {\bibinfo {volume} {346}},\ \bibinfo {pages}
  {602} (\bibinfo {year} {2014})}\BibitemShut {NoStop}%
\bibitem [{\citenamefont {Albrecht}\ \emph {et~al.}(2016)\citenamefont
  {Albrecht}, \citenamefont {Higginbotham}, \citenamefont {Madsen},
  \citenamefont {Kuemmeth}, \citenamefont {Jespersen}, \citenamefont
  {Nyg{\aa}rd}, \citenamefont {Krogstrup},\ and\ \citenamefont
  {Marcus}}]{Albrecht_2016}%
  \BibitemOpen
  \bibfield  {author} {\bibinfo {author} {\bibfnamefont {S.~M.}\ \bibnamefont
  {Albrecht}}, \bibinfo {author} {\bibfnamefont {A.~P.}\ \bibnamefont
  {Higginbotham}}, \bibinfo {author} {\bibfnamefont {M.}~\bibnamefont
  {Madsen}}, \bibinfo {author} {\bibfnamefont {F.}~\bibnamefont {Kuemmeth}},
  \bibinfo {author} {\bibfnamefont {T.~S.}\ \bibnamefont {Jespersen}}, \bibinfo
  {author} {\bibfnamefont {J.}~\bibnamefont {Nyg{\aa}rd}}, \bibinfo {author}
  {\bibfnamefont {P.}~\bibnamefont {Krogstrup}},\ and\ \bibinfo {author}
  {\bibfnamefont {C.~M.}\ \bibnamefont {Marcus}},\ }\href
  {https://doi.org/10.1038/nature17162} {\bibfield  {journal} {\bibinfo
  {journal} {Nature}\ }\textbf {\bibinfo {volume} {531}},\ \bibinfo {pages}
  {206} (\bibinfo {year} {2016})}\BibitemShut {NoStop}%
\bibitem [{\citenamefont {Deacon}\ \emph {et~al.}(2017)\citenamefont {Deacon},
  \citenamefont {Wiedenmann}, \citenamefont {Bocquillon}, \citenamefont
  {Dom{\'{\i}}nguez}, \citenamefont {Klapwijk}, \citenamefont {Leubner},
  \citenamefont {Brüne}, \citenamefont {Hankiewicz}, \citenamefont {Tarucha},
  \citenamefont {Ishibashi}, \citenamefont {Buhmann},\ and\ \citenamefont
  {Molenkamp}}]{Deacon_2017}%
  \BibitemOpen
  \bibfield  {author} {\bibinfo {author} {\bibfnamefont {R.~S.}\ \bibnamefont
  {Deacon}}, \bibinfo {author} {\bibfnamefont {J.}~\bibnamefont {Wiedenmann}},
  \bibinfo {author} {\bibfnamefont {E.}~\bibnamefont {Bocquillon}}, \bibinfo
  {author} {\bibfnamefont {F.}~\bibnamefont {Dom{\'{\i}}nguez}}, \bibinfo
  {author} {\bibfnamefont {T.~M.}\ \bibnamefont {Klapwijk}}, \bibinfo {author}
  {\bibfnamefont {P.}~\bibnamefont {Leubner}}, \bibinfo {author} {\bibfnamefont
  {C.}~\bibnamefont {Brüne}}, \bibinfo {author} {\bibfnamefont {E.~M.}\
  \bibnamefont {Hankiewicz}}, \bibinfo {author} {\bibfnamefont
  {S.}~\bibnamefont {Tarucha}}, \bibinfo {author} {\bibfnamefont
  {K.}~\bibnamefont {Ishibashi}}, \bibinfo {author} {\bibfnamefont
  {H.}~\bibnamefont {Buhmann}},\ and\ \bibinfo {author} {\bibfnamefont {L.~W.}\
  \bibnamefont {Molenkamp}},\ }\href
  {https://doi.org/10.1103/physrevx.7.021011} {\bibfield  {journal} {\bibinfo
  {journal} {Phys. Rev. X}\ }\textbf {\bibinfo {volume} {7}},\ \bibinfo {pages}
  {021011} (\bibinfo {year} {2017})}\BibitemShut {NoStop}%
\bibitem [{\citenamefont {Lindner}\ \emph {et~al.}(2012)\citenamefont
  {Lindner}, \citenamefont {Berg}, \citenamefont {Refael},\ and\ \citenamefont
  {Stern}}]{Lindner_2012_parafermions}%
  \BibitemOpen
  \bibfield  {author} {\bibinfo {author} {\bibfnamefont {N.~H.}\ \bibnamefont
  {Lindner}}, \bibinfo {author} {\bibfnamefont {E.}~\bibnamefont {Berg}},
  \bibinfo {author} {\bibfnamefont {G.}~\bibnamefont {Refael}},\ and\ \bibinfo
  {author} {\bibfnamefont {A.}~\bibnamefont {Stern}},\ }\href
  {https://doi.org/10.1103/physrevx.2.041002} {\bibfield  {journal} {\bibinfo
  {journal} {Phys. Rev. X}\ }\textbf {\bibinfo {volume} {2}},\ \bibinfo {pages}
  {041002} (\bibinfo {year} {2012})}\BibitemShut {NoStop}%
\bibitem [{\citenamefont {Clarke}\ \emph {et~al.}(2013)\citenamefont {Clarke},
  \citenamefont {Alicea},\ and\ \citenamefont {Shtengel}}]{Clarke_2013}%
  \BibitemOpen
  \bibfield  {author} {\bibinfo {author} {\bibfnamefont {D.~J.}\ \bibnamefont
  {Clarke}}, \bibinfo {author} {\bibfnamefont {J.}~\bibnamefont {Alicea}},\
  and\ \bibinfo {author} {\bibfnamefont {K.}~\bibnamefont {Shtengel}},\ }\href
  {https://doi.org/10.1038/ncomms2340} {\bibfield  {journal} {\bibinfo
  {journal} {Nat. Commun.}\ }\textbf {\bibinfo {volume} {4}},\ \bibinfo {pages}
  {1348} (\bibinfo {year} {2013})}\BibitemShut {NoStop}%
\bibitem [{\citenamefont {Cheng}(2012)}]{Cheng_2012}%
  \BibitemOpen
  \bibfield  {author} {\bibinfo {author} {\bibfnamefont {M.}~\bibnamefont
  {Cheng}},\ }\href {https://doi.org/10.1103/physrevb.86.195126} {\bibfield
  {journal} {\bibinfo  {journal} {Phys. Rev. B}\ }\textbf {\bibinfo {volume}
  {86}},\ \bibinfo {pages} {195126} (\bibinfo {year} {2012})}\BibitemShut
  {NoStop}%
\bibitem [{\citenamefont {Vaezi}(2013)}]{Vaezi_2013}%
  \BibitemOpen
  \bibfield  {author} {\bibinfo {author} {\bibfnamefont {A.}~\bibnamefont
  {Vaezi}},\ }\href {https://doi.org/10.1103/PhysRevB.87.035132} {\bibfield
  {journal} {\bibinfo  {journal} {Phys. Rev. B}\ }\textbf {\bibinfo {volume}
  {87}},\ \bibinfo {pages} {035132} (\bibinfo {year} {2013})}\BibitemShut
  {NoStop}%
\bibitem [{\citenamefont {Alicea}\ and\ \citenamefont
  {Fendley}(2016)}]{Alicea_2016}%
  \BibitemOpen
  \bibfield  {author} {\bibinfo {author} {\bibfnamefont {J.}~\bibnamefont
  {Alicea}}\ and\ \bibinfo {author} {\bibfnamefont {P.}~\bibnamefont
  {Fendley}},\ }\href
  {https://doi.org/10.1146/annurev-conmatphys-031115-011336} {\bibfield
  {journal} {\bibinfo  {journal} {Annu. Rev. Condens. Matter Phys.}\ }\textbf
  {\bibinfo {volume} {7}},\ \bibinfo {pages} {119} (\bibinfo {year}
  {2016})}\BibitemShut {NoStop}%
\bibitem [{\citenamefont {Mong}\ \emph {et~al.}(2014)\citenamefont {Mong},
  \citenamefont {Clarke}, \citenamefont {Alicea}, \citenamefont {Lindner},
  \citenamefont {Fendley}, \citenamefont {Nayak}, \citenamefont {Oreg},
  \citenamefont {Stern}, \citenamefont {Berg}, \citenamefont {Shtengel},\ and\
  \citenamefont {Fisher}}]{Mong_2014}%
  \BibitemOpen
  \bibfield  {author} {\bibinfo {author} {\bibfnamefont {R.~S.~K.}\
  \bibnamefont {Mong}}, \bibinfo {author} {\bibfnamefont {D.~J.}\ \bibnamefont
  {Clarke}}, \bibinfo {author} {\bibfnamefont {J.}~\bibnamefont {Alicea}},
  \bibinfo {author} {\bibfnamefont {N.~H.}\ \bibnamefont {Lindner}}, \bibinfo
  {author} {\bibfnamefont {P.}~\bibnamefont {Fendley}}, \bibinfo {author}
  {\bibfnamefont {C.}~\bibnamefont {Nayak}}, \bibinfo {author} {\bibfnamefont
  {Y.}~\bibnamefont {Oreg}}, \bibinfo {author} {\bibfnamefont {A.}~\bibnamefont
  {Stern}}, \bibinfo {author} {\bibfnamefont {E.}~\bibnamefont {Berg}},
  \bibinfo {author} {\bibfnamefont {K.}~\bibnamefont {Shtengel}},\ and\
  \bibinfo {author} {\bibfnamefont {M.~P.~A.}\ \bibnamefont {Fisher}},\ }\href
  {https://doi.org/10.1103/physrevx.4.011036} {\bibfield  {journal} {\bibinfo
  {journal} {Phys. Rev. X}\ }\textbf {\bibinfo {volume} {4}},\ \bibinfo {pages}
  {011036} (\bibinfo {year} {2014})}\BibitemShut {NoStop}%
\bibitem [{\citenamefont {Lee}\ \emph {et~al.}(2017)\citenamefont {Lee},
  \citenamefont {Huang}, \citenamefont {Efetov}, \citenamefont {Wei},
  \citenamefont {Hart}, \citenamefont {Taniguchi}, \citenamefont {Watanabe},
  \citenamefont {Yacoby},\ and\ \citenamefont {Kim}}]{Lee_2017}%
  \BibitemOpen
  \bibfield  {author} {\bibinfo {author} {\bibfnamefont {G.-H.}\ \bibnamefont
  {Lee}}, \bibinfo {author} {\bibfnamefont {K.-F.}\ \bibnamefont {Huang}},
  \bibinfo {author} {\bibfnamefont {D.~K.}\ \bibnamefont {Efetov}}, \bibinfo
  {author} {\bibfnamefont {D.~S.}\ \bibnamefont {Wei}}, \bibinfo {author}
  {\bibfnamefont {S.}~\bibnamefont {Hart}}, \bibinfo {author} {\bibfnamefont
  {T.}~\bibnamefont {Taniguchi}}, \bibinfo {author} {\bibfnamefont
  {K.}~\bibnamefont {Watanabe}}, \bibinfo {author} {\bibfnamefont
  {A.}~\bibnamefont {Yacoby}},\ and\ \bibinfo {author} {\bibfnamefont
  {P.}~\bibnamefont {Kim}},\ }\href {https://doi.org/10.1038/nphys4084}
  {\bibfield  {journal} {\bibinfo  {journal} {Nat. Phys.}\ }\textbf {\bibinfo
  {volume} {13}},\ \bibinfo {pages} {693} (\bibinfo {year} {2017})}\BibitemShut
  {NoStop}%
\bibitem [{\citenamefont {Sanchez-Yamagishi}\ \emph {et~al.}(2016)\citenamefont
  {Sanchez-Yamagishi}, \citenamefont {Luo}, \citenamefont {Young},
  \citenamefont {Hunt}, \citenamefont {Watanabe}, \citenamefont {Taniguchi},
  \citenamefont {Ashoori},\ and\ \citenamefont
  {Jarillo-Herrero}}]{Sanchez-Yamagishi_2016}%
  \BibitemOpen
  \bibfield  {author} {\bibinfo {author} {\bibfnamefont {J.~D.}\ \bibnamefont
  {Sanchez-Yamagishi}}, \bibinfo {author} {\bibfnamefont {J.~Y.}\ \bibnamefont
  {Luo}}, \bibinfo {author} {\bibfnamefont {A.~F.}\ \bibnamefont {Young}},
  \bibinfo {author} {\bibfnamefont {B.~M.}\ \bibnamefont {Hunt}}, \bibinfo
  {author} {\bibfnamefont {K.}~\bibnamefont {Watanabe}}, \bibinfo {author}
  {\bibfnamefont {T.}~\bibnamefont {Taniguchi}}, \bibinfo {author}
  {\bibfnamefont {R.~C.}\ \bibnamefont {Ashoori}},\ and\ \bibinfo {author}
  {\bibfnamefont {P.}~\bibnamefont {Jarillo-Herrero}},\ }\href
  {https://doi.org/10.1038/nnano.2016.214} {\bibfield  {journal} {\bibinfo
  {journal} {Nat. Nanotechnol.}\ }\textbf {\bibinfo {volume} {12}},\ \bibinfo
  {pages} {118} (\bibinfo {year} {2016})}\BibitemShut {NoStop}%
\bibitem [{\citenamefont {Ronen}\ \emph {et~al.}(2018)\citenamefont {Ronen},
  \citenamefont {Cohen}, \citenamefont {Banitt}, \citenamefont {Heiblum},\ and\
  \citenamefont {Umansky}}]{Ronen_2018}%
  \BibitemOpen
  \bibfield  {author} {\bibinfo {author} {\bibfnamefont {Y.}~\bibnamefont
  {Ronen}}, \bibinfo {author} {\bibfnamefont {Y.}~\bibnamefont {Cohen}},
  \bibinfo {author} {\bibfnamefont {D.}~\bibnamefont {Banitt}}, \bibinfo
  {author} {\bibfnamefont {M.}~\bibnamefont {Heiblum}},\ and\ \bibinfo {author}
  {\bibfnamefont {V.}~\bibnamefont {Umansky}},\ }\href
  {https://doi.org/10.1038/s41567-017-0035-2} {\bibfield  {journal} {\bibinfo
  {journal} {Nat. Phys.}\ }\textbf {\bibinfo {volume} {14}},\ \bibinfo {pages}
  {411} (\bibinfo {year} {2018})}\BibitemShut {NoStop}%
\bibitem [{\citenamefont {Gul}\ \emph {et~al.}(2020)\citenamefont {Gul},
  \citenamefont {Ronen}, \citenamefont {Lee}, \citenamefont {Shapourian},
  \citenamefont {Zauberman}, \citenamefont {Lee}, \citenamefont {Watanabe},
  \citenamefont {Taniguchi}, \citenamefont {Vishwanath}, \citenamefont
  {Yacoby},\ and\ \citenamefont {Kim}}]{Ronen_2020}%
  \BibitemOpen
  \bibfield  {author} {\bibinfo {author} {\bibfnamefont {O.}~\bibnamefont
  {Gul}}, \bibinfo {author} {\bibfnamefont {Y.}~\bibnamefont {Ronen}}, \bibinfo
  {author} {\bibfnamefont {S.~Y.}\ \bibnamefont {Lee}}, \bibinfo {author}
  {\bibfnamefont {H.}~\bibnamefont {Shapourian}}, \bibinfo {author}
  {\bibfnamefont {J.}~\bibnamefont {Zauberman}}, \bibinfo {author}
  {\bibfnamefont {Y.~H.}\ \bibnamefont {Lee}}, \bibinfo {author} {\bibfnamefont
  {K.}~\bibnamefont {Watanabe}}, \bibinfo {author} {\bibfnamefont
  {T.}~\bibnamefont {Taniguchi}}, \bibinfo {author} {\bibfnamefont
  {A.}~\bibnamefont {Vishwanath}}, \bibinfo {author} {\bibfnamefont
  {A.}~\bibnamefont {Yacoby}},\ and\ \bibinfo {author} {\bibfnamefont
  {P.}~\bibnamefont {Kim}},\ }\href {https://arxiv.org/abs/2009.07836}
  {\bibinfo {title} {Induced superconductivity in the fractional quantum hall
  edge}} (\bibinfo {year} {2020}),\ \Eprint {https://arxiv.org/abs/2009.07836}
  {arXiv:2009.07836 [cond-mat.mes-hall]} \BibitemShut {NoStop}%
\bibitem [{\citenamefont {Zhang}\ and\ \citenamefont
  {Kane}(2014)}]{Zhang_2014}%
  \BibitemOpen
  \bibfield  {author} {\bibinfo {author} {\bibfnamefont {F.}~\bibnamefont
  {Zhang}}\ and\ \bibinfo {author} {\bibfnamefont {C.~L.}\ \bibnamefont
  {Kane}},\ }\href {https://doi.org/10.1103/physrevlett.113.036401} {\bibfield
  {journal} {\bibinfo  {journal} {Phys. Rev. Lett.}\ }\textbf {\bibinfo
  {volume} {113}},\ \bibinfo {pages} {036401} (\bibinfo {year}
  {2014})}\BibitemShut {NoStop}%
\bibitem [{\citenamefont {Orth}\ \emph {et~al.}(2015)\citenamefont {Orth},
  \citenamefont {Tiwari}, \citenamefont {Meng},\ and\ \citenamefont
  {Schmidt}}]{Orth_2015}%
  \BibitemOpen
  \bibfield  {author} {\bibinfo {author} {\bibfnamefont {C.~P.}\ \bibnamefont
  {Orth}}, \bibinfo {author} {\bibfnamefont {R.~P.}\ \bibnamefont {Tiwari}},
  \bibinfo {author} {\bibfnamefont {T.}~\bibnamefont {Meng}},\ and\ \bibinfo
  {author} {\bibfnamefont {T.~L.}\ \bibnamefont {Schmidt}},\ }\href
  {https://doi.org/10.1103/physrevb.91.081406} {\bibfield  {journal} {\bibinfo
  {journal} {Phys. Rev. B}\ }\textbf {\bibinfo {volume} {91}},\ \bibinfo
  {pages} {081406(R)} (\bibinfo {year} {2015})}\BibitemShut {NoStop}%
\bibitem [{\citenamefont {Fleckenstein}\ \emph {et~al.}(2019)\citenamefont
  {Fleckenstein}, \citenamefont {Ziani},\ and\ \citenamefont
  {Trauzettel}}]{Fleckenstein_2019}%
  \BibitemOpen
  \bibfield  {author} {\bibinfo {author} {\bibfnamefont {C.}~\bibnamefont
  {Fleckenstein}}, \bibinfo {author} {\bibfnamefont {N.~T.}\ \bibnamefont
  {Ziani}},\ and\ \bibinfo {author} {\bibfnamefont {B.}~\bibnamefont
  {Trauzettel}},\ }\href {https://doi.org/10.1103/physrevlett.122.066801}
  {\bibfield  {journal} {\bibinfo  {journal} {Phys. Rev. Lett.}\ }\textbf
  {\bibinfo {volume} {122}},\ \bibinfo {pages} {066801} (\bibinfo {year}
  {2019})}\BibitemShut {NoStop}%
\bibitem [{\citenamefont {Kane}\ \emph {et~al.}(2002)\citenamefont {Kane},
  \citenamefont {Mukhopadhyay},\ and\ \citenamefont
  {Lubensky}}]{Kane2002_LaughlinWire}%
  \BibitemOpen
  \bibfield  {author} {\bibinfo {author} {\bibfnamefont {C.~L.}\ \bibnamefont
  {Kane}}, \bibinfo {author} {\bibfnamefont {R.}~\bibnamefont {Mukhopadhyay}},\
  and\ \bibinfo {author} {\bibfnamefont {T.~C.}\ \bibnamefont {Lubensky}},\
  }\href@noop {} {\bibfield  {journal} {\bibinfo  {journal} {Phys. Rev. Lett.}\
  }\textbf {\bibinfo {volume} {88}} (\bibinfo {year} {2002})}\BibitemShut
  {NoStop}%
\bibitem [{\citenamefont {Repellin}\ \emph {et~al.}(2018)\citenamefont
  {Repellin}, \citenamefont {Cook}, \citenamefont {Neupert},\ and\
  \citenamefont {Regnault}}]{Repellin_2018}%
  \BibitemOpen
  \bibfield  {author} {\bibinfo {author} {\bibfnamefont {C.}~\bibnamefont
  {Repellin}}, \bibinfo {author} {\bibfnamefont {A.~M.}\ \bibnamefont {Cook}},
  \bibinfo {author} {\bibfnamefont {T.}~\bibnamefont {Neupert}},\ and\ \bibinfo
  {author} {\bibfnamefont {N.}~\bibnamefont {Regnault}},\ }\href
  {https://doi.org/10.1038/s41535-018-0085-4} {\bibfield  {journal} {\bibinfo
  {journal} {npj Quantum Mater.}\ }\textbf {\bibinfo {volume} {3}},\ \bibinfo
  {pages} {14} (\bibinfo {year} {2018})}\BibitemShut {NoStop}%
\bibitem [{\citenamefont {Laughlin}(1983)}]{Laughlin_1983}%
  \BibitemOpen
  \bibfield  {author} {\bibinfo {author} {\bibfnamefont {R.~B.}\ \bibnamefont
  {Laughlin}},\ }\href {https://doi.org/10.1103/physrevlett.50.1395} {\bibfield
   {journal} {\bibinfo  {journal} {Phys. Rev. Lett.}\ }\textbf {\bibinfo
  {volume} {50}},\ \bibinfo {pages} {1395} (\bibinfo {year}
  {1983})}\BibitemShut {NoStop}%
\bibitem [{\citenamefont {von Delft}\ and\ \citenamefont
  {Schoeller}(1998)}]{Delft1998_BosoBegginers}%
  \BibitemOpen
  \bibfield  {author} {\bibinfo {author} {\bibfnamefont {J.}~\bibnamefont {von
  Delft}}\ and\ \bibinfo {author} {\bibfnamefont {H.}~\bibnamefont
  {Schoeller}},\ }\href
  {https://doi.org/10.1002/(SICI)1521-3889(199811)7:4<225::AID-ANDP225>3.0.CO;2-L}
  {\bibfield  {journal} {\bibinfo  {journal} {Ann. Phys.}\ }\textbf {\bibinfo
  {volume} {7}},\ \bibinfo {pages} {225} (\bibinfo {year} {1998})}\BibitemShut
  {NoStop}%
\bibitem [{\citenamefont {Francesco}\ \emph {et~al.}(1997)\citenamefont
  {Francesco}, \citenamefont {Mathieu},\ and\ \citenamefont
  {S{\'{e}}n{\'{e}}chal}}]{Di_Francesco_1997}%
  \BibitemOpen
  \bibfield  {author} {\bibinfo {author} {\bibfnamefont {P.~D.}\ \bibnamefont
  {Francesco}}, \bibinfo {author} {\bibfnamefont {P.}~\bibnamefont {Mathieu}},\
  and\ \bibinfo {author} {\bibfnamefont {D.}~\bibnamefont
  {S{\'{e}}n{\'{e}}chal}},\ }\href {https://doi.org/10.1007/978-1-4612-2256-9}
  {\emph {\bibinfo {title} {Conformal Field Theory}}}\ (\bibinfo  {publisher}
  {Springer New York},\ \bibinfo {year} {1997})\BibitemShut {NoStop}%
\bibitem [{Note1()}]{Note1}%
  \BibitemOpen
  \bibinfo {note} {This transition is reminiscent of the cascaded
  Kosterlitz-Thouless transition discussed by Podolsky et al.~\cite
  {Podolsky_2009}.}\BibitemShut {Stop}%
\bibitem [{Note2()}]{Note2}%
  \BibitemOpen
  \bibinfo {note} {This term is discussed in \cite {Kane2002_LaughlinWire}. The
  operators $\psi _{\alpha }$, in $H_{\protect \text {QH,}2D}$ must operate at
  separated yet close points to avoid cancellation due to their fermionic
  algebra.}\BibitemShut {Stop}%
\bibitem [{\citenamefont {Schulz}(1980)}]{Schulz_1980}%
  \BibitemOpen
  \bibfield  {author} {\bibinfo {author} {\bibfnamefont {H.~J.}\ \bibnamefont
  {Schulz}},\ }\href {https://doi.org/10.1103/physrevb.22.5274} {\bibfield
  {journal} {\bibinfo  {journal} {Phys. Rev. B}\ }\textbf {\bibinfo {volume}
  {22}},\ \bibinfo {pages} {5274} (\bibinfo {year} {1980})}\BibitemShut
  {NoStop}%
\bibitem [{\citenamefont {Haldane}\ \emph {et~al.}(1983)\citenamefont
  {Haldane}, \citenamefont {Bak},\ and\ \citenamefont {Bohr}}]{Haldane1983}%
  \BibitemOpen
  \bibfield  {author} {\bibinfo {author} {\bibfnamefont {F.~D.~M.}\
  \bibnamefont {Haldane}}, \bibinfo {author} {\bibfnamefont {P.}~\bibnamefont
  {Bak}},\ and\ \bibinfo {author} {\bibfnamefont {T.}~\bibnamefont {Bohr}},\
  }\href {https://doi.org/10.1103/physrevb.28.2743} {\bibfield  {journal}
  {\bibinfo  {journal} {Phys. Rev. B}\ }\textbf {\bibinfo {volume} {28}},\
  \bibinfo {pages} {2743} (\bibinfo {year} {1983})}\BibitemShut {NoStop}%
\bibitem [{Note3()}]{Note3}%
  \BibitemOpen
  \bibinfo {note} {The only remnant of the deep-bulk degrees of freedom in the
  edge problem of $H$ is accounted for by specifying the fractional charge
  accumulated in the bulk, $\DOTSI \intop \ilimits@ \mathop {}\protect \tmspace
  -\thinmuskip {.1667em}{\mathgroup \symoperators d}x\protect \tmspace
  +\thinmuskip {.1667em}\partial _x\eta _{R,-1}^{A}/2\pi $ and $\DOTSI \intop
  \ilimits@ \mathop {}\protect \tmspace -\thinmuskip {.1667em}{\mathgroup
  \symoperators d}x\protect \tmspace +\thinmuskip {.1667em}\partial _x\eta
  _{L,1}^{B}/2\pi $ mod $1/m$.}\BibitemShut {Stop}%
\bibitem [{Note4()}]{Note4}%
  \BibitemOpen
  \bibinfo {note} {Condition \protect \textup {\hbox {\mathsurround \z@
  \protect \normalfont (\ignorespaces \ref {eq:K_repulsive}\unskip
  \@@italiccorr )}} establishes that any density-density of the form $\partial
  _x\eta _{R,j}^s\partial _x\eta _{L,j+1}^s$ comes with a positive coefficient.
  For the rest of the coefficients in $V_{\alpha ,\alpha '}$ to be positive, we
  also require $u(K_b+K_b^{-1})/2\geq (m+m^{-1})v_F$.}\BibitemShut {Stop}%
\bibitem [{\citenamefont {Zamolodchikov}(1995)}]{Zamolodchikov_1995}%
  \BibitemOpen
  \bibfield  {author} {\bibinfo {author} {\bibfnamefont {A.~B.}\ \bibnamefont
  {Zamolodchikov}},\ }\href {https://doi.org/10.1142/S0217751X9500053X}
  {\bibfield  {journal} {\bibinfo  {journal} {Int. J. Mod. Phys. A}\ }\textbf
  {\bibinfo {volume} {10}},\ \bibinfo {pages} {1125} (\bibinfo {year}
  {1995})}\BibitemShut {NoStop}%
\bibitem [{\citenamefont {Kosterlitz}(1974)}]{Kosterlitz_1974}%
  \BibitemOpen
  \bibfield  {author} {\bibinfo {author} {\bibfnamefont {J.~M.}\ \bibnamefont
  {Kosterlitz}},\ }\href {https://doi.org/10.1088/0022-3719/7/6/005} {\bibfield
   {journal} {\bibinfo  {journal} {J. Phys. C}\ }\textbf {\bibinfo {volume}
  {7}},\ \bibinfo {pages} {1046} (\bibinfo {year} {1974})}\BibitemShut
  {NoStop}%
\bibitem [{Note5()}]{Note5}%
  \BibitemOpen
  \bibinfo {note} {Note that, in contrast to the IQH case, we have not used the
  conserved quantity $\DOTSB \sum@ \slimits@ _{j\leq 0} n_j^A - \DOTSB \sum@
  \slimits@ _{j\geq 0} n_j^B$ present in our model. The addition of
  backscattering terms, that explicitly break this conservation rule, to the
  model do not change our arguments in the strong proximity coupling regime of
  the FQH case.}\BibitemShut {Stop}%
\bibitem [{\citenamefont {Kapustin}(2014)}]{Kapustin_2014}%
  \BibitemOpen
  \bibfield  {author} {\bibinfo {author} {\bibfnamefont {A.}~\bibnamefont
  {Kapustin}},\ }\href {https://doi.org/10.1103/physrevb.89.125307} {\bibfield
  {journal} {\bibinfo  {journal} {Phys. Rev. B}\ }\textbf {\bibinfo {volume}
  {89}},\ \bibinfo {pages} {125307} (\bibinfo {year} {2014})}\BibitemShut
  {NoStop}%
\bibitem [{\citenamefont {Barkeshli}\ \emph
  {et~al.}(2013{\natexlab{b}})\citenamefont {Barkeshli}, \citenamefont {Jian},\
  and\ \citenamefont {Qi}}]{Barkeshli_2013_abel}%
  \BibitemOpen
  \bibfield  {author} {\bibinfo {author} {\bibfnamefont {M.}~\bibnamefont
  {Barkeshli}}, \bibinfo {author} {\bibfnamefont {C.-M.}\ \bibnamefont
  {Jian}},\ and\ \bibinfo {author} {\bibfnamefont {X.-L.}\ \bibnamefont {Qi}},\
  }\href {https://doi.org/10.1103/physrevb.88.241103} {\bibfield  {journal}
  {\bibinfo  {journal} {Phys. Rev. B}\ }\textbf {\bibinfo {volume} {88}},\
  \bibinfo {pages} {241103} (\bibinfo {year} {2013}{\natexlab{b}})}\BibitemShut
  {NoStop}%
\bibitem [{\citenamefont {Lecheminant}\ \emph {et~al.}(2002)\citenamefont
  {Lecheminant}, \citenamefont {Gogolin},\ and\ \citenamefont
  {Nersesyan}}]{Lecheminant_2002}%
  \BibitemOpen
  \bibfield  {author} {\bibinfo {author} {\bibfnamefont {P.}~\bibnamefont
  {Lecheminant}}, \bibinfo {author} {\bibfnamefont {A.~O.}\ \bibnamefont
  {Gogolin}},\ and\ \bibinfo {author} {\bibfnamefont {A.~A.}\ \bibnamefont
  {Nersesyan}},\ }\href {https://doi.org/10.1016/s0550-3213(02)00474-1}
  {\bibfield  {journal} {\bibinfo  {journal} {Nucl. Phys. B}\ }\textbf
  {\bibinfo {volume} {639}},\ \bibinfo {pages} {502} (\bibinfo {year}
  {2002})}\BibitemShut {NoStop}%
\bibitem [{\citenamefont {Podolsky}\ \emph {et~al.}(2009)\citenamefont
  {Podolsky}, \citenamefont {Chandrasekharan},\ and\ \citenamefont
  {Vishwanath}}]{Podolsky_2009}%
  \BibitemOpen
  \bibfield  {author} {\bibinfo {author} {\bibfnamefont {D.}~\bibnamefont
  {Podolsky}}, \bibinfo {author} {\bibfnamefont {S.}~\bibnamefont
  {Chandrasekharan}},\ and\ \bibinfo {author} {\bibfnamefont {A.}~\bibnamefont
  {Vishwanath}},\ }\href {https://doi.org/10.1103/physrevb.80.214513}
  {\bibfield  {journal} {\bibinfo  {journal} {Phys. Rev. B}\ }\textbf {\bibinfo
  {volume} {80}},\ \bibinfo {pages} {214513} (\bibinfo {year}
  {2009})}\BibitemShut {NoStop}%
\bibitem [{\citenamefont {Cardy}(1996)}]{Cardy_1996}%
  \BibitemOpen
  \bibfield  {author} {\bibinfo {author} {\bibfnamefont {J.}~\bibnamefont
  {Cardy}},\ }\href {https://doi.org/10.1017/cbo9781316036440} {\emph {\bibinfo
  {title} {Scaling and Renormalization in Statistical Physics}}}\ (\bibinfo
  {publisher} {Cambridge University Press},\ \bibinfo {year}
  {1996})\BibitemShut {NoStop}%
\bibitem [{\citenamefont {Wen}(2007)}]{Wen_2007}%
  \BibitemOpen
  \bibfield  {author} {\bibinfo {author} {\bibfnamefont {X.-G.}\ \bibnamefont
  {Wen}},\ }\href {https://doi.org/10.1093/acprof:oso/9780199227259.001.0001}
  {\emph {\bibinfo {title} {Quantum Field Theory of Many-Body Systems}}}\
  (\bibinfo  {publisher} {Oxford University Press},\ \bibinfo {year}
  {2007})\BibitemShut {NoStop}%
\end{thebibliography}%

\appendix

\section{RG scheme}\label{apdx:RG scheme}

In the RG flow equations discussed in the main text, we employ a minimal-subtraction scheme to regularize the quantum field theory. We review this technique below (for a more extensive discussion see \cite{Cardy_1996}). We will use the action formalism which is equivalent to the Hamiltonian formalism. 

Consider the Euclidean action of a conformally invariant system $S_0$, with respect to the velocity $u$. Adding a set of local perturbations, the action becomes
\begin{equation}
    S[\theta;a] = S_0[\theta] +\sum_p \int\diff{}x\diff{}\tau\,\frac{u y_p}{a^{2-d_p}}\mathcal{O}_p(x,\tau),
\end{equation}
where $\mathcal{O}_p$ are local fields, $d_p$ are their scaling dimensions, $y_p$ are their dimensionless couplings and $a$ is the short length scale cutoff. 

We begin by describing the general conformal structure of $S_0$ \cite{Di_Francesco_1997}. The conformal structure can be better understood in holomorphic and antiholomorphic coordinates 
\begin{equation}
    z=u\tau-ix,\qquad \bar{z}=u\tau+ix.
\end{equation}
The scaling dimension and conformal spin of the operator $\mathcal{O}_p$ are $d_p$ and $s_p$. These can be extracted from the (imaginary) time-ordered two-point correlators of the conformal system, $\langle\mathcal{T}[\mathcal{O}_p(z,\bar{z})\mathcal{O}_q(0,0)]\rangle_{S_0}$. This correlator vanishes unless $d_p=d_q$ and $s_p=s_q$ in which case it takes the form 
\begin{equation}\label{eq:two-pt normalization}
    \langle\mathcal{T}[\mathcal{O}_p(z,\bar{z})\mathcal{O}_q(0,0)]\rangle_{S_0} =
    \frac{C_{p q}}{(z\bar{z})^{d_p}}\left(\frac{z}{\bar{z}}\right)^{s_p},
    \quad\lvert z\rvert > a
\end{equation}
where $C_{pq}$ is a dimensionless normalization coefficient.

More conformal data is encoded in the so-called operator product expansion (OPE). This is a formal series expansion of a product of fields of the form
\begin{widetext}
\begin{equation}\label{eq:OPE general}
    \mathcal{O}_p(z_1,\bar{z}_1) \mathcal{O}_q(z_2,\bar{z}_2) = \sum_{r}C_{p q}^{r}
        \frac{
            (z_{12}/\bar{z}_{12})^{(s_p+s_q-s_r)/2}
        }{
            (z_{12}\bar{z}_{12})^{(d_p+d_q-d_r)/2}}
        \mathcal{O}_r\left(\frac{z_1+z_2}{2},\frac{\bar{z}_1+\bar{z}_2}{2}\right),
\end{equation}
\end{widetext}
where $z_{i j}=z_i-z_j$, $\bar{z}_{i j}=\bar{z}_i-\bar{z}_j$ and $C_{p q}^{r}$ are called the fusion coefficients. A product of fields within a time-ordered correlator of the conformal model, $S_0$, can be substituted by their OPE. It is customary to denote the OPE by $\mathcal{O}_p\star\mathcal{O}_q=\sum_r C_{p q}^{r}\mathcal{O}_{r}$. 

To regularize the model we subtract the divergence of correlations of nearby fields by demanding
\begin{equation}
    \langle\mathcal{T}[\mathcal{O}_{p_1}(z_1,\bar{z}_1)\cdots\mathcal{O}_{p_n}(z_n,\bar{z}_n)]\rangle_{S_0,a} = 0
\end{equation}
if any pair of locations satisfy $|z_{i}-z_{j}|<a$. Otherwise, this correlator takes its conformal expectation value. 

The RG step proceeds by changing the cutoff $a\to\tilde{a}=a e^{\diff\ell}$. The RG flow equations are then obtained by expanding $Z=\int\mathcal{D}_a\theta\,e^{-S[\theta]}$ perturbatively in powers of $y$,
\begin{equation}
\begin{aligned}
    Z   & =\int\mathcal{D}_a\theta\,
            e^{-S_{0}[\theta]}\left[
            1-
            \int\diff{}^{2}r\,
                \sum_{p}\frac{y_{p}}{a^{2-d_{p}}}\mathcal{O}_{p}(z,\bar{z})\right.\\
        &   + \frac{1}{2}\intop_{\lvert z_{12}\rvert>a}
                \diff{}^{2}r_{1}\diff{}^{2}r_{2}\,
                \sum_{p,q}\frac{y_{p}y_{q}}{a^{4-d_{p}-d_{q}}}\mathcal{O}_{p}(z_{1},\bar{z}_{1})\mathcal{O}_{q}(z_{2},\bar{z}_{2}) \\
        &   +\left.\vphantom{\sum}O(y^{3})\vphantom{\sum_n}\right].
\end{aligned}\label{eq:Z expand}
\end{equation}
To change the cutoff to $\tilde{a}$ and keep the partition function identical we change the dimensionless couplings to $y_p\to\tilde{y}_p$ and obtain a correction at each order of perturbation theory such that the partition function remains the same.
The first order correction yields $\tilde{y}_r\approx e^{(2-d_r)\diff \ell}y_r$. The second order correction can be obtained by using the OPE and is given by
\begin{equation}
    ({\diff y_r})_{\text{2nd order}} = -\pi\sum_{p,q} C_{p q}^r y_p y_q \delta_{s_p+s_q,s_r}.
\end{equation}
Overall, the flow equations of the couplings become
\begin{equation}
    \frac{\diff y_r}{\diff\ell} = (2-d_r) y_r-\pi\sum_{p,q}C_{p q}^r y_p y_q \delta_{s_p+s_q,s_r} + O(y^3).\label{eq:RG flow general}
\end{equation}
In principle higher order terms can be obtained from the scaling dimension, conformal spin and the OPE.

\section{Renormalization group for coupled wires}\label{apdx:coupled wires RG}

Here we discuss the renormalization group flow of a model of real coupled chiral movers, $\vec{\Phi}(x)=\vec{\Phi}^\dagger(x)$, with a general $K$-matrix (see also Ref.~\cite[Ch.~7.4]{Wen_2007}), i.e.,\ 
\begin{equation}\label{eq:K-matrix comm}
    \left[\partial_x\Phi_\alpha(x),\Phi_\beta(y)\right]=2\pi i(\vec{K}^{-1})_{\alpha\beta}\delta(x-y).
\end{equation}
We consider the Hamiltonian $H(a)=H_0+H_\text{pert}$, given by 
\begin{equation}\label{eq:H0 and Hpert}
\begin{aligned}
    H_{0} &= \int\diff{}x\,\frac{1}{4\pi}  \partial_x\vec{\Phi}^{\transpose}\vec U \,\partial_x\vec{\Phi},\\
    H_{\text{pert}} &= \int\diff{}x\, \sum_r \frac{u y_r}{\pi a^2} \cos(\vec{\lambda}_r\cdot \vec{\Phi}),
\end{aligned}
\end{equation}
where the perturbations all have vanishing conformal spin. In this section we will use the RG scheme described in Appendix \ref{apdx:RG scheme} to obtain the RG equations of this model [Eqs.~\eqref{eq:FQH RG unify}]. The imaginary time used in Appendix \ref{apdx:RG scheme} is related to real time by a Wick rotation $it=\tau$.

The quadratic part of the Hamiltonian $H_0$ corresponds to a conformally symmetric model if it has only a single velocity scale $u$. We begin by showing that this holds if and only if Eq.~\eqref{eq:conf coupled wires} is satisfied. We diagonalize $H_0$ via a linear transformation of the fields $\widetilde{\Phi}_\alpha=\sum_\beta S_{\alpha\beta}\Phi_\beta$, by demanding that $\widetilde{\vec{U}}=\vec{S}^{\transpose{}-1}\vec{U}\vec{S}^{-1}$ is a diagonal matrix (note that $\vec{S}$ is not necessarily orthogonal). Furthermore, we require that the transformation decouples the commutation relations of the different fields to simple chiral fields, i.e.,\ the new fields $\widetilde{\vec{\Phi}}$ have a diagonal $K$-matrix, $\widetilde{\vec{K}}=\vec{S}^{\transpose{}-1}\vec{K}\vec{S}^{-1}$, with $\pm1$ entries on the diagonal and zeros elsewhere. Using $\widetilde{\vec{\Phi}}$, the Hamiltonian $H_0$ is written
\begin{equation}
    H_0 = \int\diff{}x\,\sum_\alpha\frac{\widetilde{U}_{\alpha\alpha}}{4\pi}(\partial_x\widetilde{\Phi}_\alpha)^2.
\end{equation}

From this we find the time evolution of the fields 
\begin{multline}\label{eq:Heisenberg}
    \widetilde{\Phi}_\alpha(x,t) = e^{iH_0 t}\widetilde{\Phi}_\alpha(x)e^{-iH_0 t} \\
    = \widetilde{\Phi}_\alpha\bigl(x-(\widetilde{\vec{K}}^{-1}\widetilde{\vec{U}})_{\alpha\alpha}t\bigr).
\end{multline}
It is interesting to note the matrix diagonalization relation  $\vec{S}\vec{K}^{-1}\vec{U}\vec{S}^{-1}=\widetilde{\vec{K}}^{-1}\widetilde{\vec{U}}$. The condition \eqref{eq:conf coupled wires} follows from $u\mathbbm{1}=\widetilde{\vec{U}}$.

\subsection{Scaling behavior}

Following common convention (for example see Ref.\ \cite[Sec.~9.B]{Delft1998_BosoBegginers}), we consider the operators
\begin{equation}\label{eq:Or}
    \mathcal{O}_r=\cos(\vec{\lambda}_r\cdot\vec{\Phi})/a^{d_r}=\cos(\widetilde{\vec{\lambda}}_r\cdot\widetilde{\vec{\Phi}})/a^{d_r}
\end{equation}
where $a$ is the UV cutoff defined such that the two-point correlators \eqref{eq:two-pt normalization} are normalized as
\begin{equation}
    \langle\mathcal{T}[\mathcal{O}_r(x,t)\mathcal{O}_r(0,0)]\rangle_{H_0} =
    \frac{1/2}{(z\bar{z})^{d_r}}\left(\frac{z}{\bar{z}}\right)^{s_r},
\end{equation}
where the expectation values are computed with respect to the ground state of $H_0$. The operators $\mathcal{O}_r$ in Eq.~\eqref{eq:Or} have scaling dimension and conformal spin
\begin{align}
    d_r &= \sum_\alpha \widetilde{\lambda}_{r,\alpha}^2/2=\vec{\lambda}_r^\transpose{} (u\vec{U}^{-1}) \vec{\lambda}_r/2, \label{eq:scaling dims}\\
    s_r &= \sum_\alpha \widetilde{\lambda}_{r,\alpha}^2/(2\widetilde{K}_{\alpha\alpha})=\vec{\lambda}_r^\transpose{} \vec{K}^{-1} \vec{\lambda}_r/2,\label{eq:conformal spins}
\end{align}
where $\widetilde{\lambda}_{r,\alpha}$ are defined via $\widetilde{\vec{\lambda}}_r\cdot \widetilde{\vec{\Phi}} \equiv \vec{\lambda}_r\cdot \vec{\Phi}$.

\subsection{Fusion of different cosine terms}

We use the leading contributions to the OPE [see Eq.~\eqref{eq:OPE general}] of $\mathcal{O}_p\star\mathcal{O}_q$ (for $p\neq q$) to derive the corresponding RG flow equations. The OPE gives
\begin{multline}
    \frac{\cos(\vec{\lambda}_p\cdot \vec{\Phi})}{a^{d_p}}\star
    \frac{\cos(\vec{\lambda}_q\cdot \vec{\Phi})}{a^{d_q}} = \\
    \frac{1}{2}\frac{\cos(\vec{\lambda}_+\cdot \vec{\Phi})}{a^{d_+}}+
    \frac{1}{2}\frac{\cos(\vec{\lambda}_-\cdot \vec{\Phi})}{a^{d_-}}+\cdots
\end{multline}
where $\vec{\lambda}_\pm=\vec{\lambda}_p\pm\vec{\lambda}_q\neq\vec{0}$ and the $(\cdots)$ includes contributions of less relevant terms. Together with the scaling dimensions \eqref{eq:scaling dims} we obtain the RG equation for the coefficient $y_r$, Eq.~\eqref{eq:FQH RG unify y}.

\subsection{Fusion of a cosine term with itself} 

The OPE $\mathcal{O}_p\star\mathcal{O}_p$ has contribution from the identity operator $\mathbbm{1}$ and $\cos(2\vec{\lambda}_p\cdot \vec{\Phi})$. The former only renormalizes the vacuum energy, which we do not keep track of, and the latter is less relevant than $\cos(\vec{\lambda}_p\cdot\vec{\Phi})$ and so is usually neglected. The next relevant operators in the OPE are
\begin{multline}\label{eq:OPE cos with self}
    \frac{\cos(\vec{\lambda}_p\cdot \vec{\Phi})}{a^{d_p}}\star
    \frac{\cos(\vec{\lambda}_p\cdot \vec{\Phi})}{a^{d_p}} = \\
    -\frac{1}{4} \sum_{\alpha,\beta} \widetilde{\lambda}_{p,\alpha} \widetilde{\lambda}_{p,\beta}\, \widetilde{\partial}_\alpha\widetilde{\Phi}_\alpha\, \widetilde{\partial}_\beta\widetilde{\Phi}_\beta
    +\cdots, 
\end{multline}
where the $(\cdots)$ includes contributions of $\mathbbm{1}$, $\cos(2\vec{\lambda}_p\vec{\Phi})$ and of terms that are less relevant than either $\cos(2\vec{\lambda}_p\vec{\Phi})$ or $\widetilde{\partial}_\alpha\widetilde{\Phi}_\alpha\, \widetilde{\partial}_\beta\widetilde{\Phi}_\beta$ (we do not keep track of the $\cos(2\vec{\lambda}_p\vec{\Phi})$ term, since it is less relevant then the $\cos(\vec{\lambda}_p\vec{\Phi})$ term). In Eq.~\eqref{eq:OPE cos with self} we use the notations 
\begin{equation}
    \widetilde{z}^\alpha \equiv i(ut-\widetilde{K}_{\alpha\alpha}x),\qquad
    \widetilde{\partial}_\alpha \equiv \partial/\partial \widetilde{z}^\alpha.
\end{equation}
Using the relation $\widetilde{\partial}_\alpha\widetilde{\Phi}_\alpha=i\partial_x\widetilde{\Phi}_\alpha/\widetilde{K}_{\alpha\alpha}$ [that follows from Eq.~\eqref{eq:Heisenberg}], we can rewrite 
\begin{equation}
    -\widetilde{K}_{\alpha\alpha} \widetilde{K}_{\beta\beta}\, \widetilde{\partial}_\alpha\widetilde{\Phi}_\alpha\, \widetilde{\partial}_\beta\widetilde{\Phi}_\beta = 
    \partial_x\widetilde{\Phi}_\alpha\, \partial_x\widetilde{\Phi}_\beta .
\end{equation} 
If we further restrict to conformally invariant perturbations, i.e.,~those with $s_p=0$, we can use Eq.~\eqref{eq:RG flow general} to find corrections to the fixed point $H_0$ of the form
\begin{equation}\label{eq:H0-fix-point-correction}
    \frac{\diff H_0}{\diff \ell} = \sump_{\alpha,\beta}\,
    \int \diff x\,\sum_p \frac{u }{4\pi} y_p^2 \widetilde{\lambda}_{p,\alpha} \widetilde{\lambda}_{p,\beta}
    \partial_x\widetilde{\Phi}_\alpha \partial_x\widetilde{\Phi}_\beta ,
\end{equation} 
where the primed sum goes over all pairs $(\alpha,\beta)$ such that $\widetilde{K}_{\alpha\alpha} \widetilde{K}_{\beta\beta}=-1$. This yields the RG flow equation
\begin{equation}
    \left((\vec{S}^{-1})^{\transpose{}}\frac{\diff \vec{U}}{\diff\ell} \vec{S}^{-1}\right)_{\alpha\beta} = u \sum_p y_p^2 \, \widetilde{\lambda}_{p,\alpha}  \widetilde{\lambda}_{p,\beta} \, \frac{1-\widetilde{K}_{\beta\beta}\widetilde{K}_{\alpha\alpha}}{2}
\end{equation}
or equivalently Eq.~\eqref{eq:FQH RG unify U}. Note that the condition for Lorentz invariance, Eq.~\eqref{eq:conf coupled wires}, is maintained under the flow.

\subsection{RG flow of the scaling dimension}

The RG flow equations of the scaling dimensions follow from Eq.~\eqref{eq:FQH RG unify U} and are
\begin{equation}
    \frac{\diff d_r}{\diff \ell} = \frac{1}{4}\sum_p y_p^2\left(
    (\vec{\lambda}_r^\transpose{}\vec{K}^{-1}\vec{\lambda}_p)^2
    -\bigl(\vec{\lambda}_r^\transpose{}(u\vec{U}^{-1})\vec{\lambda}_p\bigr)^2
    \right).
\end{equation}
Recalling the commutation relation
\begin{multline}
    \left[\partial_x(\vec{\lambda}_r\cdot\vec{\Phi})(x),\, (\vec{\lambda}_p\cdot\vec{\Phi})(y)\right] \\
    = 2\pi i (\vec{\lambda}_r^\transpose{} \vec{K}^{-1} \vec{\lambda}_p)\delta(x-y).
\end{multline}
we find that $d_r$ is decreasing along the RG flow if $\mathcal{O}_r$ commutes with all the perturbations, $\mathcal{O}_p$'s.

\section{Tracing out massive fields in the coupled-wires model} \label{apdx:coupled wires projection}

In this Appendix we discuss the low-energy effective description of a wire construction with several massive fields. We describe the procedure for tracing out the massive fields and obtaining the low-energy effective theory for the massless ones. Consider a model of real bosonic fields and their dual fields,
\begin{equation}
    \vec{\Phi} = 
        \left( \begin{array}{ccc|ccc}
        \varphi_1, & \ldots & \varphi_N & \theta_1, & \ldots & \theta_N
        \end{array}\right)^\transpose
\end{equation}
with commutation relations as in Eq.~\eqref{eq:K-matrix comm} with $K$-matrix
\begin{equation}
    \vec K =
        \left( \begin{array}{c|c}
            \vec{0}_N               & 2\cdot\mathbbm{1}_N \\
            \hline
            2\cdot\mathbbm{1}_N     & \vec{0}_N
        \end{array}\right).
\end{equation}
The Hamiltonian of the model is $H_0+H_M$, where $H_0$ is given in Eq.~\eqref{eq:H0 and Hpert} and 
\begin{equation}
    H_{M} = \int\diff{}x\,\frac{1}{4\pi} \vec{\Phi}^{\transpose} \vec{M} \vec{\Phi},
\end{equation}
where $\vec M$ a real symmetric positive semi-definite matrix in block form 
\begin{equation}\label{eq:mass form}
    \vec M = 
    \left( \begin{array}{c|c}
        \vec{m}     & \vec{0}_N \\
        \hline
        \vec{0}_N   & \vec{0}_N
    \end{array}\right).
\end{equation}
The Hamiltonian part $H_M$ can be thought of as a strong coupling description of commuting cosine terms where we expand the cosine terms to second order in their arguments. Condition \eqref{eq:mass form} assures that a set of commuting fields are massive. Note the fields and model used here are not necessarily in one-to-one correspondence to those in used the main text.

We assume that the model has Lorentz invariance with respect to the velocity $u$, i.e.,~that $\vec U$ satisfies condition \eqref{eq:conf coupled wires}. For simplicity we denote the $N\times N$ blocks of $\vec U$ by
\begin{equation}
    \vec U = 
    \left( \begin{array}{c|c}
        \vec{U}^{\varphi\varphi}    & \vec{U}^{\varphi\theta} \\
        \hline
        \vec{U}^{\theta\varphi}     & \vec{U}^{\theta\theta}
    \end{array} \right).
\end{equation}

\subsection{Exact diagonalization}

We start analyzing the model by reviewing its exact solution. The model is translation invariant so we can use Fourier transformed fields
\begin{equation}
    B_\alpha (p) = \int_{-\infty} ^\infty \diff x\ \Phi_\alpha(x)e^{-ipx}
\end{equation}
to write
\begin{equation}
    H=\int\frac{\diff p}{2\pi} \frac{1}{4\pi} \vec{B}^\dagger(p) \,(p^2 \vec{U}+\vec{M})\, \vec{B}(p).
\end{equation}
We can decouple the interaction between pairs of fields and their duals by a linear transformation
\begin{equation}
    \vec{B}(p)=\vec{S}(p)\, \widetilde{\vec{B}}(p)
\end{equation}
where $\vec{S}(p)$ is an $N\times N$ matrix with complex entries that is subject to the following conditions:
\renewcommand{\labelenumi}{\arabic{enumi})}
\begin{enumerate}
    \item the fields $\widetilde{B}_\alpha(p)$ correspond to a real field in real space. This means that they satisfy $\widetilde{B}_\alpha^\dagger(p)=\widetilde{B}_\alpha(-p)$, or equivalently
    \begin{equation}
        \vec{S}(p)=\vec{S}^*(-p).
    \end{equation}
    
    \item the transformation preserves the commutation relations, $[B_\alpha(p),B_\beta^\dagger(q)]=[\widetilde{B}_\alpha(p),\widetilde{B}_\beta^\dagger(q)]$, or equivalently
    \begin{equation}\label{eq:gen-commutation}
        \vec{K}^{-1}=\vec{S}(p) \vec{K}^{-1} \vec{S}^\dagger(p).
    \end{equation}
    
    \item in terms of $\widetilde{B}$-fields the Hamiltonian $H$ couples only $\widetilde{B}_\alpha$ to itself, its dual $\widetilde{B}_{\alpha+N\,\text{mod}\,2N}$ and their Hermitian conjugate, i.e.,\ by denoting
    \begin{equation} \label{eq:gen-diagonalization}
    \begin{aligned}
        \vec{D}(p) &\equiv \vec{S}^\dagger(p) (p^2\vec{U}+\vec{M})\vec{S}(p) \\
        & \equiv
        \left( \begin{array}{c|c}
        \vec{D}_1(p) & \vec{D}_2(p) \\
        \hline
        \vec{D}_3(p) & \vec{D}_4(p)
        \end{array} \right),
    \end{aligned}
    \end{equation}
    the $\vec{D}_i(p)$ are diagonal $N\times N$ matrices and the matrix $\vec{D}(p)$ is Hermitian. 
\end{enumerate}

For $p\neq0$ the matrix $\vec{S}(p)$ can always be chosen to be a real matrix satisfying these three conditions. The coupled quartet $\widetilde{B}_{\alpha}(\pm p)$, $\widetilde{B}_{\alpha+N}(\pm p)$ can be further decoupled to single particle eigenmodes, but in the following we will not need this further decoupling.

\subsection{Low-energy effective description}

Here we show that at small-momenta, the Hamiltonian should be described in a basis of fields for which the $\vec U$ term does not couple any non-massive fields $\varphi_j, \theta_j$ to highly fluctuating $\partial_x \theta_i$ that are conjugate to any massive $\varphi_i$ fields. 

At $p=0$, the kinetic term coefficients are those in the matrix $\widetilde{\vec{U}} = \vec{S}^\dagger(0)\vec{U}\vec{S}(0),$ and the mass term is written
\begin{equation}
    \widetilde{\vec{M}} = 
        \vec{S}^\dagger(0)\vec{M}\vec{S}(0) =
        \diag{
            \begin{pmatrix}
            m_1, & \ldots & m_n, & 0, & \ldots & 0
            \end{pmatrix}},
\end{equation}
i.e.,\ the fields labeled $1,\ldots n$ are massive, while the $n^c=N-n$  fields labeled $n+1,\ldots N$ are massless.

At small non-zero momenta, $p\neq0$, we expand 
\begin{equation}
    \vec S (p)\approx (\mathbbm{1}_{2N} + p^2 \vec{R}+O(\epsilon^4))\,\vec{S}(0),
\end{equation}
where the small dimensionless parameter is $\epsilon=p/(m_i u)$, with $m_i>0$ is the smallest mass term in $\widetilde{\vec{M}}$.

Due to Eq.~\eqref{eq:gen-commutation}, the matrix $\vec{R}$ satisfies 
\begin{equation}
    \vec{R}\vec{K}^{-1}+\vec{K}^{-1}\vec{R}^\dagger=\vec{0}_{2N},
\end{equation}
and together with Eq.~\eqref{eq:gen-diagonalization} this yields
\begin{equation}
    \left.\frac{\diff \vec{D}}{\diff (p^2)}\right\vert_{p=0} = \widetilde{\vec{U}} +\vec{K}[\vec{K}^{-1}\widetilde{\vec{M}}\,,\ \vec{R}].
\end{equation}
Together with Eq.~\eqref{eq:gen-diagonalization} this implies  
\begin{equation}\label{eq:U in diag basis}
    \widetilde{U}^{\theta\varphi}_{jk} = 0 
    \quad\text{and}\quad 
    \widetilde{U}^{\theta\theta}_{jk} = 0
\end{equation}
if $j\leq n$ and $k>n$. In terms of the fields 
\begin{equation}
    \widetilde{\varphi}_i \equiv (\vec S(0)^{-1} \vec{\Phi})_i,
    \  
    \widetilde{\theta}_i \equiv (\vec S(0)^{-1} \vec{\Phi})_{i+N},\quad i\leq N, 
\end{equation}
Eq.~\eqref{eq:U in diag basis} shows that $H_0$ does not couple the duals of the massive fields $\widetilde{\theta}_j$ to any of the massless fields, $\widetilde{\varphi}_k$ and $\widetilde{\theta}_k$.

At high momentum the full $p$-dependence of $\vec{S}(p)$ is needed to accurately describe the dynamics. However, at low-energies we can use the effective description by projecting-out the massive fields and their duals, leaving the low-energy modes $\widetilde{\varphi}_k$ and $\widetilde{\theta}_k$ for $k>n$ with the effective Hamiltonian
\begin{equation}\label{eq:effective H apdx}
    H^\text{eff} = \int\diff{}x\,\frac{1}{4\pi} \
        \sum_{j,k>n}\partial_x\widetilde{\Phi}_j\,
        \widetilde{U}_{jk}\,
        \partial_x\widetilde{\Phi}_k.
\end{equation} 

In particular, if the original kinetic term $H_0$ satisfies the conformal condition \eqref{eq:conf coupled wires}, then the transformation $\vec{S}(0)$ takes the kinetic matrix to the block form
\begin{equation}\label{eq:conforml Utilde}
    \widetilde{\vec U} = 
        \left( \begin{array}{cc|cc}
            \elemdontcare & \vec{0} &  \elemdontcare &  \vec{0}\\
            \vec{0}& \vec{V}^{\varphi\varphi} &  \vec{0}& \vec{V}^{\varphi\theta} \\
            \hline
            \elemdontcare & \vec{0} & \elemdontcare & \vec{0} \\
            \vec{0}& \vec{V}^{\theta\varphi} & \vec{0} & \vec{V}^{\theta\theta}\\
        \end{array} \right),
\end{equation}
where $\elemdontcare$ are $n\times n$ blocks of the matrix that will drop after we trace out the fields $\varphi_j,\theta_j$ with $j\leq n$. As a consequence, the decoupling of the massive and massless degrees of freedom is exact, and the effective model \eqref{eq:effective H apdx} satisfies condition \eqref{eq:conf coupled wires} for Lorentz invariance
\begin{equation}
    \left(
    \begin{pmatrix}
        \vec{0}_{n^c}                 & 2\cdot\mathbbm{1}_{n^c} \\
        2\cdot\mathbbm{1}_{n^c}  & \vec{0}_{n^c}
    \end{pmatrix}^{-1}
    \begin{pmatrix}
        \vec{V}^{\varphi\varphi}    & \vec{V}^{\varphi\theta} \\
        \vec{V}^{\theta\varphi}     & \vec{V}^{\theta\theta}
    \end{pmatrix}
    \right)^2
    =
    u^2 \mathbbm{1}_{2n^c}.
\end{equation}

\subsection{Tracing out a single massive field}

We can give a closed form for the case of a single massive field 
\begin{equation}
    \vec{M} = \diag{
        \begin{pmatrix}
            m, & 0, & \ldots & 0
        \end{pmatrix}}.
\end{equation}
This simple case is used in the main text twice to trace out two massive fields and obtain the effective model \eqref{eq:eff strong cp lim Hamiltonian}.

The appropriate $\vec{S}(0)$ transformation, c.f.~Eqs.~\eqref{eq:gen-commutation} and \eqref{eq:conforml Utilde}, is given by
\begin{equation}
\begin{aligned}
    \varphi_1 &= \widetilde{\varphi}_1, \\
    \theta_1 &= \widetilde{\theta}_1 - \Big( \textstyle{\sum_{k>1}} U^{\varphi\theta}_{k,1} U^{\theta\theta}_{k,1} \Bigr) \widetilde{\varphi}_1 / \bigl( U^{\theta\theta}_{1,1} \bigr)^2 \\ 
    & \hphantom{=}-\textstyle{\sum_{k>1}} \left(\bigl( U^{\varphi\theta}_{k,1}/U^{\theta\theta}_{1,1}\bigr) \widetilde{\varphi}_k
     +\bigl(U^{\theta\theta}_{k,1}/U^{\theta\theta}_{1,1}\bigr) \widetilde{\theta}_k\right),\\
    \varphi_j &= \widetilde{\varphi}_j + \bigl(U^{\theta\theta}_{j,1}/U^{\theta\theta}_{1,1}\bigr) \widetilde{\varphi}_1, \\
    \theta_j &= \widetilde{\theta}_j + \bigl(U^{\varphi\theta}_{j,1}/U^{\theta\theta}_{1,1}\bigr) \widetilde{\varphi}_1,
\end{aligned}
\end{equation}
for $j>1$, and the kinetic matrix of the effective model is given by
\begin{equation}
    \widetilde{U}^{\alpha\beta}_{j,k}=U^{\alpha\beta}_{j,k} - U^{\alpha\theta}_{j,1} \,U^{\theta\beta}_{1,k} / U^{\theta\theta}_{1,1}\quad \text{for }j,k>1,
\end{equation}
where $\alpha,\beta\in\{\varphi,\theta\}$.

We can treat a class of perturbations of the original Hamiltonian, $H_0+H_M$, as perturbations of this effective Hamiltonian $H^\text{eff}$ [see Eq.~\eqref{eq:effective H apdx}]. Consider the perturbation
\begin{equation}
    H_\text{pert} = \int\diff{}x\ \frac{u y}{\pi a^2} \cos(\vec{\lambda}\cdot \vec{\Phi}),
\end{equation}
with zero conformal spin, $\vec{\lambda}^\transpose\vec{K}^{-1}\vec{\lambda}/2=0$,  that commutes with the massive degree of freedom $\varphi_1$ and its dual $\theta_1$, and $\vec{\lambda}\cdot \vec{\Phi}$ is not proportional to $\varphi_1$. These restrictions allows us to choose a basis $\vec\Phi$ such that the second bosonic field is $\varphi_2 = \vec{\lambda}\cdot \vec{\Phi}$.  
For a small $y$ the effective model can be approximated by $H^\text{eff}+H^\text{eff}_\text{pert}$ with the effective perturbation
\begin{equation}
    H^\text{eff}_\text{pert} = \int\diff{}x\ \frac{u y}{\pi a^2}  \Bigl\langle e^{i (U^{\theta\theta}_{2,1}/U^{\theta\theta}_{1,1})\widetilde{\varphi}_1}\Bigr\rangle_1 e^{i\widetilde{\varphi}_2} + \text{h.c.},
\end{equation}
where $\langle\cdot\rangle_1$ denotes expectation value taken over only the sector of the fields $\widetilde{\varphi}_1,\widetilde{\theta}_1$. The perturbation $e^{i\widetilde{\varphi}_2}$ has the scaling dimension
\begin{equation}
    d^\text{eff}=d-(U^{\theta\theta}_{2,1})^2/(2u\, U^{\theta\theta}_{1,1})
\end{equation}
where $d$ is the scaling dimension of the original perturbation, $\cos(\vec{\lambda}\cdot\vec{\Phi})$ [see Eq.~\eqref{eq:scaling dims}]. In particular the perturbation becomes more relevant after tracing out the massive field $\varphi_1$, $d^\text{eff}<d$.

\section{Relevant operators in the strong proximity coupling regime} \label{apdx: rel ops in strong pairing}

Here we show that the only relevant operators that agrees with the strong proximity coupling regime of the fractional $\nu=1/3$ wire model are those discussed in the main paper. Namely, the terms labeled by $\Delta_1$, $\Delta_2$, $\Delta_B$, $J_g$ and the identity are the only terms that conserve momentum, commute with the $\Delta_B$ term and are relevant for repulsive Luttinger parameters 
\begin{equation}\label{eq:repulsive apdx D}
    K_e,K_b\leq 1
\end{equation}
of the fixed point \eqref{eq:3_LuttLiquids}. The different cosine terms are defined in Eqs.~
\eqref{eq:H_pairing}, \eqref{eq:DB_term} and \eqref{eq:strong pair gapping term}. 

A general local interaction term takes the form 
\begin{multline}
	\mathcal{O} = \exp\bigl[i( 3 n_1 \eta_{R,-1}^{A} + n_2 \phi_{L,0}^{A} + n_3 \phi_{R,0}^{A} \\
	+ n_4 \phi_{L,0}^{B} + n_5 \phi_{R,0}^{B} + 3 n_6 \eta_{L,1}^{B})\bigr]
\end{multline}
where $n_j$ are integers.

Requiring that this operator is momentum conserving and recalling the bosonization identity
\begin{equation}
    \psi_{j}(x)\sim e^{i(b j + k_{F})x} e^{+i\phi_{R,j}(x)} + e^{i(b j - k_{F})x} e^{-i\phi_{L,j}(x)},
\end{equation}
implies that the term $\mathcal{O}$ should naturally appear with an oscillating coefficient $e^{ikx}$, where
\begin{equation}
    k = (3n_{1}+n_{2}+n_{3}+n_{4}+n_{5}+3n_{6})k_{F}-(n_{1}+n_{6})b.
\end{equation}
Together with the filling fraction being $2k_F/b=1/3$, and requiring $k=0$, we find $n_2+n_3+n_4+n_5 = 3(n_1+n_6)$. Requiring that $\mathcal{O}$ commutes with the  $\Delta_B$ term, implies $n_2+n_5 = n_3+n_4$.

We denote by $A,B,C,D$ the following integers
\begin{equation}
\begin{gathered}
    6D = 2(n_2+n_5) = 2(n_3+n_4) = 3(n_1+n_6), \\
    2A+3D = n_4-n_3, \\
    2B+3D = n_2-n_5, \\
    2C-2D = n_6-n_1.
\end{gathered}
\end{equation}
The scaling dimension of $\mathcal{O}$ is
\begin{widetext}
\begin{equation}
\begin{aligned}
    d & 
     =\frac{3K_{e}}{4}D^{2}+\frac{1}{12K_{e}}(4A+2B+9D)^{2} +\frac{K_{b}}{12}\left[(A+2B-3C+12D)^{2}+(A+2B-3C+3D)^{2}\right] \\
     &\hphantom{=}+\frac{1}{12K_{b}}\left[(A+2B+3C)^{2}+(A+2B+3C+3D)^{2}\right]
\end{aligned}
\end{equation}
\end{widetext}
Below we will find the conditions on $A,B,C,D$ for which the operator is relevant and thus $d<2$. 

A useful inequality is the following: if $a,b,K>0$ then 
\begin{equation}\label{eq:apdx D useful}
    a K + b/K \geq 2 \sqrt{ab}.
\end{equation}
This inequality yields $d\geq(3D/2)^2$, so for the operator $\mathcal{O}$ to be relevant we must have $D=0$.

We continue by separating to cases:
\begin{itemize}
    \item For $A+2B=3C$ the scaling dimension is 
        \begin{equation}
            d=6C^2/K_b+3(2C-B)^2/K_e,
        \end{equation}
        so repulsion condition \eqref{eq:repulsive apdx D} only admits the coefficients $A=B=C=0$, i.e.,~the identity operator.

    \item For $A+2B=-3C$ the scaling dimension is  
        \begin{equation}
            d=3(B+2C)^{2}/K_{e}+6C^{2}K_b.
        \end{equation}
        The repulsive condition $K_e\leq 1$ limits the  coefficients to $B=-2C=-2A$, or alternatively $\mathcal{O}=e^{iC(6\varphi_3+2\varphi_1-4\varphi_2)}$. This operator is the $C$-th power of the $J_g$ term in Eq.~\eqref{eq:strong pair gapping term}.

    \item For $|A+2B|\geq3|C|+1$ the inequality \eqref{eq:apdx D useful} yields a bound 
        \begin{equation}
            d \geq 2|C|+1/3.
        \end{equation}
        This implies $C=0$, which in turn, together with the inequality \eqref{eq:apdx D useful} and $K_e\leq 1$, yields
        \begin{equation}
            d \geq (2A+B)^{2}/3 + (A+2B)^{2}/3.
        \end{equation}
        This last bound admits only the four possible terms: $A=B=0$ (the identity), $A=\pm1,B=0$ [$\Delta_1$ term in Eq.~\eqref{eq:pairing density}], $A=0,B=\pm1$ [$\Delta_2$ term in Eq.~\eqref{eq:pairing density}] or $A=-B=\pm1$ [$\Delta_B$ term in Eq.~\eqref{eq:DB_term}].

    \item For $|A+2B|\leq3|C|-1$ the inequality \eqref{eq:apdx D useful} yields the bounds 
        \begin{equation}
        \begin{aligned}
            d   & \geq\left[(3C)^{2}-(A+2B)^{2}\right]/3 \\
                & \geq 2|C|-1/3.
        \end{aligned}
        \end{equation}
        Thus, $\mathcal{O}$ is relevant only if $|C|=1$ and $|A+2B|=2$. Next, the inequalities \eqref{eq:apdx D useful} and $K_e\leq 1$ yield $d\geq(2A+B)^2/3+5/3$. This last bound implies $A=0$ and $|B|=1$. These values of the coefficients together with \eqref{eq:repulsive apdx D} admit no relevant terms (curiously, the $B=-C=\pm1$ case yields a marginal operator, but only for $K_e=1$ and $K_b=1/5$).
\end{itemize}

\section{Cosine terms in different field bases} \label{apdx:basis-dictionary}

Here we list the different forms of the different cosine terms encountered in the main text in terms of the field bases introduced. The different terms are denoted by their coefficients, $J_\alpha$ and $\Delta_\beta$ with $\alpha=A,B,2,g$ and $\beta=1,2,B,g$ [see Eqs.~\eqref{eq:H_QH density}, \eqref{eq;J2 term}, \eqref{eq:strong pair gapping term}, \eqref{eq:H_pairing},  \eqref{eq:DB_term} and \eqref{eq:weak pairing gapping term}]. The listing is included in Table \ref{table:2}.

\begin{table*}[!ht]
\centering  
\begin{tabular}{| c |  c  c  c  c  c |} 
    \hline
    Term & $\vec{\Phi}$ &  & $\varphi_\alpha,\,\theta_\alpha$,  $\alpha=A,B,e$ &  & $\varphi_\alpha,\,\theta_\alpha$,   $\alpha=1,2,3$ \\ [0.5ex]
    \hline 
    
    $J_A$ & $3\eta_{R,-1}^A+2\phi_{L,0}^A+\phi_{R,0}^A$ & = & $6\theta_A$ & = &  $3(\theta_3+\varphi_3)+2(\theta_2-\varphi_2)+(\theta_1+\varphi_1)$ \\ [0.2ex]
    
    $J_B$ & $3\eta_{L,1}^B+2\phi_{R,0}^B+\phi_{L,0}^B$ & = & $6\theta_B$ & = &  $3(\theta_3-\varphi_3)+2(\theta_2+\varphi_2)+(\theta_1-\varphi_1)$ \\ [0.2ex]
    
    $\Delta_1$ & $\phi_{R,0}^A-\phi_{L,0}^B$ & = & $4\varphi_e+\varphi_A-\theta_A+\varphi_B+\theta_B$ & = &  $2\varphi_1$ \\ [0.5ex]
    
    $\Delta_2$ & $\phi_{R,0}^B-\phi_{L,0}^A$ & = & $2(\varphi_e+\varphi_A-\theta_A+\varphi_B+\theta_B)$ & = &  $2\varphi_2$ \\ [0.5ex]
    
    $\Delta_B$ & $\phi_{R,0}^B+\phi_{L,0}^B-\phi_{R,0}^A-\phi_{L,0}^A$ & = & $-2\varphi_e+\varphi_A-\theta_A+\varphi_B+\theta_B$ & = &  $2\varphi_2-2\varphi_1$ \\ [0.5ex]
    
    $\Delta_g$ & $2\phi_{R,0}^A+\phi_{L,0}^A-2\phi_{L,0}^B-\phi_{R,0}^B$ & = & $6\varphi_e$ & = &  $4\varphi_1-2\varphi_2$ \\ [0.5ex]
    
    $J_2$ & $3\eta_{R,-1}^A+2\phi_{L,0}^A+\phi_{R,0}^A+\phi_{L,0}^B+2\phi_{R,0}^B+3\eta_{L,1}^B$ & = & $6\theta_A+6\theta_B$ & = &  $6\theta_3+4\theta_2+2\theta_1$ \\ [0.5ex]
    
    $J_g$ & $3\eta_{R,-1}^A+2\phi_{L,0}^A+\phi_{R,0}^A-\phi_{L,0}^B-2\phi_{R,0}^B-3\eta_{L,1}^B$ & = & $6\theta_A-6\theta_B$ & = &  $6\varphi_3-4\varphi_2+2\varphi_1$ \\ [0.5ex]
    
    \hline 
\end{tabular}
\caption{
    Table summarizing the different cosine terms mentioned throughout the paper in different bases of fields for the case of FQH at filling fraction $\nu=1/3$. The left column specifies the term by its coefficient, while the rest are present the argument of the corresponding cosine term in a specific basis. The different basis definition can be found in Table \ref{table:1}.}
\label{table:2}
\end{table*}

\section{Examples of RG flow of proximity coupled fractional quantum Hall edges}\label{apdx:RG flow examples}

\begin{figure*}[!ht]
    \subfloat[][]{
        \includegraphics[scale=0.66]
            {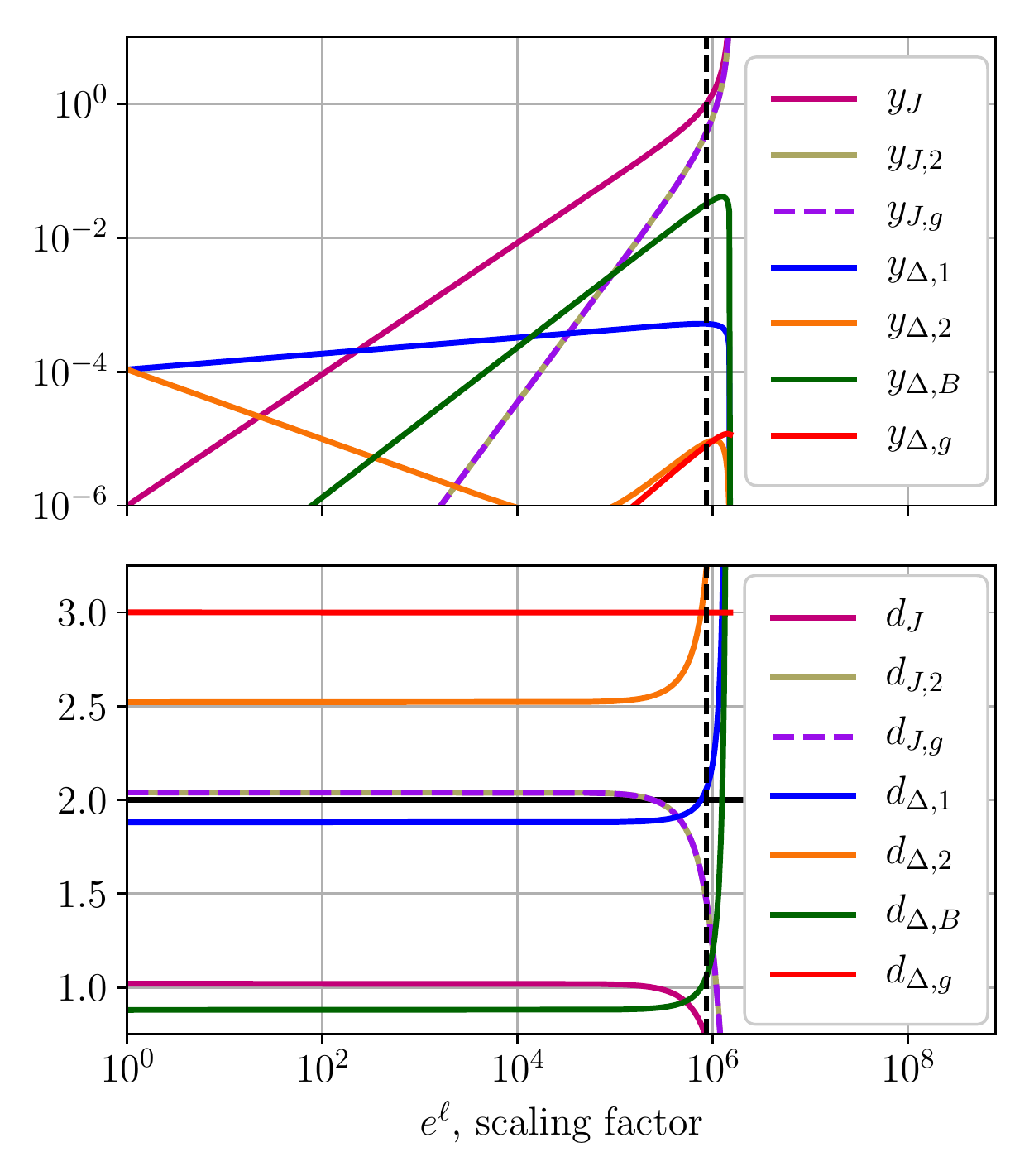}
        \label{subfig:weak cp flow gapless}
    }
    \hfill
    \subfloat[][]{
        \includegraphics[scale=0.66]
            {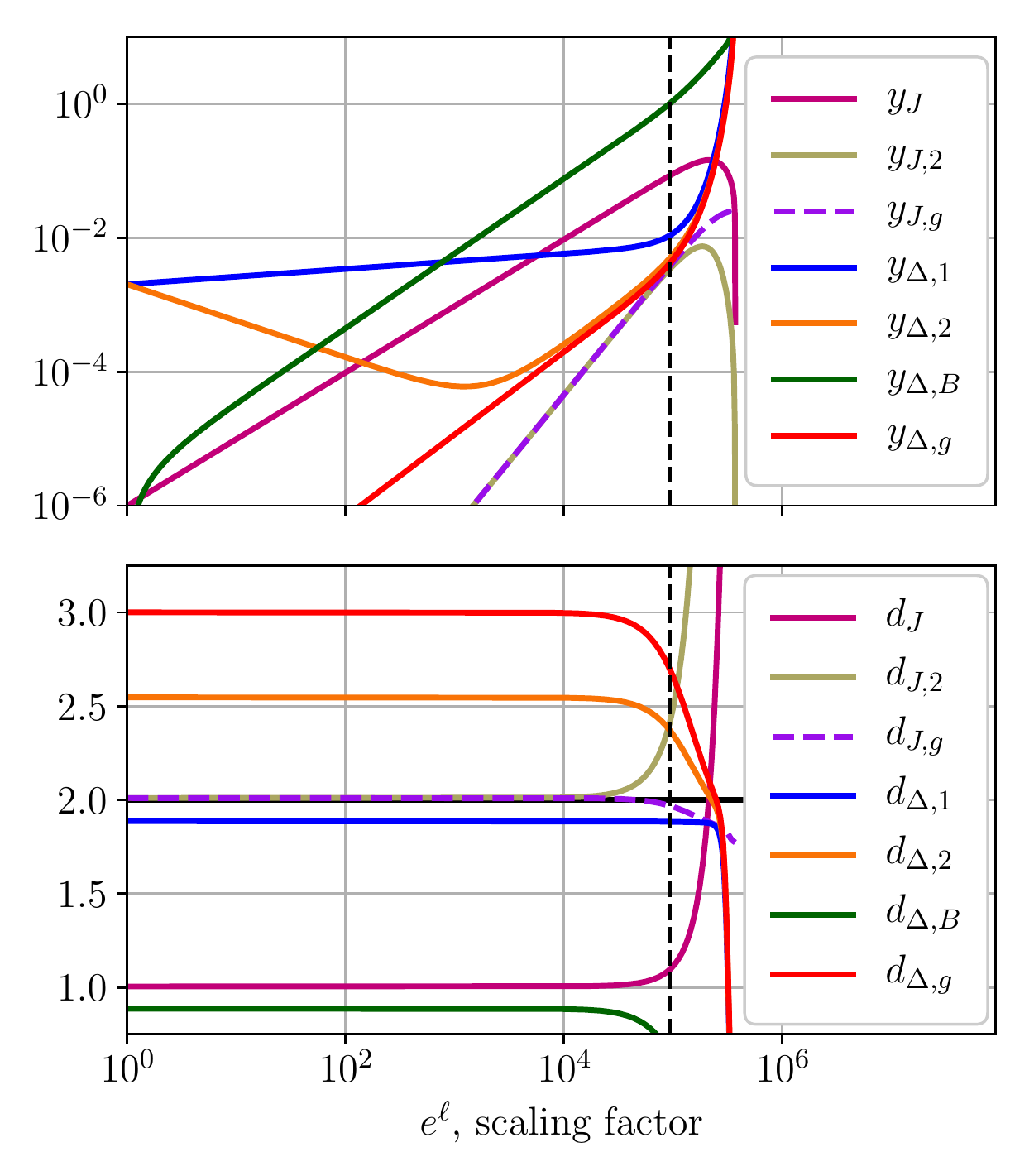}
        \label{subfig:strong cp flow gapped}
    }
    \caption{
        \label{fig:apdx-flow-plots}
         RG flow solution, the dashed lines correspond to the stopping condition of the flow, $|y_r(\ell^*)|=1$. \protect\subref{subfig:weak cp flow gapless} corresponds to the square marker in Fig.~\ref{subfig:yD-Kb} and depicts a flow of gapless phase in the weak proximity coupling regime. \protect\subref{subfig:strong cp flow gapped} corresponds to the star marker in Fig.~\ref{subfig:yD-Kb} and depicts a flow of gapped phase in the strong proximity coupling regime.
    }
\end{figure*}

This appendix includes several plots of the RG flow Equations \eqref{eq:FQH RG unify} for Hamiltonian density $\mathcal{H}_0 + \mathcal{H}_\text{pert}$ where $\mathcal{H}_0$ and $\mathcal{H}_\text{pert}$ are \eqref{eq:model_fixed_point} and \eqref{eq:Hpert} with the terms with dimensionless couplings $y_{J,\alpha}$ and $y_{\Delta,\beta}$ where $\alpha=A,B,2,g$ and $\beta=1,2,B,g$ [given in Eqs.~\eqref{eq:H_QH density}, \eqref{eq;J2 term}, \eqref{eq:strong pair gapping term}, \eqref{eq:H_pairing},  \eqref{eq:DB_term} and \eqref{eq:weak pairing gapping term}]. The plots are in shown Fig.~\ref{fig:apdx-flow-plots}.

\end{document}